\documentclass[11pt, a4paper]{article}
\pdfoutput=1
\usepackage{jheppub}
\usepackage{latexsym, graphicx}
\usepackage{amsmath, amssymb}
\usepackage{color}

\title{Entanglement Dynamics of Detectors in an Einstein Cylinder}

\author[a,b,c]{Shih-Yuin Lin}
\author[d]{Chung-Hsien Chou}
\author[e]{B. L. Hu}

\affiliation[a]{Department of Physics, National Changhua University of Education, Changhua 50007, Taiwan}

\affiliation[b]{Department of Physics and Astronomy, University of Waterloo, Waterloo, Ontario N2L 3G1, Canada}

\affiliation[c]{Perimeter Institute for Theoretical Physics, Waterloo, Ontario N2L 2Y5, Canada}

\affiliation[d]{Department of Physics, National Cheng Kung University, Tainan 70101, Taiwan}

\affiliation[e]{Maryland Center for Fundamental Physics and Joint Quantum Institute,
University of Maryland, College Park, Maryland 20742-4111, USA}

\emailAdd{sylin@cc.ncue.edu.tw}
\emailAdd{chouch@mail.ncku.edu.tw}
\emailAdd{blhu@umd.edu}

\abstract{We investigate how nontrivial topology affects the entanglement dynamics between a detector and a quantum field and between two detectors mediated by a quantum field. Nontrivial topology refers to both that of the \textit{base space} and that of the \textit{bundle}.
Using a derivative-coupling Unruh-DeWitt-like detector model interacting with a quantum scalar field in an Einstein cylinder
${\bf S}^1$ (space) $ \times {\bf R}_1$ (time), we see the beating behaviors in the dynamics of the detector-field entanglement
and the detector-detector entanglement, which distinguish from the results in the non-compact (1+1) dimensional Minkowski space.
The beat patterns of entanglement dynamics in a normal %an untwisted  %%%1/17
and a twisted field with the same parameter values are different  because
of the difference in the spectrum of the %eigen-modes  %%% 9/19
field modes. In terms of the kinetic momentum of the detectors, we find that the contribution
by the zero mode in a normal field to entanglement dynamics has no qualitative difference from those by the nonzero modes.}

\arxivnumber{1508.06221}

\begin{document}

\maketitle

\section{Introduction}
%%%%%%%%%%%%%%%%%%%%%%%%%%%%%%%%%%%%%%%%%%%%%%%%%%%%%%%%%%%%%%%%%%%%%%%%%%%%%%%%%%%%%%%%%%%%%%%%%%%%%%%%

The last decade has seen a rapid increase of research in relativistic quantum information (RQI) (see, e.g., \cite{ISRQI}),
which studies the relativistic features of quantum information with ``relativistic" referring to the
\textit{localized quantum objects in relativistic motion}, the \textit{relativistic nature of quantum fields} and the
\textit{properties of spacetime structures}.
In particular, the third aspect concerning spacetime properties ranges from the rudimentary yet necessary, such as time-dilation and hypersurface-slicing considerations \cite{LCH08, %LH09,
LCH15} at the level of special relativity, to spacetime curvature and topology effects \cite{La05, ZBLH13} often considered in the context of general relativity.
The present work will focus on how nontrivial topology affects the entanglement dynamics of a system of $N$ ($=1$ or $2$ here)
detectors, which are localized atom-like quantum objects with internal degrees of freedom
sensitive to the field they interact with.
Here, nontrivial topology refers to the \textit{base space}, turning a flat, noncompact two dimensional Minkowski spacetime
${\bf R}^1_1 = {\bf R}^1$ (space) $\times {\bf R}_1$ (time) into a spatially-compact  ${\bf S}^1$ (space) $\times {\bf R}_1$ (time), and also
the \textit{bundle}, whose fibre is the range of the field and a field configuration is a cross section -- we are interested in seeing
how entanglement dynamics for a {\it twisted field} is different from a normal (untwisted) field.

Before proceeding further we hasten to point out that although the mathematical nature of these investigations is obvious, they are physically relevant to actual experimental setups:  boundary effects pertain to the effects of a mirror or a dielectric slab \cite{Bo03}
on entanglement, while a ${\bf S}^1$ spatial topology refers to a toroidal cavity \cite{toroidal}, both are obviously essential components
in  quantum experiments.
On the theoretical side, quantum field theory in spacetimes with boundaries or with nontrivial topology was investigated in the late 70s by Isham, Dowker, DeWitt, and co-workers \cite{Is78, DB78, DHI79}. The prototype spacetime ${\bf S}^1 \times {\bf R}_1$ was referred to as the Einstein cylinder (presumably inspired by the ``Einstein Universe" with topology ${\bf S}^3 \times {\bf R}_1$, a static closed Robertson-Walker Universe).  The new aspect in our present investigation is entanglement dynamics in this topologically nontrivial spacetime \cite{Edu15}, for both untwisted and twisted quantum fields \footnote{The effect of the topology of the {\it state-vector space} on the decoherence and entanglement processes has been investigated for some time. See, e.g., \cite{Karol, Joynt}}.

Let us ponder upon how the overall features of entanglement dynamics between detectors and fields or between two detectors in a common field would be modified in a circle (e.g., a narrow micro-toroidal cavity) as compared to a straight line, as investigated before in
e.g., \cite{LH09}.

\paragraph{Topology of the base space} In ${\bf R}^1$ space, dynamics of the detector-field  or detector-detector entanglement shows no periodicity at large time scales, we expect entanglement would just grow or decay and saturate.
In ${\bf S}^1$, the space is finite and compact, so the retarded field emitted by the detector will after one period around the circle return to affect the detector and do so periodically. In a fully enclosed space there is no true dissipation, only apparent one on a time scale short compared to the recurrence time, as well illustrated in the Kac ring \cite{Kac}, a 1D closed harmonic chain, but the energy apparently ``lost" to the rest of the chain will be replenished after one period
\footnote{Entanglement propagation in a quantum Kac ring has been studied by \cite{RingEnt}.}.
This of course is the advantage for information processing using micro-toroidal cavities (the space in realistic situations is of course not fully enclosed, as there will be input and output laser activities).
For quantum systems the time lag between successive cycles generates interference which shows up as beats.
Unlike in ${\bf R}^1$, where equilibrium or steady state can be found to exist at late times, in {${\bf S}^1$}, beating in entanglement distinguishes its behavior.

%[ In a fancier language, this is because the zero mode is not a condensation point
%in the spectrum of the eigen-mode of the interacting detector \cite{Po11}. ]

\paragraph{Topology of the bundle} Here the nontrivial topology exists in twisted fields, thus the comparison we aim at is between the entanglement dynamics of normal (untwisted) and twisted fields.
In ${\bf S}^1$, the untwisted and twisted fields possesses different eigen-modes, and hence different beat patterns. The normal field possesses a zero mode whereas for twisted fields there is no zero-mode.  Note that the existence of the zero mode is a topological, not a geometric effect, in the sense that for normal fields different sizes of ${\bf S}^1$ as in a toroidal cavity will all possess a zero mode. Likewise there is no zero mode for twisted fields for all sizes of ${\bf S}^1$ cavity.

We mention two related work \cite{La05, Edu15} which uses the time-dependent perturbation theory with a switching function type of interaction to study the finite-time response of a single 2-level Unruh-DeWitt detector and the quantum entanglement of two
such detectors \cite{Edu15} in locally flat (3+1) dimensional spacetimes with different topologies.
The lowest-order results do depend on the field spectrum, which in turn depend on the spatial topology, though the
differences in different topologies are small. However, their lowest-order results  do not show beating behavior. The main reason
is that the influences by the echoes of a detector itself, as well as the retarded influences by the other detectors separated
at a distance, arise from the higher order terms in the coupling expansion.

Below, we use a nonperturbative method to study the nonequilibrium dynamics of a harmonic-oscillator detector-field system
with all the mutual influences included in our consideration, in order to capture the full extent of the topological effect in entanglement dynamics.
The setup in our study is introduced in Section 2. We consider a Unruh-DeWitt-like (UD') detector theory with derivative-coupling
 \cite{Unr76, DeW79, UZ89, RSG91, RHA96, RH00} because a usual detector minimally coupled to a scalar field in (1+1) dimensional Minkowski space has undesirable non-positive-definite energy and higher-derivative radiation-reaction.
%and the infrared ambiguity of the Wightman functions in perturbation theory \cite{Lo14}
\footnote{
An additional advantage when using time-dependent perturbation theory is that the infrared behavior of the response function of a detector with derivative-coupling is better than that with minimal-coupling in (1+1) dimensions, see \cite{Lo14}.}
In Section 3 we consider one detector interacting with a quantum field in an Einstein cylinder and  compare its entanglement behavior with earlier results obtained for Minkowski space ${\bf R}^1_1$. In Section 4 we consider the entanglement dynamics of two identical UD' detectors in the presence of a quantum field in the same ${\bf S}^1\times {\bf R}_1$ spacetime for both normal and twisted fields. Our findings are summarized in Section 5.  More details about the two-point correlators of a UD' detector in ${\bf R}^1_1$ are presented in Appendix A.
In our numerical results, we set $c=\hbar=1$.

%%%%%%%%%%%%%%%%%%%%%%%%%%%%%%%%%%%%%%%%%%%%%%%%%%%%%%%%%%%%%%%%%%%%%%%%%%%%%%%%%%%%%%%%%%%%%%%%%%%%%%%
\section{Model}
\label{themodel}

\begin{figure}
\includegraphics[width=3cm]{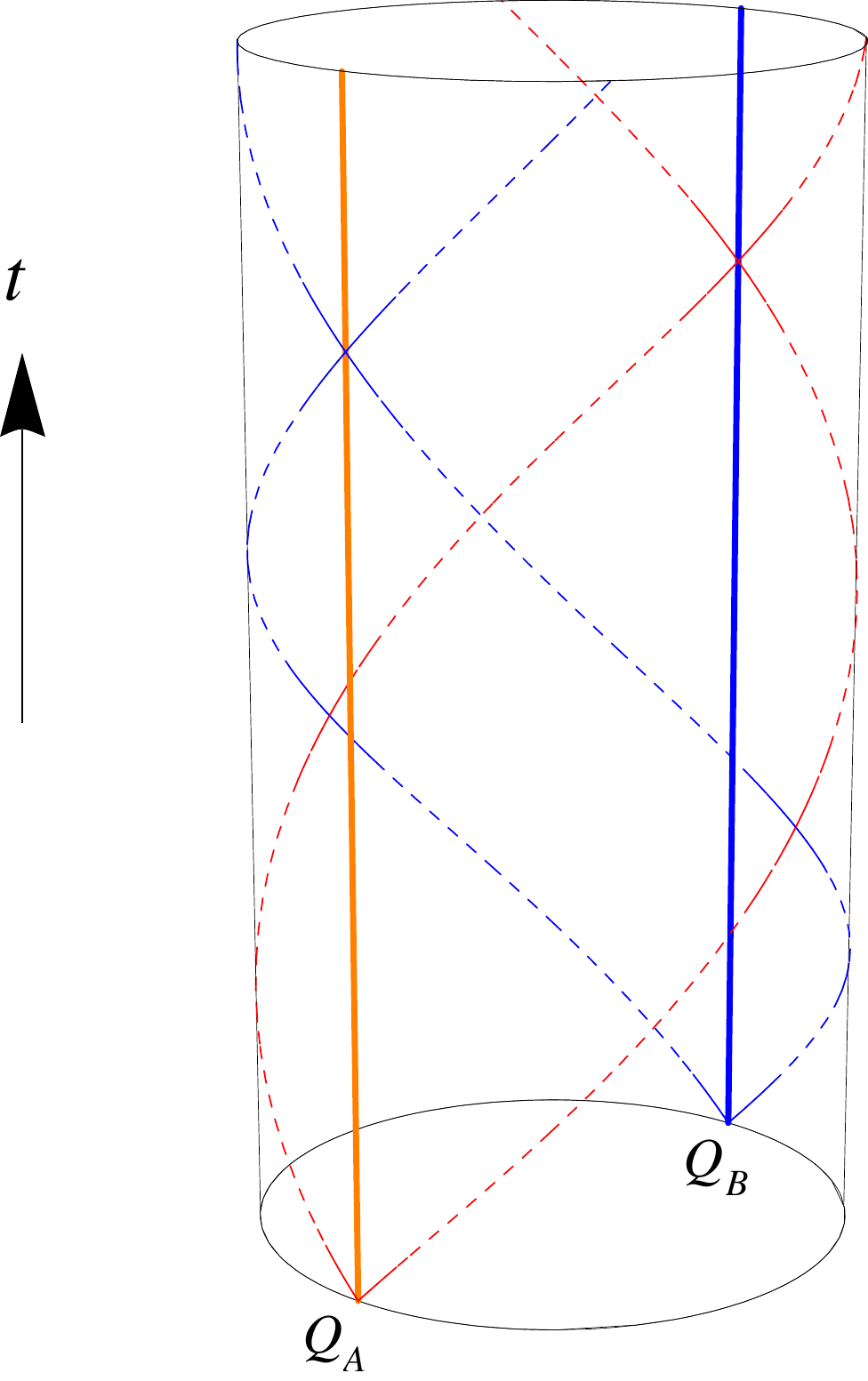}\hspace{.5cm}
\includegraphics[width=5.5cm]{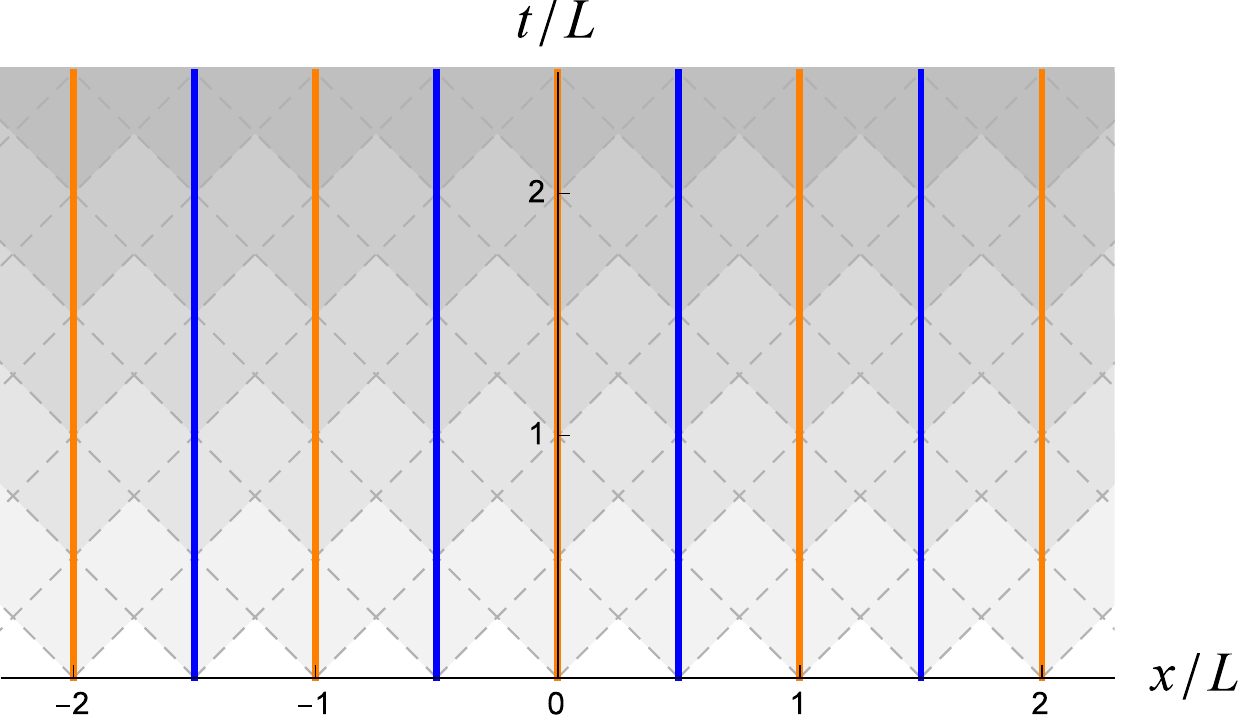}
\includegraphics[width=5.5cm]{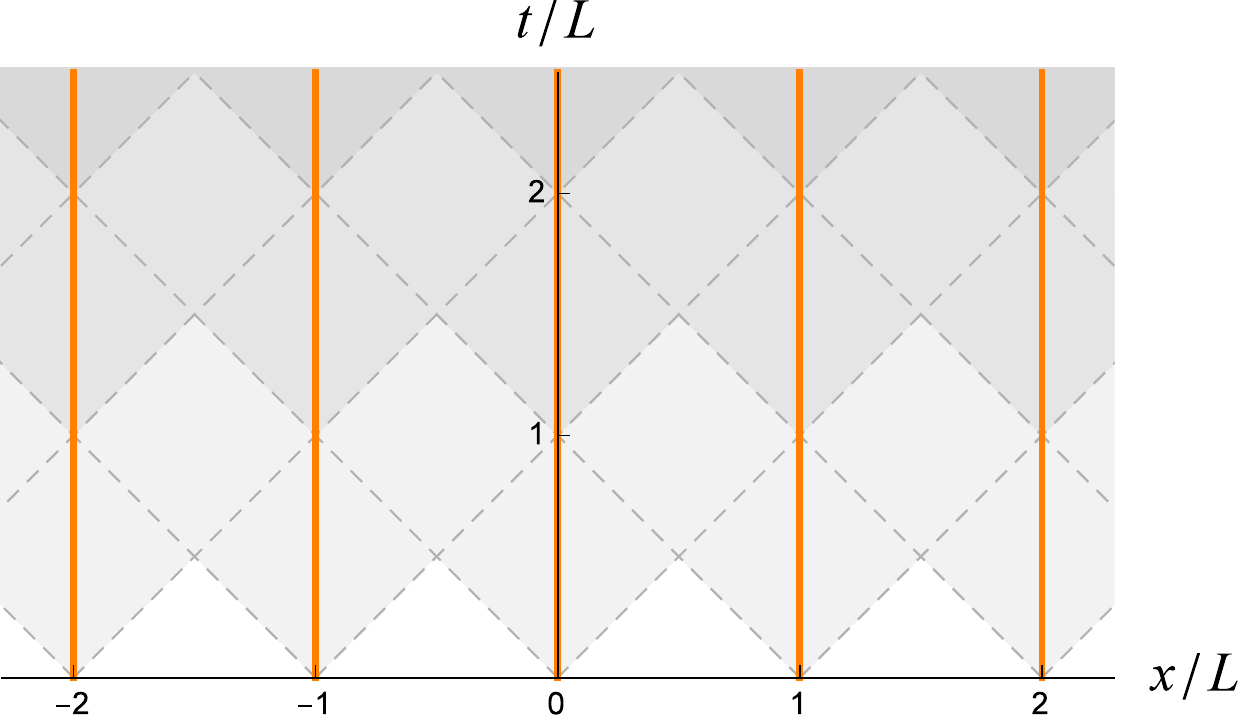}
\caption{(Left) The setup. The harmonic oscillators $Q_A$ and $Q_B$ are the internal degrees of freedom of detectors $A$ and $B$ located at $x=0$ and $x=L/2 \equiv R\pi$, respectively.  At $t=0$ we let the UD' detectors  begin to couple with the field. The orange and blue solid lines are the worldlines of the two detectors, and the dashed lines represent the fronts of the retarded influences sourced by the detectors after the interaction is switched on.
(Middle) An equivalent setup in the extended coordinates in space ${\bf R}^1$ by identifying the points $x$ to $x$ mod $L$ . (Right) Similar diagram for the cases with a single detector $Q_A$ only.}
\label{setup}
\end{figure}

Consider a (1+1) dimensional flat spacetime with topology ${\bf S}^1$ in space and ${\bf R}_1$ in time, namely, the Einstein cylinder.
The metric is given by
\begin{equation}
  ds^2 = -dt^2 + dx^2, \label{S1R1}
\end{equation}
where $x=R\varphi$ with radius $R$ a positive real constant and the azimuthal angle $\varphi \in (-\pi, \pi]$. The circumference of the circle is  thus $L\equiv 2\pi R$.  We will also refer to the extended space  $x, x' \in {\bf R}^1$ obtained by identifying the points $x$ to $x$ mod $L$ (see Figure \ref{setup}).
Consider placing a finite number of the derivative-coupling Unruh-DeWitt-like (UD') \cite{Unr76,DeW79, UZ89, RSG91, RH00} detectors
with the internal harmonic oscillators $Q_{\bf d}$ coupled to a common massless scalar field $\Phi$ in the above spacetime,
described by the action
\begin{equation}
  S = -{1\over 2}\int d^2 x \partial_{\alpha} \Phi \partial^{\alpha} \Phi + \sum_{\bf d} {1\over 2} \int d\tau^{}_{\bf d}
	  \left[ (\partial^{}_{\bf d} Q^{}_{\bf d})^2 - \omega^2_{\bf d} Q_{\bf d}^2\right] + S^{}_I %\right. \nonumber\\
\label{theaction}
\end{equation}
where the interaction action is
\begin{equation}
  S_I = -\sum_{\bf d} \lambda \int d\tau^{}_{\bf d} Q^{}_{\bf d} \partial^{}_{\bf d}
    \int d^2x \Phi(t,x) \delta^2(x^\alpha- z^\alpha_{\bf d}(\tau^{}_{\bf d})),
\label{SI1}
\end{equation}
with $x^\alpha = (t,x)$, $\alpha=0,1$, the detector label ${\bf d} = A$ for one-detector case, ${\bf d}=A,B$ for two-detector
case, and $\partial^{}_{\bf d} \equiv d/d\tau^{}_{\bf d}$. $z^\alpha_{\bf d}(\tau^{}_{\bf d})$ are the prescribed
trajectory of the detector ${\bf d}$. The canonical momenta conjugate to $Q^{}_{\bf d}$ and $\Phi$ are
\begin{eqnarray}
  & & P^{}_{\bf d}(\tau^{}_{\bf d}) = {\delta S\over \delta \partial^{}_t Q^{}_{\bf d} (\tau_{\bf d})} =
	    \partial^{}_{\bf d} Q^{}_{\bf d}\left(\tau^{}_{\bf d}\right),
			\label{Pdef} \\
  & &\Pi(t,x) = {\delta S\over \delta \partial^{}_t \Phi(t,x)} = %-\sqrt{-g}\partial^0 \Phi(t,x) =
	  \partial^{}_t \Phi(t,x) - \sum_{\bf d} \lambda \int d\tau^{}_{\bf d} Q^{}_{\bf d}(\tau^{}_{\bf d}) v^0_{\bf d}(\tau^{}_{\bf d})
		\delta^2(x^\alpha- z^\alpha_{\bf d}(\tau^{}_{\bf d})), \label{Pidef}
\end{eqnarray}
respectively, where $v^0_{\bf d}\equiv \partial^{}_{\bf d}z^0_{\bf d}$. Then, after a Legendre transformation,
one can write down the Hamiltonian as
\begin{eqnarray}
  && H(t) %&=& \sum_{\bf d} P^{}_{\bf d} \partial^{}_t Q^{}_{\bf d} + \int dx \Pi_{x} \partial^{}_t\Phi_{x} -L \nonumber\\
  = \sum_{\bf d} {1\over 2v^0_{\bf d}(t)}\left\{ P^{2}_{\bf d}(t) +
		\omega_{\bf d}^2 Q^2_{\bf d}(\tau^{}_{\bf d}(t))\right\} + \nonumber\\&&
		{1\over 2}\int dx \left\{ \left[  \Pi^{}_{x}(t) +
 		\sum_{\bf d} \lambda \int d\tau^{}_{\bf d} Q^{}_{\bf d}(\tau^{}_{\bf d}) v^0_{\bf d}(\tau^{}_{\bf d})
		\delta^2(x^\alpha- z^\alpha_{\bf d}(\tau^{}_{\bf d})) \right]^2
		+ \left[\partial^{}_x\Phi^{}_{x}(t)\right]^2\right\},
\label{Hami}
\end{eqnarray}
which is parametrized by the time variable $x^0=t$ of the observer's frame and defined on the whole time-slice $x \in {\bf S}^1$
associated with $t$.

Alternatively, one may adopt the interaction action
\begin{equation}
  S'_I = \sum_{\bf d} \lambda \int d\tau^{}_{\bf d} \partial^{}_{\bf d} Q^{}_{\bf d}
    \int d^2x \Phi(t,x) \delta^2(x^\alpha- z^\alpha_{\bf d}(\tau^{}_{\bf d})),
\label{SI2}
\end{equation}
plus surface terms evaluated at the initial and final moments from (\ref{SI1}), which gives the same Euler-Lagrange equations
for the dynamical variables $Q^{}_{\bf d}$ and $\Phi_{x}$.
Starting with (\ref{SI2}), the canonical momenta conjugate to $Q^{}_{\bf d}$ and $\Phi$ become
\begin{eqnarray}
  & & P'_{\bf d}(\tau^{}_{\bf d}) = \partial^{}_{\bf d} Q^{}_{\bf d}\left(\tau^{}_{\bf d}\right) +
      \lambda\Phi\left(z^\alpha_{\bf d}(\tau^{}_{\bf d})\right), \label{Pdef2} \\
  & &\Pi'(t,x) = \partial^{}_t \Phi(t,x),\label{Pidef2}
\end{eqnarray}
respectively, and so the Hamiltonian reads
\begin{eqnarray}
  H'(t) %&=& \sum_{\bf d} P'_{\bf d} \partial^{}_t Q^{}_{\bf d} + \int dx^1 \Pi'_{x^1} \partial^{}_t\Phi_{x^1} -L \nonumber\\
  &=& \sum_{\bf d} {1\over 2v^0_{\bf d}(t)}\left\{ \left[ P'_{\bf d}(t) -
    \lambda \Phi^{}_{z^1_{\bf d}(t)}\left(t\right)\right]^2 +
		\omega_{\bf d}^2 Q^2_{\bf d}(\tau^{}_{\bf d}(t))\right\} + \nonumber\\ & &
		{1\over 2}\int dx \left\{ \Pi'^{2}_{x}(t)
		+ \left[\partial^{}_x\Phi^{}_{x}(t)\right]^2\right\}.
\label{Hami2}
\end{eqnarray}
%The interaction actions (\ref{SI1}) and (\ref{SI2}) are analogous to the $e\,{\bf x}\cdot{\bf E}$ and $e\,{\bf v}\cdot {\bf A}$ couplings
%in electrodynamics, respectively, with the coupling constant $e$. Recall that the canonical momentum of a charged particle with mass $m$
%in the electromagnetic (EM) fields, ${\bf p}=m{\bf v}+e{\bf A}$, for the $e\, {\bf v}\cdot {\bf A}$ coupling
%{depends on the choice of gauge condition
%and not uniquely defined, thus not physically measurable, while the kinetic momentum $m{\bf v}$ does not. Similarly,
While the values of the Hamiltonian $H'$ and the momentum $P^{}_{\bf d} =
\dot{Q}^{}_{\bf d}$ are invariant under a shift $\Phi_x \to \Phi_x + {\cal C}$ with a constant ${\cal C}$, such a global symmetry is enough to make the value of $P'_{\bf d} = \dot{Q}^{}_{\bf d} + \lambda\Phi\left(z^\alpha_{\bf d}(\tau^{}_{\bf d})\right)$
not ``gauge" invariant.
Since we are looking at the reduced state of the detectors in this paper, we prefer to work with (\ref{SI1}) to get rid of some weird dynamical behaviors of $P'^{}_{\bf d}$ which could be ``gauged away" by the surface terms mentioned below (\ref{SI2}).
Later we will see that the two point correlators $\langle \hat{P}'^2_{\bf d} \rangle$ and the corresponding uncertainty functions will
have indefinite growths in time, while $\langle \hat{P}^2_{\bf d} \rangle$ of the kinetic momentum is well-behaved.

Suppose the combined system of the detectors and the field is initially in a Gaussian state,
which could be pure or mixed. Then the quantum state will always be Gaussian in its history because
of the linearity of the interaction. A Gaussian state is fully determined by the two-point correlators of the dynamical variables
of the combined system, which can be obtained by taking the expectation values of two operators, evolved in the Heisenberg picture,
with respect to the initial state (see, e.g., \cite{LH09}).

The Heisenberg equations of motion for the operators read
\begin{eqnarray}
  & &\partial^{2}_{\bf d} \hat{Q}^{}_{\bf d}(\tau^{}_{\bf d}) + \omega_{\bf d}^2 \hat{Q}^{}_{\bf d}(\tau^{}_{\bf d}) =
	-\lambda \partial^{}_{\bf d} \hat{\Phi}(z^{}_{\bf d}(\tau^{}_{\bf d})), \label{HOeom}\\
  & & %\sqrt{-g}
	    -\Box\hat{\Phi}(x) = \lambda \sum_{\bf d} \int d\tau^{}_{\bf d} \partial^{}_{\bf d} \hat{Q}^{}_{\bf d}(\tau^{}_{\bf d})
  \delta^2\left(x^\alpha-z^\alpha_{\bf d}(\tau^{}_{\bf d})\right), \label{fieldeq}
\end{eqnarray}
where $\Box\equiv %\sqrt{-g}^{-1}\partial_{\alpha} \sqrt{-g}g^{\alpha\beta}\partial_{\beta} =
-\partial_t^2 + \partial_x^2$. Suppose the
detector-field coupling is suddenly switched on at the initial moment $t=0$ (when $\tau^{}_{\bf d} \equiv 0$ for all detectors). By
virtue of the linearity of the system, the operators will evolve into linear combinations of the operators initially defined at $t=0$ as
\begin{eqnarray}
  & & \hat{Q}_{\bf d}(\tau^{}_{\bf d}) = \nonumber\\& & \sum_{{\bf d}'}\sqrt{\hbar\over 2\omega_{\bf d}}\left[
    q_{\bf d}^{{\bf d}'}(\tau^{}_{\bf d})\hat{a}^{}_{{\bf d}'} +q_{\bf d}^{{\bf d}'*}(\tau^{}_{\bf d})
    \hat{a}_{{\bf d}'}^\dagger \right]
   +\sum_{k} \sqrt{\hbar\over 2 \tilde{\omega}_k} \left[q_{\bf d}^{k}(\tau^{}_{\bf d})\hat{b}^{}_{k} +
    q_{\bf d}^{k*}(\tau^{}_{\bf d})\hat{b}_{k}^\dagger\right], 		\label{Qmod}\\
  & &\hat{\Phi}_x(t) = \sum_{{\bf d}'} \sqrt{\hbar\over 2\omega_{\bf d}}\left[
    \phi^{{\bf d}'}_x(t) \hat{a}^{}_{{\bf d}'}+ \phi^{{\bf d}'*}_x(t) \hat{a}_{{\bf d}'}^\dagger \right]
    +\sum_{k}  \sqrt{\hbar\over 2 \tilde{\omega}_k}\left[\phi^{k}_x(t)\hat{b}^{}_{k} +\phi^{k*}_x(t)\hat{b}_{k}^\dagger \right],
		\label{Fmod}
\end{eqnarray}
with $\tilde{\omega}_k \equiv \omega_k  L= |k| L$ for $k \not= 0$ and $\tilde{\omega}_k \equiv 1$ for $k=0$.
Here $q_{\bf d}^{{\bf d}'}$, $q_{\bf d}^{k}$, $\phi_x^{{\bf d}'}$, and $\phi_x^{k}$ are the (c-number)
detector-detector, detector-field, field-detector, and field-field mode functions, respectively,
$\hat{a}^{}_{\bf d}$ and $\hat{a}_{\bf d}^\dagger$ are the lowering and raising operators for the free detector ${\bf d}$,
while $\hat{b}^{}_{k}$ and $\hat{b}_{k}^\dagger$ are the annihilation and creation operators for the free field mode with wave number $k$.
For the normal (untwisted) field, we take $k=k_n \equiv n/R = 2\pi n/L$, $n\in {\bf Z}$ to satisfy the periodic boundary condition
$\phi_{\; x}^{{}^{[0]}k_n}(t) = \phi_{x+L}^{{}^{[0]}k_n}(t)$,
and for the twisted field, we take $k=k'_n \equiv [n-(1/2)]/R$, $n\in {\bf Z}$ to satisfy
the anti-periodic boundary condition $\phi_{\; x}^{{}^{[0]}k'_n}(t) = -\phi_{x+L}^{{}^{[0]}k'_n}(t)$
(analogous to the M\"obius band) \cite{Is78, DHI79}. There is one zero mode ($k^{}_0=0$ and $\omega_{k_0}=0$ when $n=0$) for the
untwisted field, and no zero mode for the twisted field. For the zero mode, we define $\hat{b}_0 \equiv (\hat{\Phi}_{k_0} +
i\hat{\Pi}_{k_0})/\sqrt{2\hbar}$ with the initial zero-mode field operator and its conjugate momentum operator satisfying the
equal-time commutation relation $[\hat{\Phi}_{k_0}, \hat{\Pi}_{k_0}] =i\hbar$.

Since the detectors and the field are free before the detector-field coupling is switched on at $t=\tau_{\bf d}= 0$, the initial
conditions for the mode functions are set to be $\phi^{k}_x(0) = e^{ikx}$, $\partial^{}_t \phi^{k}_x(0)=-i\omega_k e^{ikx}$ for $k\not=0$,
$q_{\bf d}^{{\bf d}'}(0)=\delta_{\bf d}^{{\bf d}'}$, $\partial^{}_t q_{\bf d}^{{\bf d}'}(0)=-i\omega^{}_{\bf d}\delta_{\bf d}^{{\bf d}'}$,
and $\phi_x^{\bf d} (0) =\partial^{}_t \phi_x^{\bf d} (0) = q^{k}_{\bf d}(0)= \partial^{}_t q^{k}_{\bf d}(0) =0$.
For the zero-mode component of the free field \cite{ML14}, one has $\hat{\Phi}_{k_0}(t) = \hat{\Phi}_{k_0}(0) + (t/L)\hat{\Pi}_{k_0}(0)
= \sqrt{\hbar/2} \left[ (1-(i/L)t)\hat{b}^{}_{k_0} + (1+ (i/L)t)\hat{b}^\dagger_{k_0}\right]$, so the initial conditions are set to be
$\phi^{k_0}_x(0) = 1$ and $\partial^{}_t \phi^{k_0}_x(0)=-i/L$. The momentum operators conjugate to
$\hat{Q}^{}_{\bf d}$ and $\hat{\Phi}^{}_x$ at $t>0$ can be obtained straightforwardly by inserting (\ref{Qmod}) and (\ref{Fmod}) to
the operator version of (\ref{Pdef}) and (\ref{Pidef}).

In terms of the above expansion, Eq. $(\ref{fieldeq})$ yields
\begin{equation}
  %\sqrt{-g}
	-\Box\phi^\mu_{x}(t) = \lambda \sum_{\bf d} \int d\tau^{}_{\bf d} \partial^{}_{\bf d}
  q^\mu_{\bf d}(\tau^{}_{\bf d})\delta^2\left(x^\alpha-z^\alpha_{\bf d}(\tau^{}_{\bf d})\right)
\label{FEqSource}
\end{equation}
where $\mu = {\bf d}, k$, and the mode functions with different $\mu$'s are decoupled.
Eq. (\ref{FEqSource}) implies $\phi^\mu_{x}(t)=\phi^{{}^{[0]}\mu}_{\;x}(t)+\phi^{{}^{[1]}\mu}_{\;x}(t)$, where the homogeneous
solutions are $\phi^{{}^{[0]}{\bf d}}_{\;x} =0$, $\phi^{{}^{[0]}k}_{\;x}(t) = e^{-i\omega_k t + ikx}$ for $k\not=0$, and
$\phi^{{}^{[0]}k_0}_{\;x}(t) = 1-(i/L)t$ for the zero mode, and the inhomogeneous solutions read
\begin{equation}
  \phi^{{}^{[1]}\mu}_{\;x^{1}}(t) =  \sum_{\bf d} \lambda \int_{0}^\infty d\tau^{}_{\bf d} G_{\rm ret}(x^\alpha;
  z^\alpha_{\bf d}(\tau^{}_{\bf d})) \partial^{}_{\bf d} q^{\mu}_{\bf d}(\tau^{}_{\bf d}) \label{inhphi}
\end{equation}
with the retarded Green's function $G_{\rm ret}$ of the field. When all of the field modes are considered, the retarded Green's function
$G_{\rm ret}$ for the massless scalar field can be written as
\begin{equation}
  G_{\rm ret}(t,x; t',x') = %{i\over\hbar}\theta(t-t')\langle 0_L| [ \phi^{}_x(t), \phi^{}_{x'}(t')] |0_L \rangle =
	{1\over 2}\sum_{n\in {\bf Z}}
	\varepsilon^n \theta\left[ t+x -(t'+x'+nL) \right]\theta\left[ t-x -(t'-x'-nL ) \right] ,
\label{retG}
\end{equation}
for $x, x' \in (-L/2, L/2]$. We take $\varepsilon=1$ for the untwisted field, and $\varepsilon = -1$ for the twisted field.
In either case, the retarded Green's function looks the same as the one in Minkowski space around the source point
$(t, x) = (t', x')$.
One can verify that $-\Box G_{\rm ret}(t,x; t',x') = \sum_{n\in{\bf Z}} \varepsilon^n \delta(t-t')\delta(x-x'-nL)$ in the extended
coordinates with $x,x'\in{\bf R}^1$ and identifying the points $x$ to $x$ mod $L$ (see Figure \ref{setup}).
Now $\phi^{{}^{[0]}\mu}_{\;x}(t)$ can be interpreted as vacuum
fluctuations of the field state, while $\phi^{{}^{[1]}\mu}_{x}(t)$ behave like the retarded fields sourced by the point-like detectors.
Note that, to be consistent with the expressions (\ref{retG}), later we have to include the contributions by enough number of the
the field modes, namely, the UV cutoff cannot be too small.

For the untwisted field, Eq. (\ref{retG}) includes the contribution by the zero mode.
If we exclude the zero mode entirely, the retarded Green's function for the untwisted field will become \cite{ML14}
\begin{eqnarray}
  & & G^{\rm nz}_{\rm ret}(t,x; t',x') \nonumber\\ &=& \lim_{\epsilon\to 0+} {1\over 2\pi}\theta(t-t') {\rm Im}
	  \left\{ \ln\left( 1-e^{-i(t-t'+(x-x')-i\epsilon)/R}\right) + \ln \left( 1-e^{-i(t-t'-(x-x')-i\epsilon)/R}\right)\right\}\nonumber\\
	&=& {1\over 2}\sum_{n\in {\bf Z}}
	\theta\left[ t+x -(t'+x'+nL) \right]\theta\left[ t-x -(t'-x'-nL ) \right] - {t-t'\over L} \theta(t-t').
\label{retGN}
\end{eqnarray}
Compared with $-\Box G^{}_{\rm ret}$, here $-\Box G^{\rm nz}_{\rm ret}$ has an extra term $-\Box\left[-(t-t')\theta(t-t')/L\right]
=-(2/L)$ $\delta(t-t')$.
This is because the field modes cannot form a complete basis without the zero mode. Moreover, $G^{\rm nz}_{\rm ret}$ is nonzero
when $|t-t'| < |x-x'|$, which implies that two spacelike separated events located at $(t,x)$ and
$(t',x')$ can have superluminal signaling, which violates causality, even at the classical level.

%%%%%%%%%%%%%%%%%%%%%%%%%%%%%%%%%%%%%%%%%%%%%%%%%%%%
\section{One-detector case}

Let us start with the simplest case with only one single detector $A$ with natural frequency $\omega^{}_A=\Omega_0$,  located at $x=0$.
To study the influence of the quantum field in the ${\bf S}^1\times {\bf R}_1$ spacetime on the detector we look at
the reduced dynamics of $A$ by tracing out the field.
For this purpose we need to calculate the two-point correlators of the detector.

Inserting the solutions of $\phi^\mu_{x}$ with (\ref{inhphi}) into $(\ref{HOeom})$, one obtains
the equations of motion for the mode functions
\begin{eqnarray}
  \left( \partial_t^2 +2 \gamma \partial^{}_t + \Omega_0^2 \right) q^{\mu}_A(t) &=& -\lambda \partial^{}_t \phi_{\;0}^{^{[0]}\mu}(t)
    -\lambda^2 \sum_{n'=1}^\infty \varepsilon^{n'} %e^{in'\delta}
		\theta(t- n' L) \partial^{}_t q^{\mu}_A(t-n' L) \label{EOMqp1} \\ &=& \left\{
  \begin{array}{lcl}
     -\lambda \partial^{}_t \phi_{\;0}^{^{[0]}\mu}(t) & {\rm for} & t<L \\
		\varepsilon\left( \partial_t^2 -2 \gamma \partial^{}_t + \Omega_0^2 \right) q^{\mu}_A(t-L)
		& {\rm for} & t\ge L,
  \end{array}\right. \label{EOMqp2}
\end{eqnarray}
with $\mu = A$, $k_n$ (untwisted field) or $k'_n$ (twisted field), $n\in {\bf Z}$,
and the coupling strength $\gamma \equiv \lambda^2/4$.
When $t\ge L$, the above expressions for different $\mu$'s have the same appearance,
while the solutions at $0 \le t < L$ before the first echo hits the detector are generated by different driving forces
$\propto \partial_t \phi_{\; 0}^{^{[0]}\mu}(t)$.
Note that the above equations for the twisted field on ${\bf S}^1$ are equivalent to those for a UD'
detector in the same scalar field and located at the center of an 1D cavity made of two perfectly-conducting mirrors separated at a
distance of $L$, while the ones for the untwisted field are equivalent to those in a cavity with infinitely-permeable mirrors
\cite{Bo03}, or in a toroidal cavity \cite{toroidal}.

Although the left-hand side (LHS) of (\ref{EOMqp1}) describes the continuous evolutions of the mode functions,
on the right-hand side (RHS) of the equation the influences by the echoes come in a discrete fashion.
The analytical solutions for these delayed differential equations can be obtained order by order
from those at very early times in principle, but the lengths of the expressions for the solutions grow rapidly
as the order of the included echoes increase. Very soon the analytical solutions will get too complicated to read off any useful information. In this case numerical computation would come in handy and give more transparent results.

Below, the delayed differential equations (\ref{EOMqp2}) will be solved with the proper initial conditions.
For $t<L$, the solutions for (\ref{EOMqp2}) have the closed form
\begin{eqnarray}
  \left.q^{A}_A(t)\right|_{0\le t<L} &=& {e^{-\gamma t}\over 2\Omega}\left[\left(\Omega-\left(\Omega_0+i \gamma\right)\right)
	  e^{i \Omega t} + \left(\Omega+ \left( \Omega_0+ i \gamma\right)\right) e^{-i \Omega t}\right],\label{qAA0}\\
  \left.q_A^{k\not=0}(t)\right|_{0\le t<L} &=& -\lambda \int_0^t d\tau \Omega^{-1} K(t-\tau) \partial^{}_t \phi_{\;0}^{^{[0]}k}(\tau)
	\nonumber\\ &=& {\lambda \omega \over 2 \Omega}\left( {e^{-i \omega t}- e^{(-\gamma + i\Omega)t}\over
	 \gamma - i(\omega + \Omega)} - {e^{-i \omega t}- e^{(-\gamma -i\Omega)t}\over \gamma - i(\omega - \Omega)} \right),\label{qAk0}\\
  \left.q^{k_0}_A(t)\right|_{0\le t<L} &=& -\lambda \int_0^t d\tau \Omega^{-1} K(t-\tau) \partial^{}_t \phi_{\;0}^{^{[0]}k_0}(\tau)
	\nonumber\\ &=&
	{i\lambda\over \Omega_0^2 L}\left[ 1 - e^{-\gamma t}\left(\cos\Omega t + {\gamma\over\Omega} \sin\Omega t\right)\right],\label{qA00}
\end{eqnarray}
with $\omega \equiv |k|$ ({$k$ can be $k_n$ or $k'_n$), $\Omega\equiv\sqrt{\Omega_0^2-\gamma^2}$, and $K(x)\equiv e^{-\gamma x}
\sin\Omega x$. The above early-time solutions for $q_A^A(t)$ and $q_A^{k\not=0}(t)$ are identical to those in Minkowski space \cite{LH06}.

%%%%%%%%%%%%%%%%%%%%%%%%%%%%%%%
\subsection{Eigen-frequencies}
\label{eigenf1}

\subsubsection{Untwisted field}

Assume at late times ($t\gg L$, $1/\gamma$) the detector would evolve into a stationary state, when the mode function $q^{A}_A$
could be written as
\begin{equation}
  q^{A}_A (t) \approx \int d\omega \tilde{q}^{A}_A(\omega)e^{i \omega t}. \label{LTans}
\end{equation}
For $\varepsilon=1$, inserting the above ansatz into (\ref{EOMqp1}) or (\ref{EOMqp2}) with $\mu=A$ yields
\begin{equation}
  -\omega^2 + \Omega_0^2 = -2\gamma \omega \cot (\omega L /2). \label{EFcond}
\end{equation}
The solutions of $\omega$ are the eigen-frequencies.
One can immediately see that $|\omega| = \Omega_0$ are solutions
but only when $R = (1+2n)/(2\Omega_0)$ (i.e. $L = \pi(1+2n)/\Omega_0$), $n=0,1,2,3,\ldots$.
More general solutions to (\ref{EFcond}) can be obtained by numerical methods. In Figure \ref{EmergeEigen} (upper-right)
one can see that when $\Omega_0\gg 2\pi/L$, for those $|\omega| < \Omega_0$ ($|\omega| > \Omega_0$),
one has $|\omega| < n/R$ ($|\omega| > n/R$), $n=1,2,3,\ldots.$.

When $\Omega_0 \approx n/R$ with some positive integer $n$,
while $q_A^{k_n}$ in (\ref{qAk0}) is on resonance at early times (which, together with other modes, produces some resonant
oscillation on top of the decaying behavior of the two-point correlators and functions of them, see Appendix \ref{2ptM2}),
at a larger time scale the detector mode with $\Omega_0$ and the field mode with $\omega_{k_n}=|k_n|=n/R$ will mix together and generate
two dominant eigen-modes
\footnote{In this paper, our ``eigen-modes" refer to those $e^{i\omega t}$ in the stationary ansatz for a mode function, e.g.,
(\ref{LTans}) or (\ref{LTknans}), rather than the mode function itself. We call the eigen-modes ``dominant" if
the amplitude $|\tilde{q} (\omega)|$ in the ansatz has the maximum values at the corresponding eigen-frequencies.}
with frequencies $\omega \approx \Omega_0 \pm \Delta$, where
\begin{equation}
  \Delta \approx \sqrt{\gamma/\pi},
  \label{Delbeat}
\end{equation}
in the weak coupling limit.
These two eigen-modes will dominate the late-time behavior of $q^{A}_A$. In particular, the frequency difference produces the beat of
$|q^{A}_A|^2$ at frequency $2\Delta$, and so $2\pi/(2\Delta) \sim \gamma^{-1/2}$ is the largest significant time scale in the evolution
of the single detector system with $\Omega_0\approx n/R$ in the weak coupling limit.
This is very different from the time scale $\gamma^{-1}$ in the detector theories in Minkowski space.

\begin{figure}
\includegraphics[width=7.2cm]{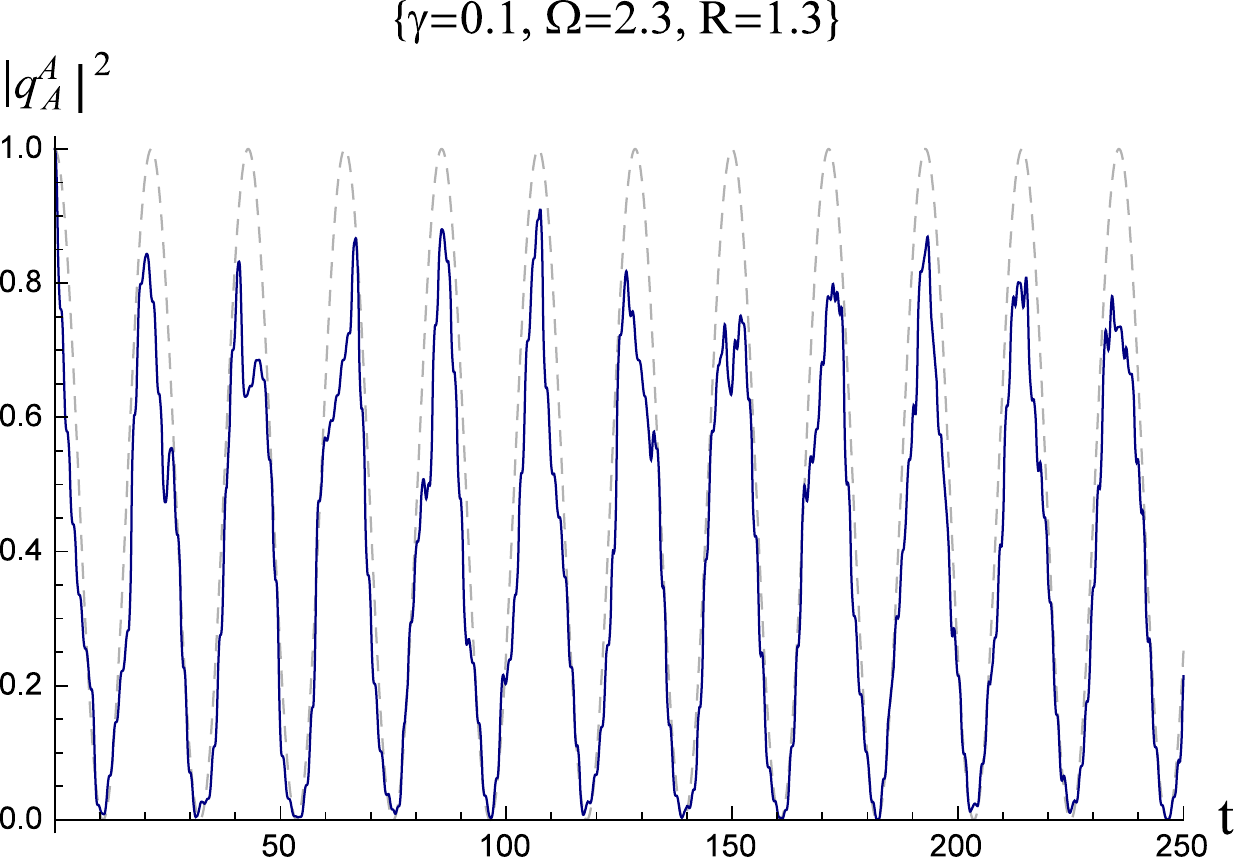}
\includegraphics[width=7.2cm]{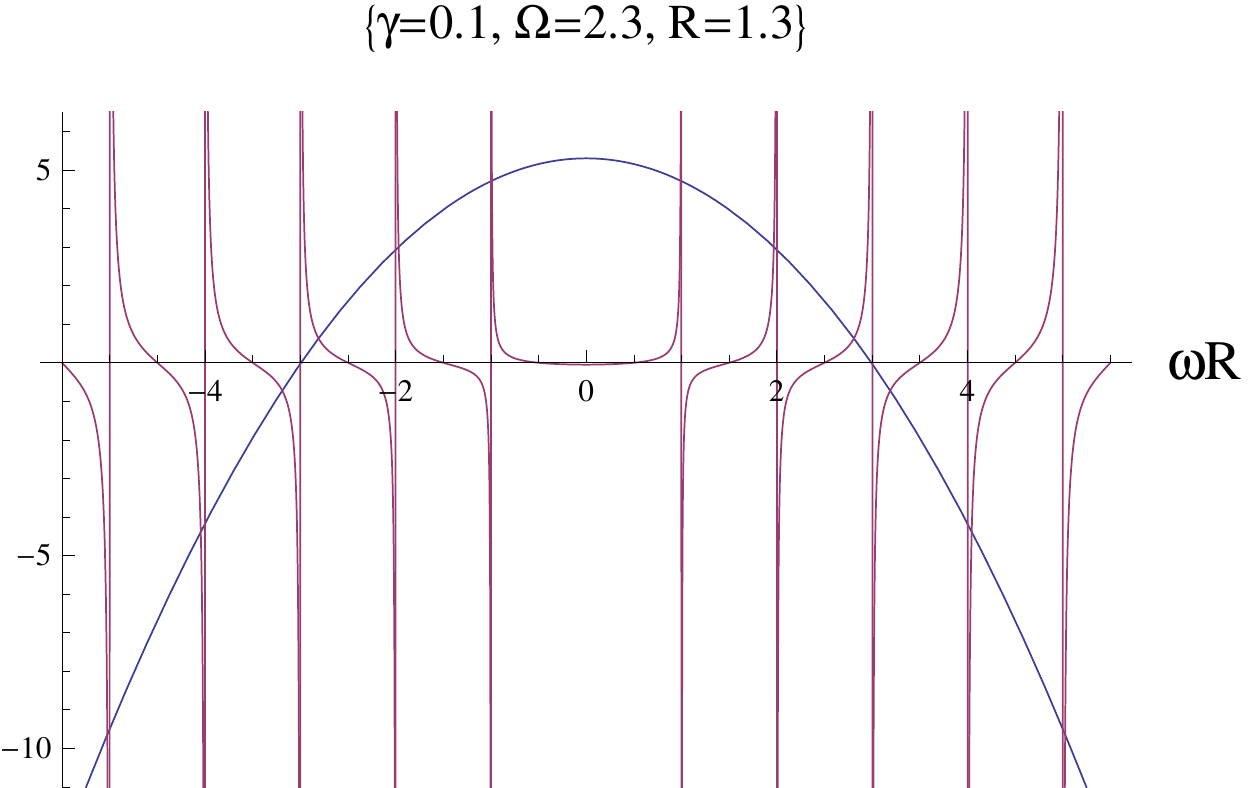}\\
\includegraphics[width=7.2cm]{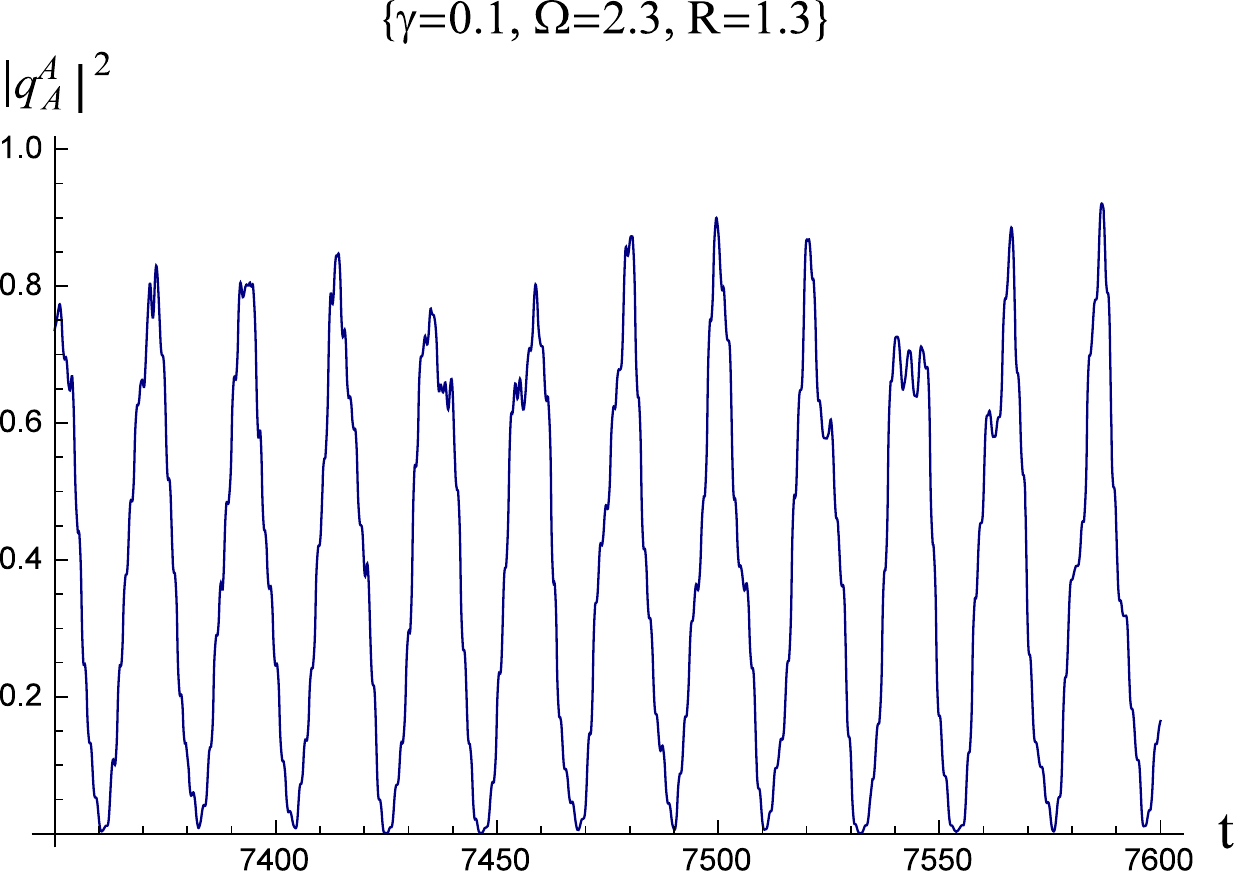}
\includegraphics[width=7.2cm]{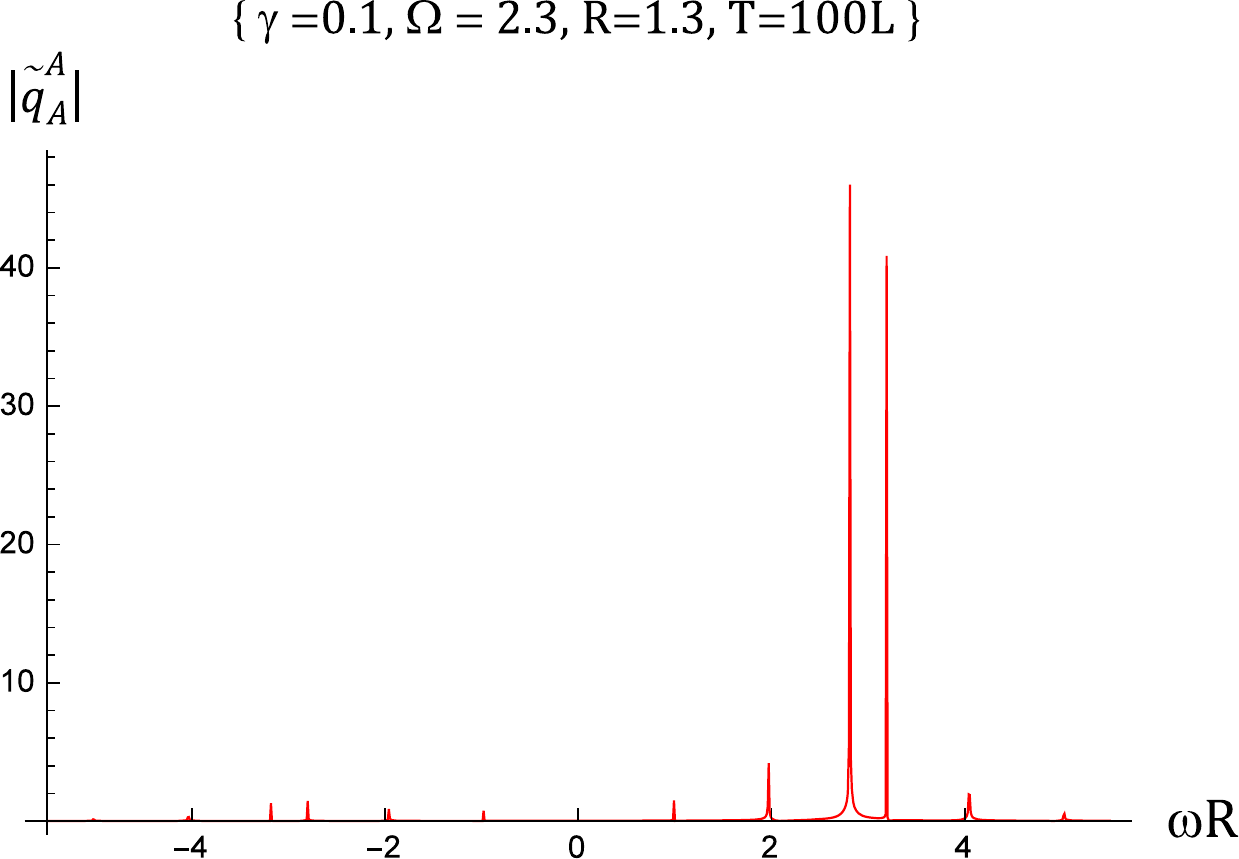}
\caption{Emergence of the frequency spectrum, after a Fourier transform of $q_A^A$ from $t=0$ to $t=T \gg L$ (lower-right). In the
upper-right plot, the blue and purple curves represent the LHS and RHS of Eq.(\ref{EFcond}) as functions of $\omega R$, respectively. Here the LHS $=0$ at the free detector's natural frequency $\Omega_0 =\pm 2.9928/R\approx 3/R$, where a free-field mode is located.
These two frequencies mix together and generate two new eigen-frequencies
$\omega R \approx 3.1901$ and $2.8089$, which are the closest two among all of the eigen-frequencies in the frequency space.
In the lower-right plot one can see that these two eigen-frequencies dominate the long-time behavior of $q^A_A$, and give the beat of
$|q_A^A|^2$ in the left plots. The period of the beat can be estimated from the difference of these two eigen-frequencies: $T_{\rm beat}
\approx 2\pi/(2\Delta) = 2\pi/[(3.1901-2.8089)/R] \approx 21.4297$ (the gray dashed curve in the upper-left plot represents
$(1+\cos 2\pi t/T_{\rm beat})/2 \approx (1+\cos 2\Delta t)/2$). Compare the upper- and lower-left plots one can see that, while the
frequency spectrum gets sharper and sharper as the duration of interaction $T$ increases, the mode function have similar behaviors from
early times all the way to late times, if observed in the same small time scale.}
\label{EmergeEigen}
\end{figure}

Even if the radius $R$ or the circumference $L$ of ${\bf S}^1$ is larger than the time scale $1/\gamma$, which is the largest time scale of the detector in (1+1) dimensional Minkowski space ${\bf R}^1_1$, only the early-time ($0< t < L$) behavior of the mode function
will be similar to those in ${\bf R}^1_1$.
Once the echoes start to affect the detector ($t> L$), the higher-order corrections from the echoes will be the same order
of magnitude as the zeroth order solution.
At large time scales one can also see the beat feature when $\Omega_0 \approx n/R$ for some integer $n$.

When $\Omega_0$ is not close to any $n/R$, $n \in {\bf Z}$, the beating behavior may be suppressed. In the example shown in Figure
\ref{Specr001W23R2}, when $\Omega_0 \approx [n-(1/2)]/R$ for some $n$, the only dominant eigen-mode has a frequency very close to
$\Omega_0$, and the largest time scale in the evolution of the mode function is about $L$, which is the period of the massless field
traveling over the space ${\bf S}^1$. However, if there are more than two dominant modes, the small differences between the
frequency-differences of the dominant modes may produce beats at an even larger time scale (e.g., Figure \ref{SL1Pi}).

\begin{figure}
\includegraphics[width=4.8cm]{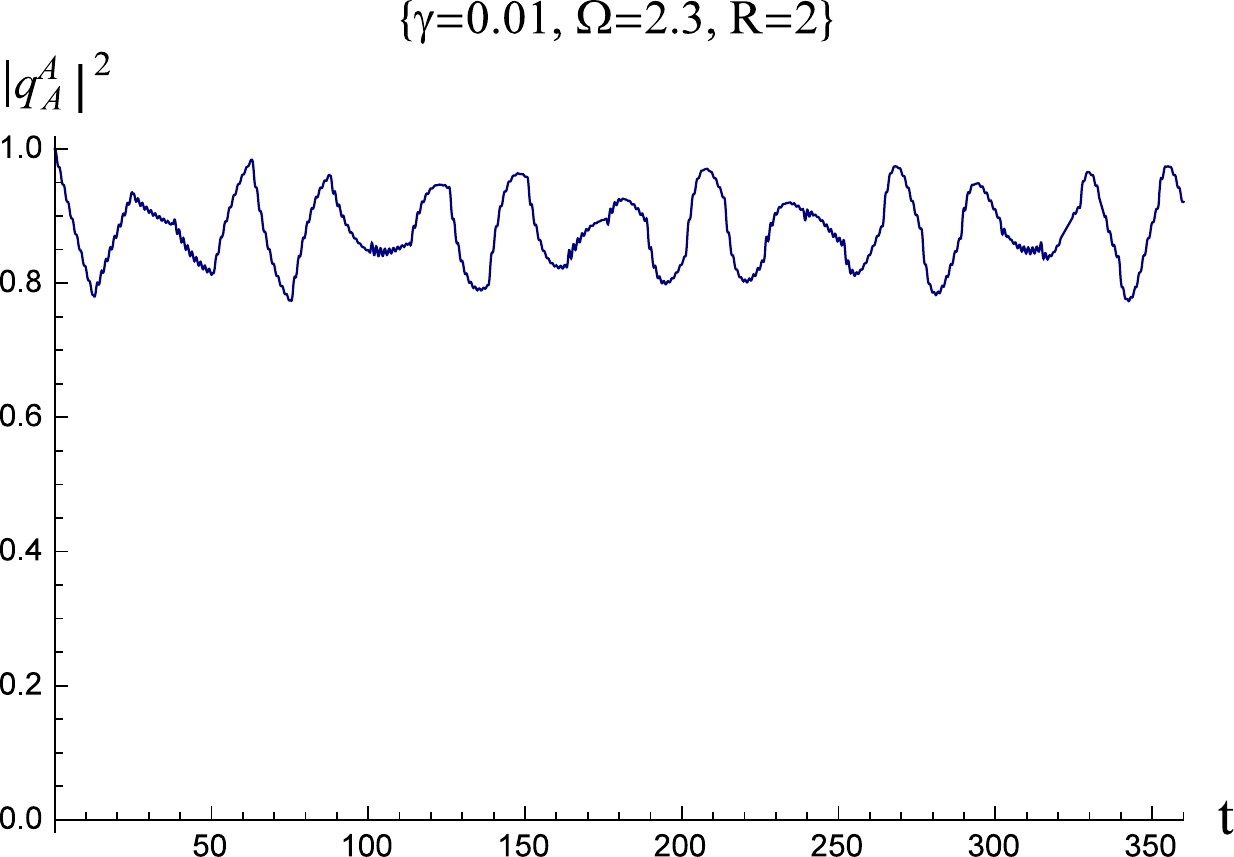}
\includegraphics[width=4.8cm]{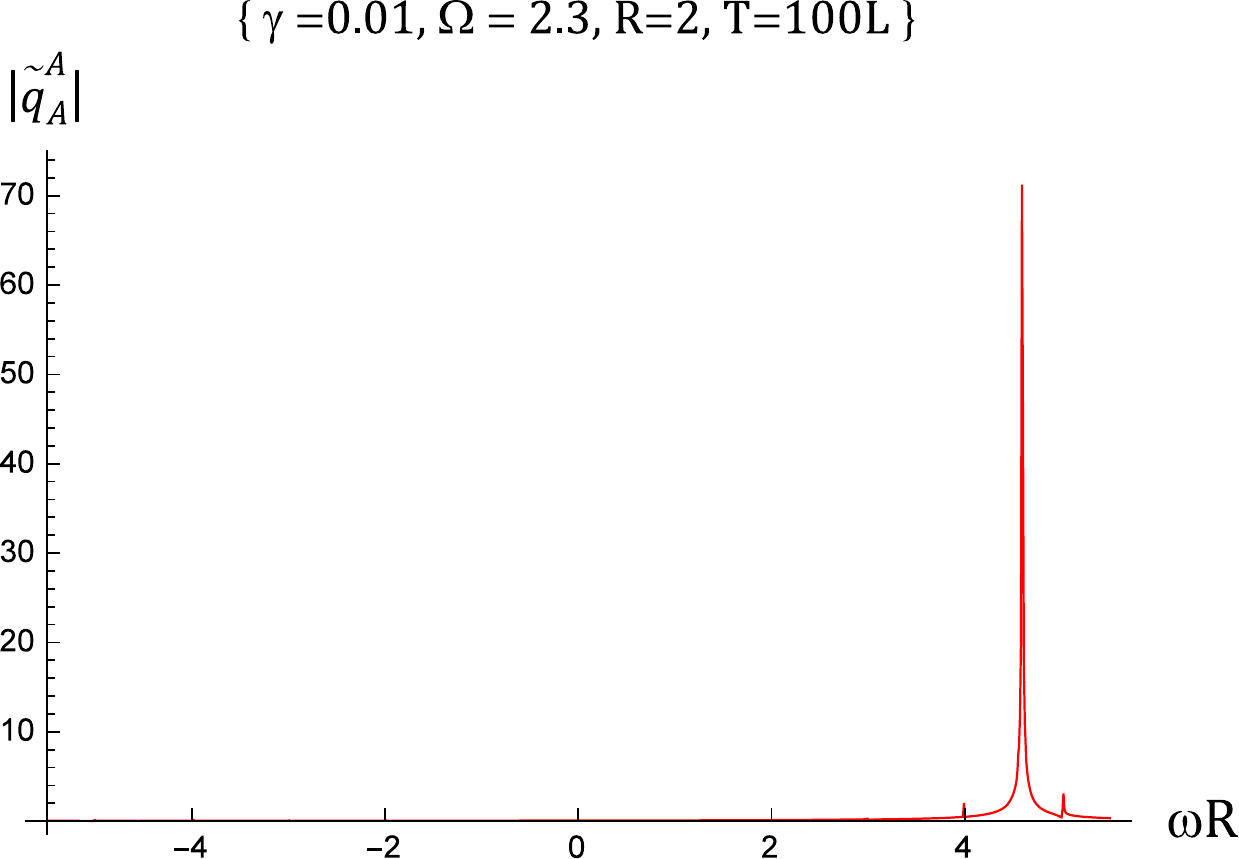}
\includegraphics[width=4.8cm]{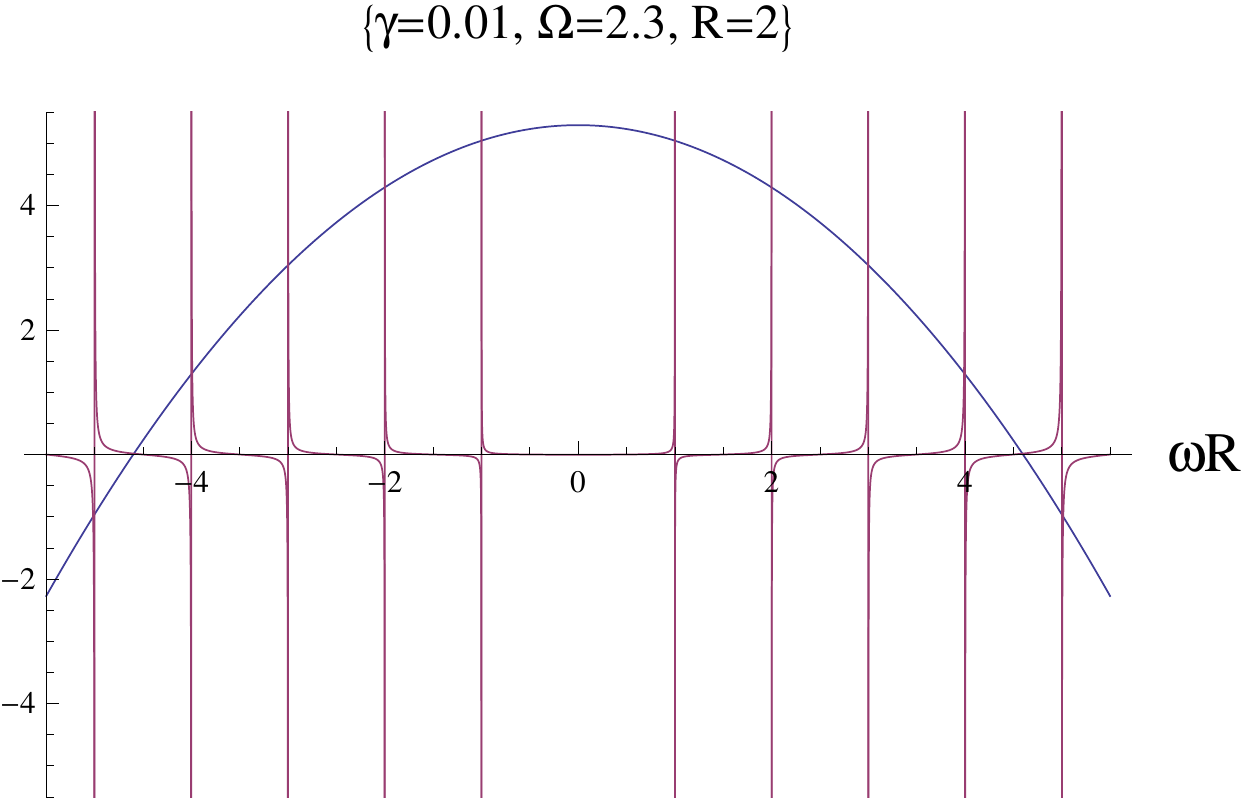}
\caption{When $\gamma$ is small and $\Omega_0$ is about $[n-(1/2)]/R$ ($n\in {\bf Z}$), which is not close to any $n/R$, no significant
beat can be observed. Here $\gamma=0.01$, $\Omega=2.3$, and $R=2$. The most significant eigen-mode has the frequency closest to
$\Omega_0$.}
\label{Specr001W23R2}
\end{figure}

For the detector-field mode function $q^{k_n}_A$, inserting a similar ansatz
\begin{equation}
  q^{k_n}_A (t) \approx \int d\omega \tilde{q}^{k_n}_A(\omega)e^{i \omega t}. \label{LTknans}
\end{equation}
into (\ref{EOMqp1}) or (\ref{EOMqp2}) with $\mu = k_n$ yields
\begin{equation}
  -\omega^2 	+ \Omega_0^2 = -2\gamma \omega \cot (\omega L/2)
	-\lambda 2\pi \delta(\omega_{k_n} - \omega).
\end{equation}
The solutions of the eigen-frequencies for $q^{k_n}_A$ are exactly the same as those for $q^{A}_A$ with the same values of the
parameters $\Omega_0$, $\lambda$, and $L$.
This is evident when comparing Figure \ref{mofnr01W23R13} with Figure \ref{EmergeEigen},
and Figure \ref{mofnr001W23R2} with Figure \ref{Specr001W23R2}.
One can clearly see the beats in Figure \ref{mofnr001W23R2} while there is no significant beat in Figure \ref{Specr001W23R2}.
This is because there are two dominant eigen-modes (with frequencies $\omega\approx \Omega_0$ and $|k_n|$) for
$q^{k_n}_A$ in Figure \ref{mofnr001W23R2}, rather than one for $q^{A}_A$ in Figure \ref{Specr001W23R2}.

\subsubsection{Twisted field}
\label{1DetTwFreq}

For $\varepsilon=-1$, inserting ansatz similar to (\ref{LTans}) or (\ref{LTknans}) into (\ref{EOMqp1}) or (\ref{EOMqp2}) yields
\begin{eqnarray}
  && -\omega^2 + \Omega_0^2 = 2\gamma \omega \tan (\omega L/2), \label{EFcondTw} \\
	&& -\omega^2 + \Omega_0^2 = 2\gamma \omega \tan (\omega L/2) -\lambda 2\pi \delta(\omega_{k'_n} - \omega), \label{EFkncondTw}
\end{eqnarray}
for $\mu=A$ and $\mu=k'_n$, respectively. Again the eigen-frequencies of $q^A_A(t)$ and $q_A^{k'_n}(t)$ obtained from the above two
equations are the same. Since $\tan(\omega L/2) = -\cot (\omega L/2 - \pi/2)$, for $\omega$ large enough and not very close to $\Omega_0$,
the solutions of (\ref{EFcondTw}) and (\ref{EFkncondTw}) are roughly those solutions for (\ref{EFcond}) shifted by $\pi/L = 1/(2R)$.
(Recall that the modes for the free twisted field have $k'_n = [n - (1/2)]/R$.)

The difference between the spectra of the free fields can alter the beating behavior drastically in different fields
while the values of the parameters $\Omega_0$, $\lambda$, and $L$ are the same. For example,
for $\Omega_0 \approx n/R$, $n\in {\bf Z}$, the beating behavior of $|q^{A}_A|^2$ is not significant in the twisted field, in contrast to
the clear beats in the cases with the same parameter values but in the untwisted field, as those in Figure \ref{EmergeEigen}.
For $\Omega_0 \approx [n-(1/2)]/R$ with large $|n|$, on the other hand, $|q^{A}_A|^2$ in the twisted field has beats at a frequency about
$2\Delta$ with the approximated value of $\Delta$ given in (\ref{Delbeat}) in the weak coupling limit, while
there is no significant beat in the untwisted field with the same parameters, as those in Figure \ref{Specr001W23R2}.

\begin{figure}
\includegraphics[width=3.6cm]{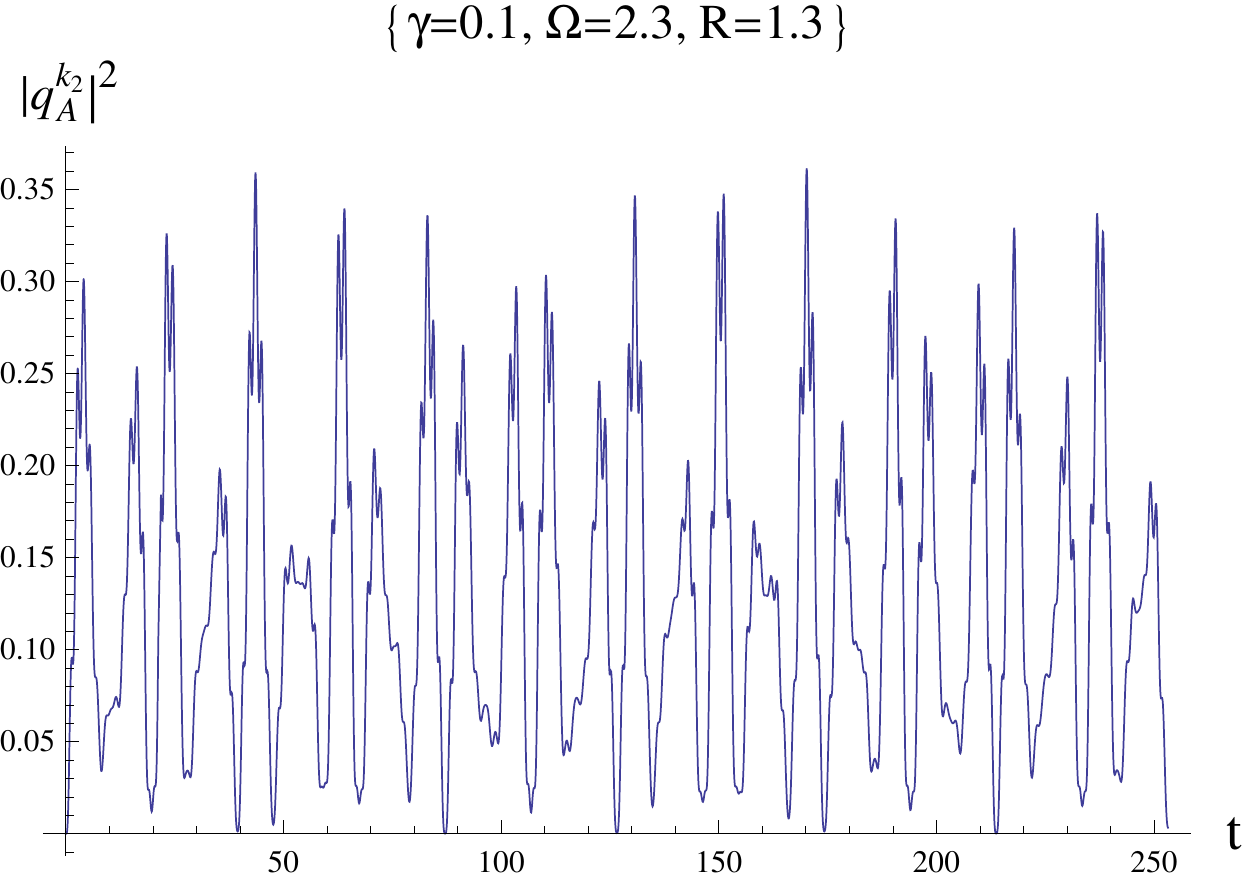}
\includegraphics[width=3.6cm]{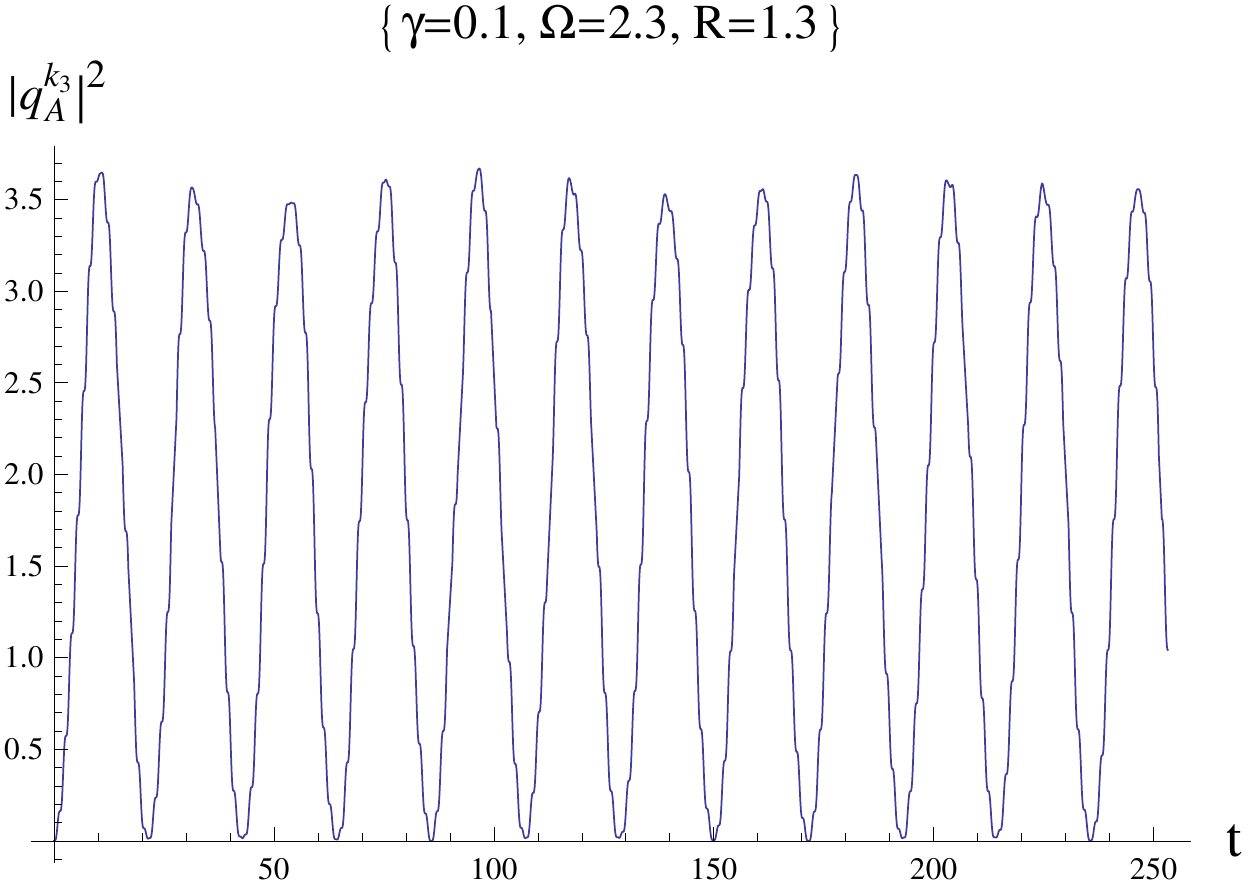}
\includegraphics[width=3.6cm]{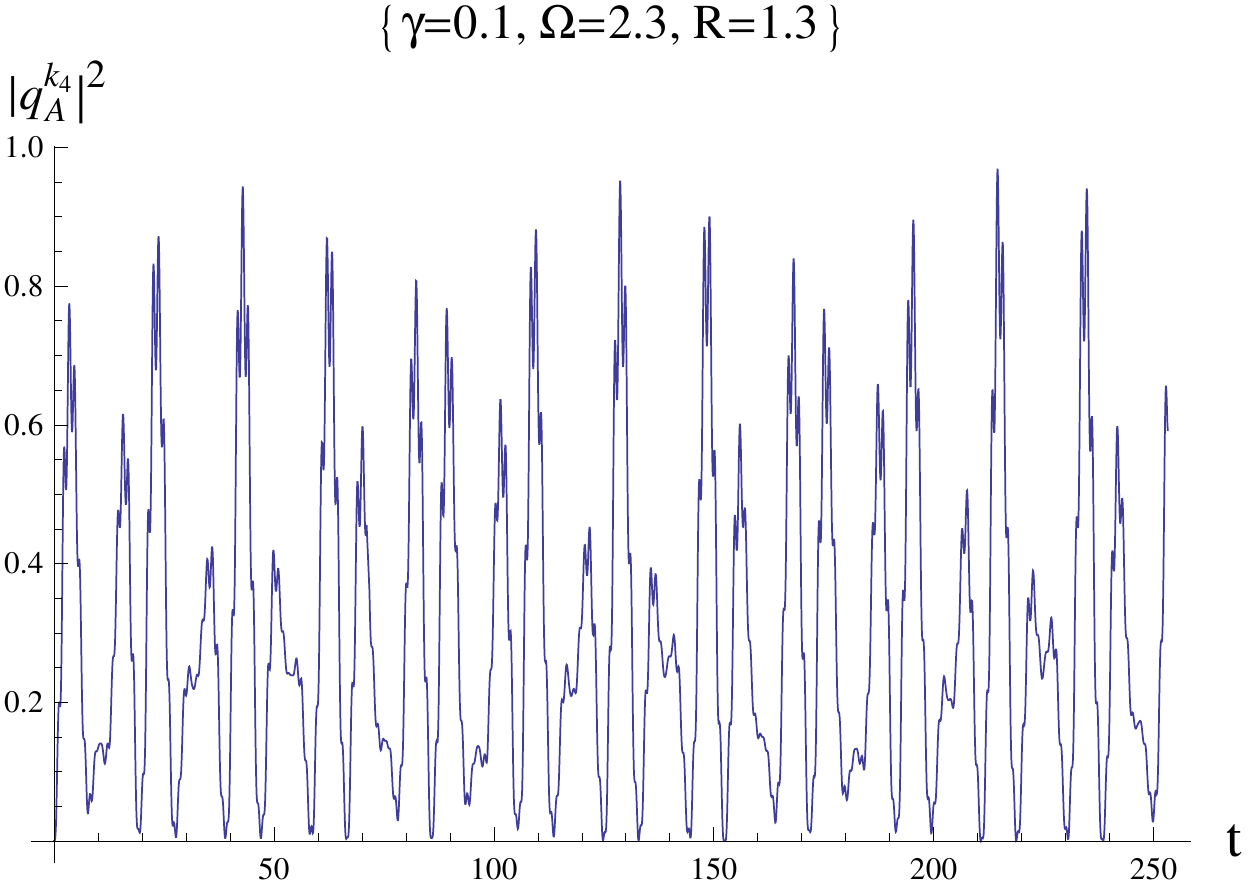}
\includegraphics[width=3.6cm]{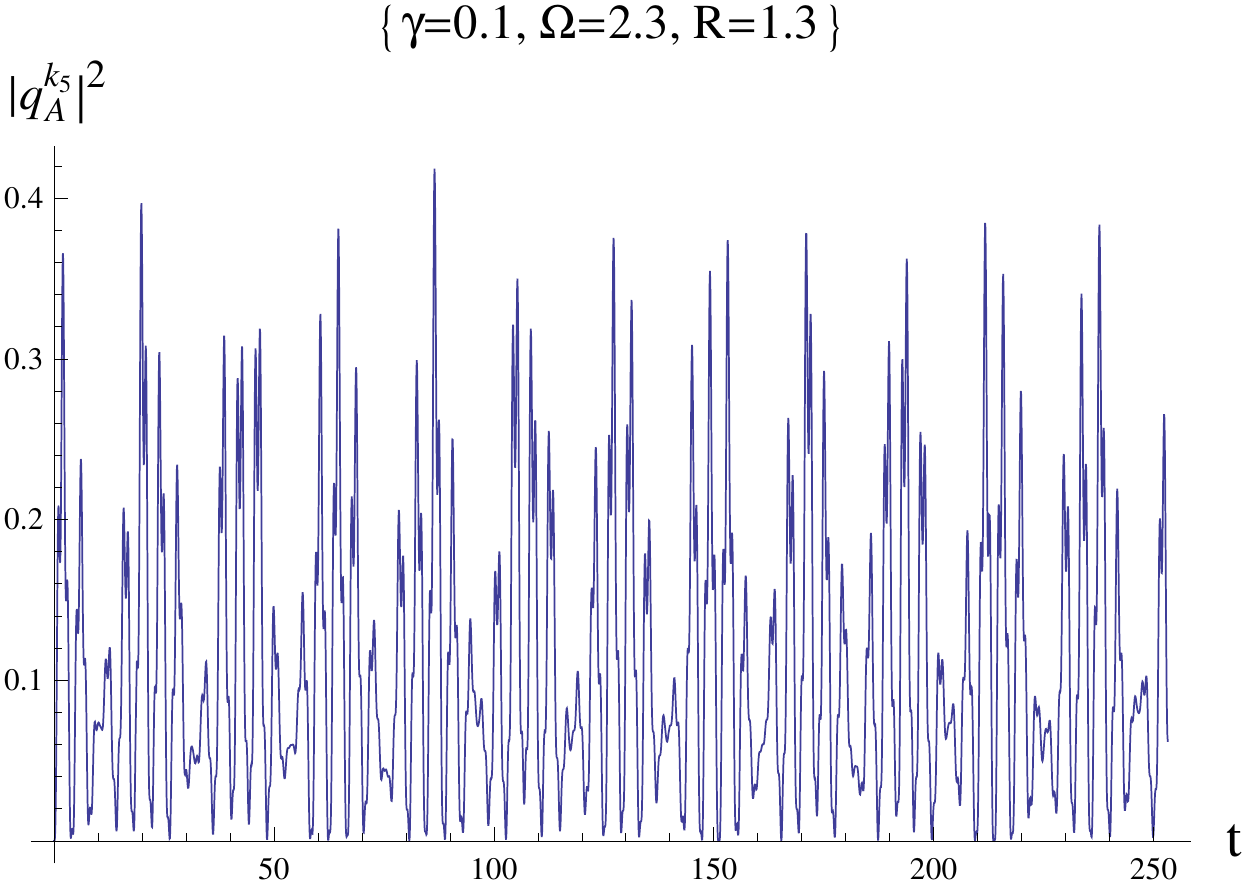}\\
\includegraphics[width=3.6cm]{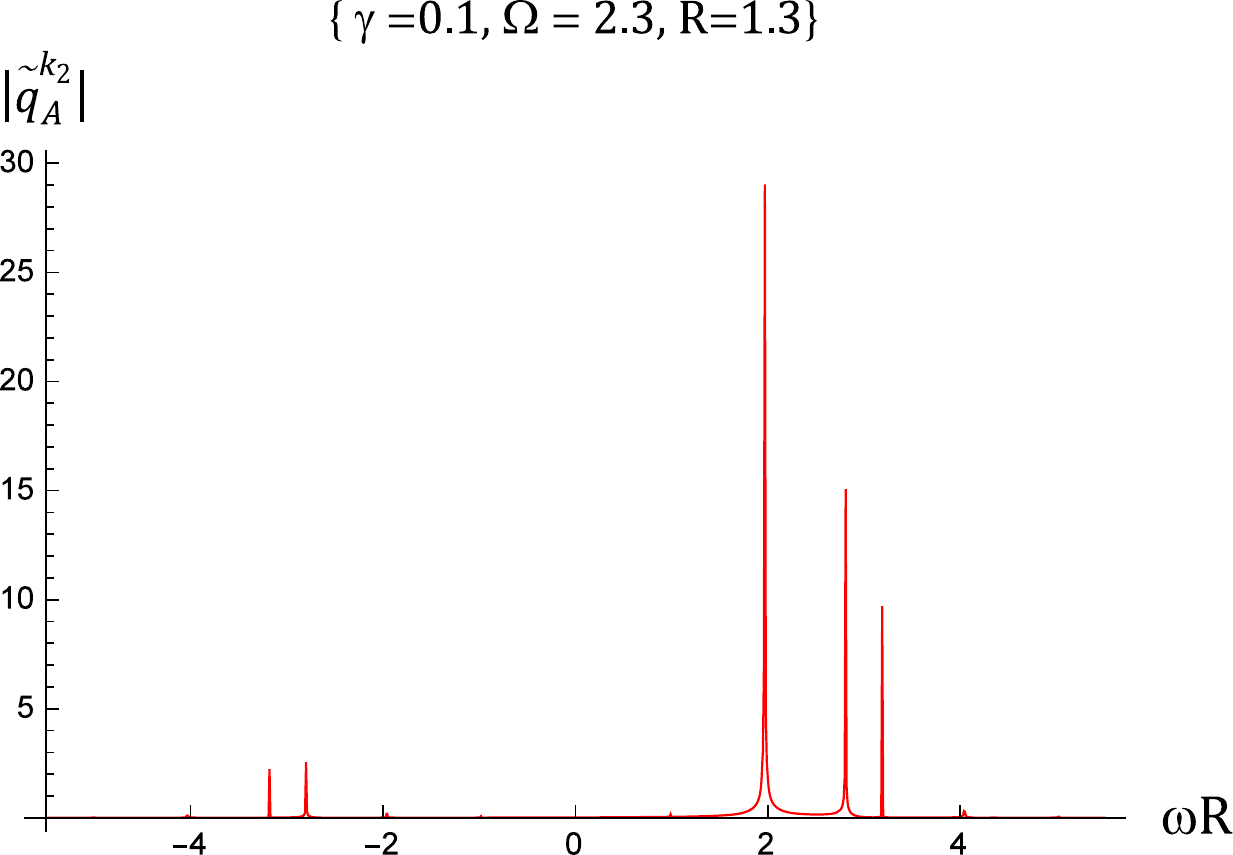}
\includegraphics[width=3.6cm]{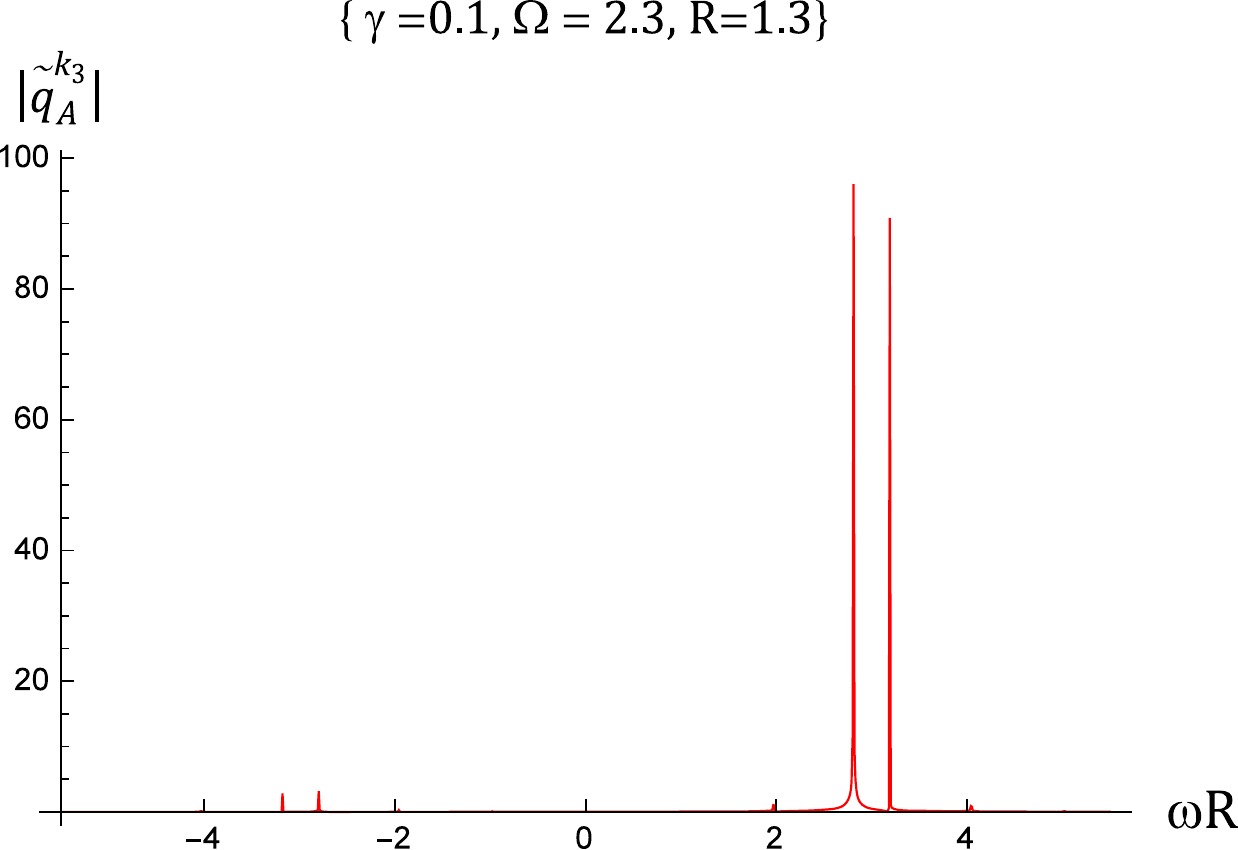}
\includegraphics[width=3.6cm]{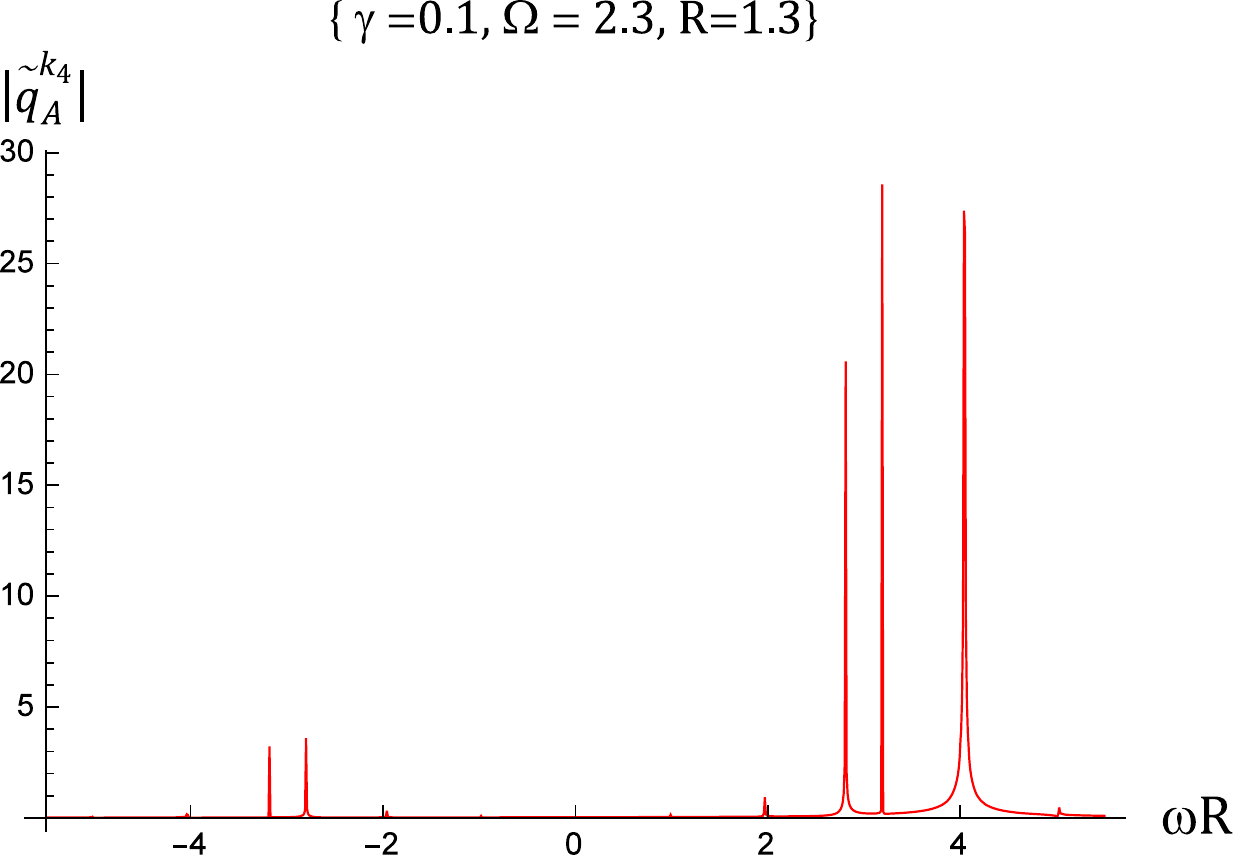}
\includegraphics[width=3.6cm]{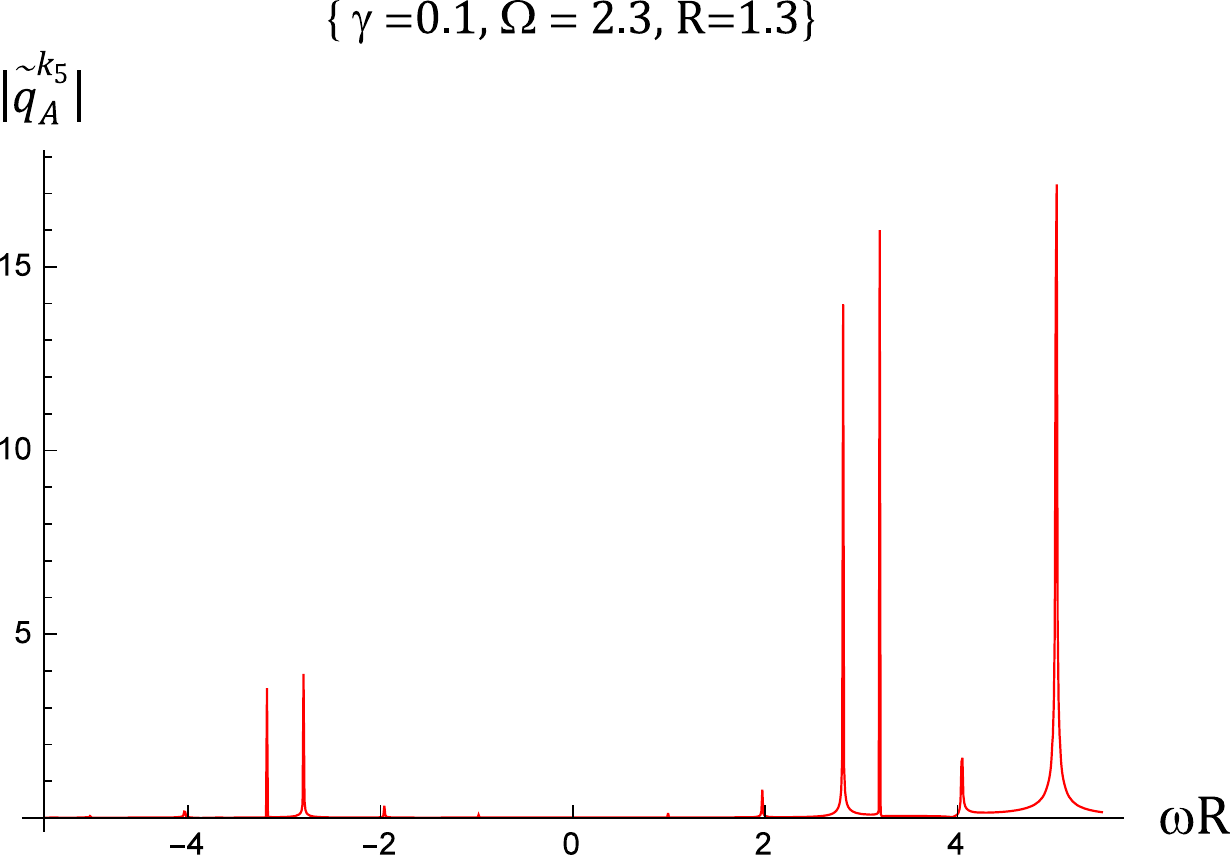}
\caption{$q^{k_n}_A (t) $ (upper row) and their frequency spectra $\tilde{q}^{k_n}_A(\omega)$ (lower row), $n=2,3,4,5$
(from left to right), with the same values of the parameters as those in Figure \ref{EmergeEigen} ($\gamma=0.1$, $\Omega=2.3$,
$R=1.3$). One can see that the eigen-frequencies are exactly the same as those for the $q^A_A$ there.}
%with the same values of the parameters.}
\label{mofnr01W23R13}
\end{figure}

\begin{figure}
\includegraphics[width=3.6cm]{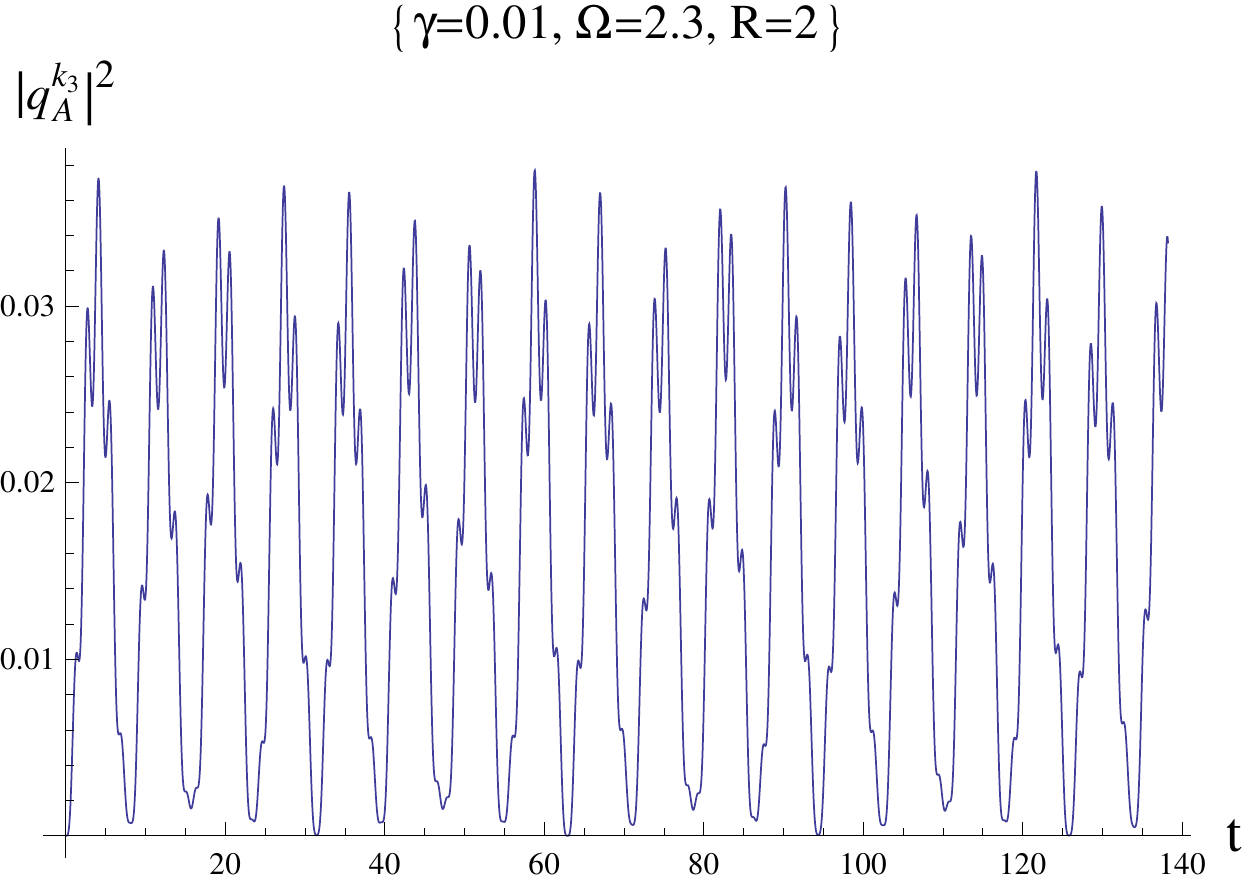}
\includegraphics[width=3.6cm]{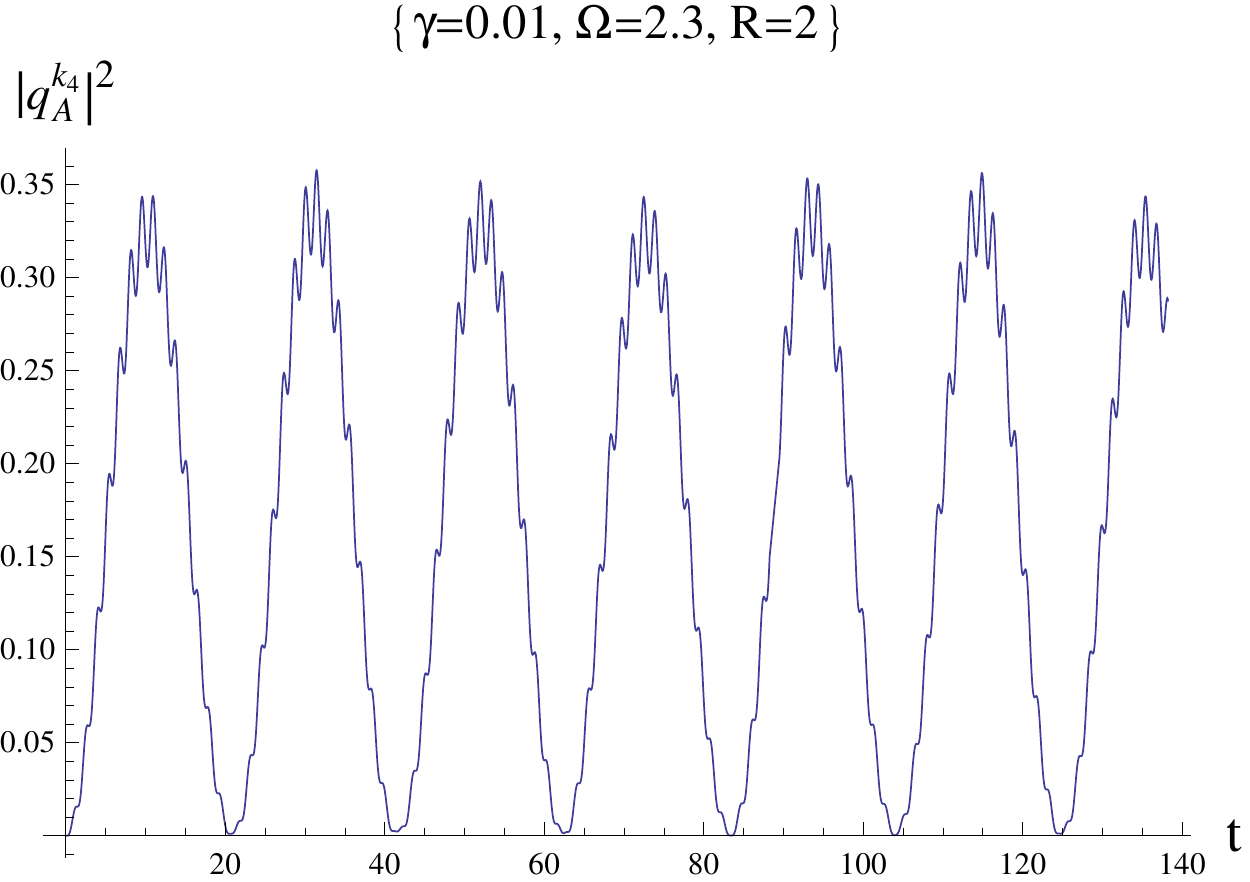}
\includegraphics[width=3.6cm]{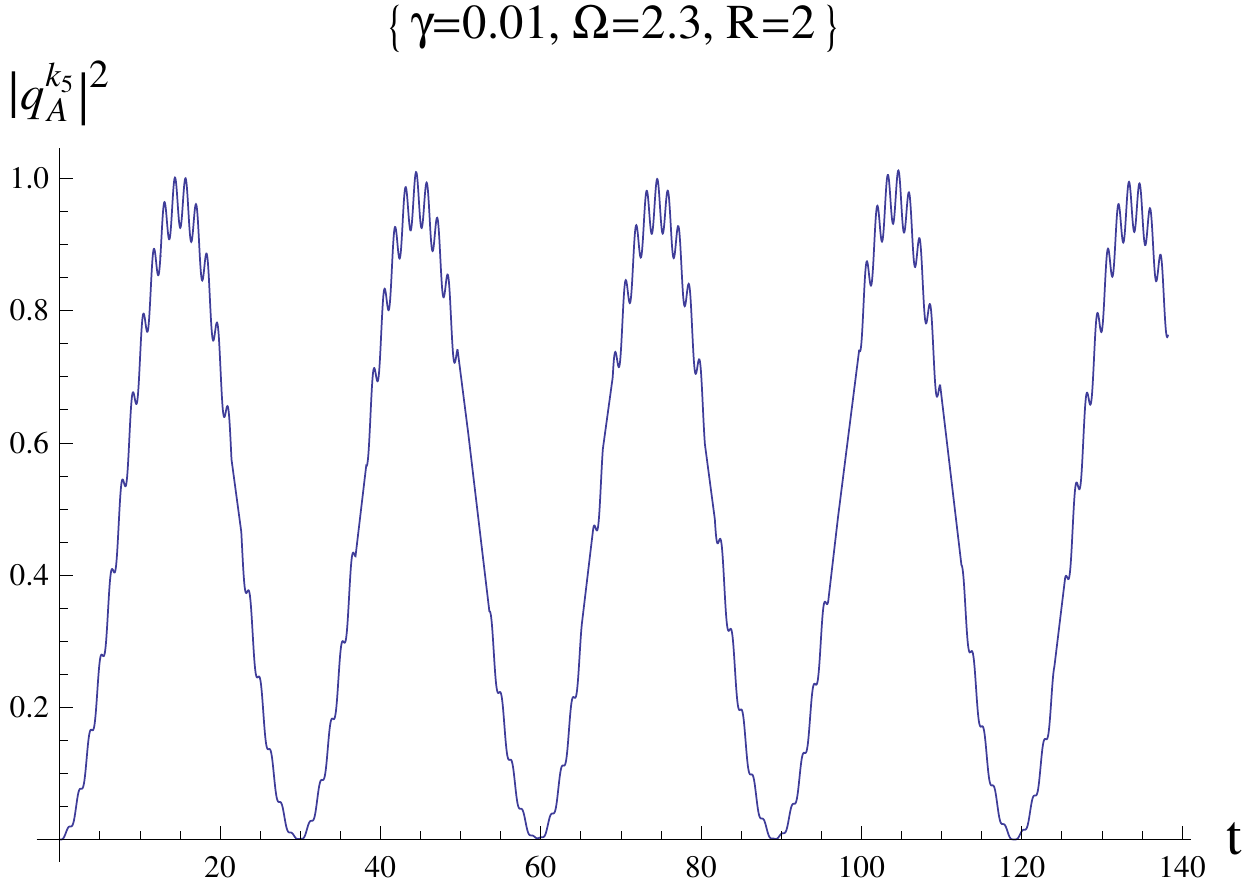}
\includegraphics[width=3.6cm]{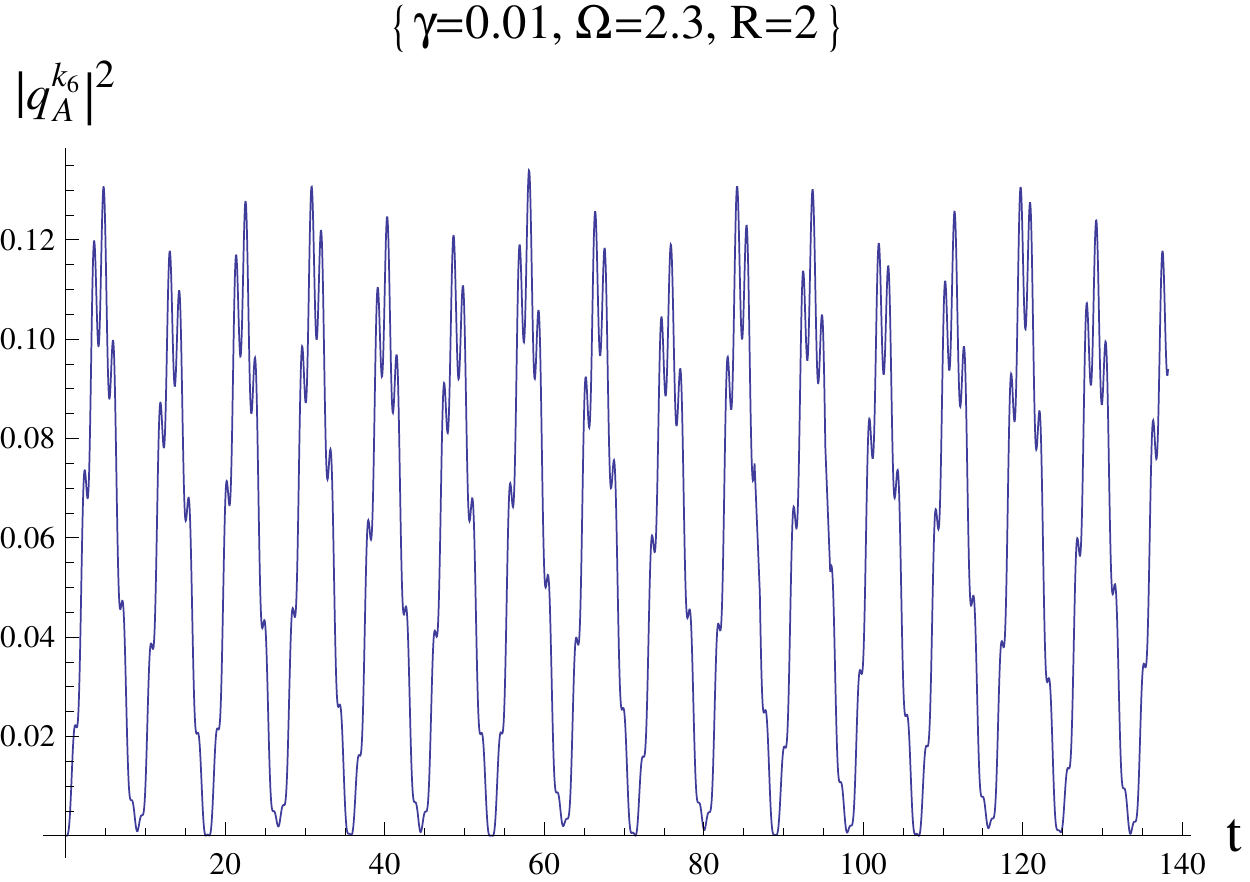}\\
\includegraphics[width=3.6cm]{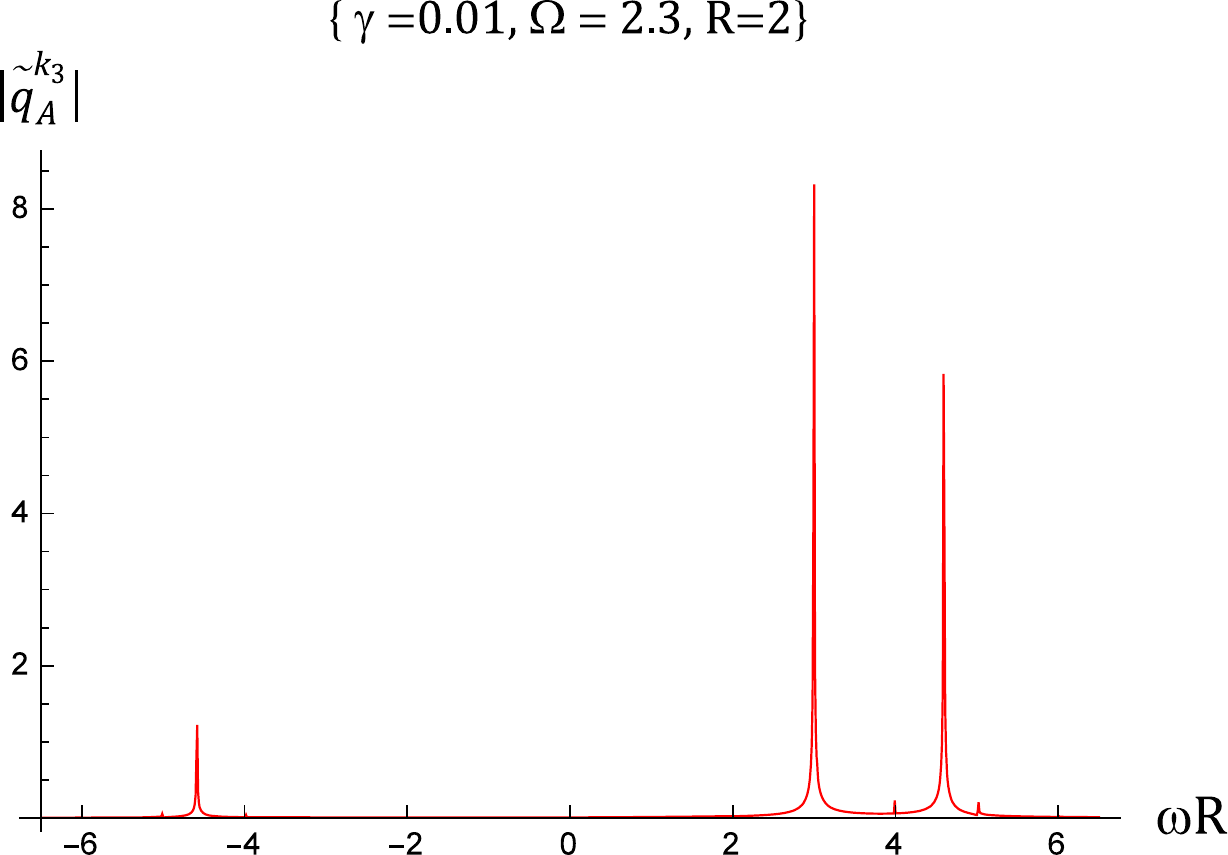}
\includegraphics[width=3.6cm]{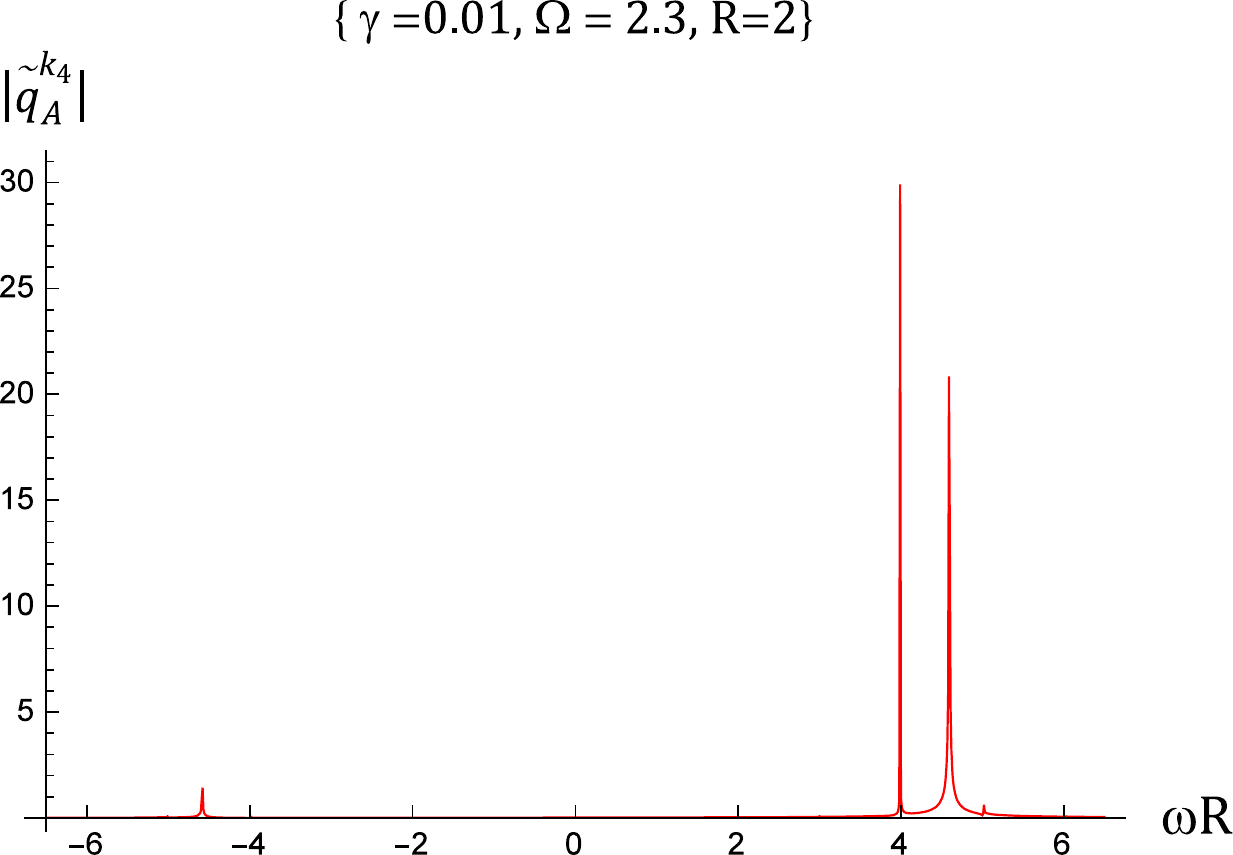}
\includegraphics[width=3.6cm]{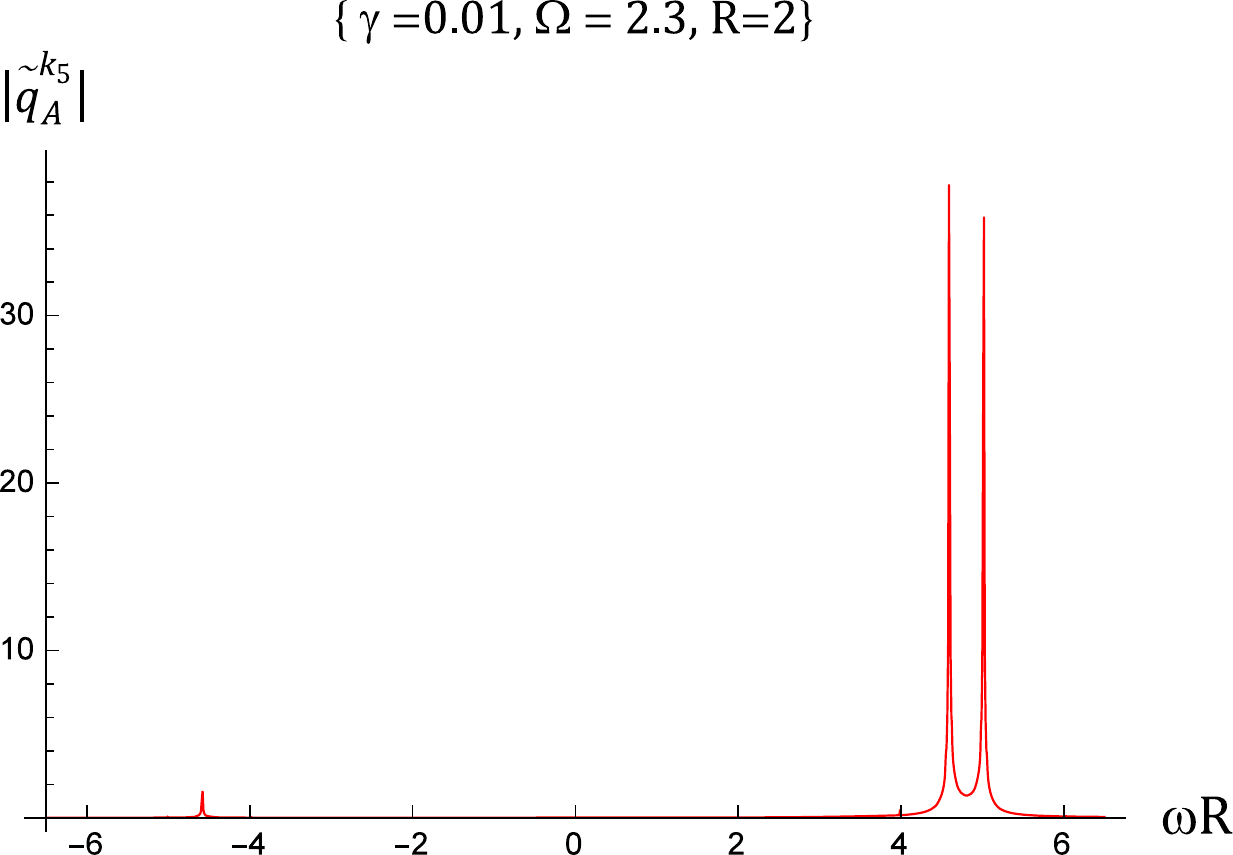}
\includegraphics[width=3.6cm]{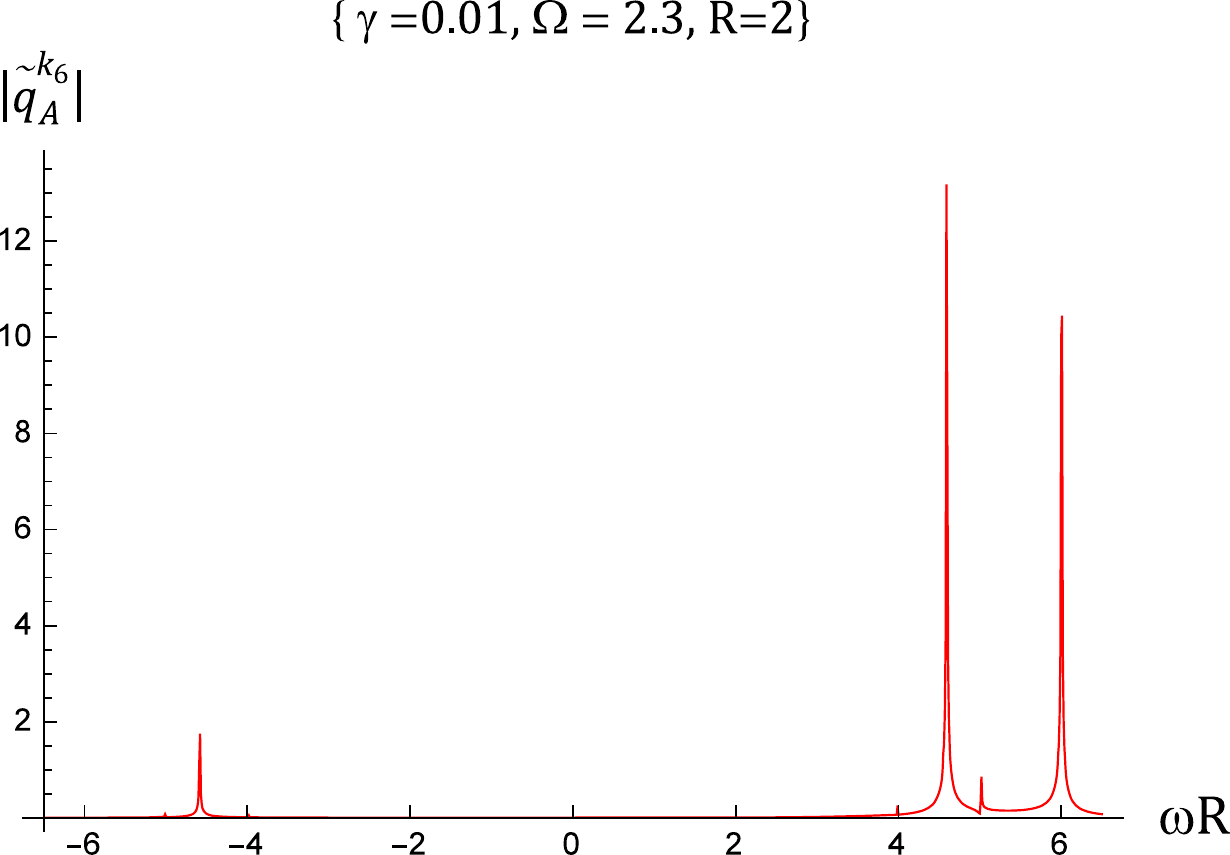}
\caption{$q^{k_n}_A (t)$ (upper) and their frequency spectra $\tilde{q}^{k_n}_A$ %with $T=100\times2\pi R$
(lower), $n=3,4,5,6$ (from left to right), with the same parameter values in Figure \ref{Specr001W23R2} ($\gamma=0.01$, $\Omega=2.3$,
$R=2$). The eigen-frequencies are exactly the same as those for the $q^A_A$ there, %in Figure \ref{Specr001W23R2},
but now one can see the beats because there are always two dominant eigen-frequencies ($\omega\approx \Omega_0$ and $|k_n|$) for
the detector-field mode functions $q^{k_n}_A (t)$.}
\label{mofnr001W23R2}
\end{figure}

%%%%%%%%%%%%%%%%%%%%%%%%%%%%%%%%%%
\subsection{Two-point correlators and detector-field entanglement}

Suppose the initial state of the combined system defined on the $t=0$ slice is a factorized state,
\begin{equation}
  | \psi(0) \rangle = |g\rangle \otimes |0_L\rangle
\label{IS1dec}
\end{equation}
which is a product of the ground state of the free detector $|g\rangle$ and the vacuum state of the free field $| 0_L \rangle$ in
the Einstein cylinder ${\bf S}^1 \times {\bf R}_1$ \cite{BD82}.
For the untwisted field, the vacuum state is further factorized to $| 0_{\rm L} \rangle  = | 0_L\rangle^{}_{\rm nz} \otimes
|0_L\rangle^{}_{\rm z}$, where $| 0_L\rangle^{}_{\rm nz}$ is the lowest energy state of the field with non-zero wave vector, and
$|0_L\rangle^{}_{\rm z}$ is the initial state of the zero mode of the free field, chosen to satisfy $\hat{b}^{}_{k_0}|0_L\rangle^{}_{\rm z}
=0$, which gives the minimal uncertainty ${}_{\rm z}\langle 0_L|(\hat{\Phi}_{k_0}(0))^2 |0_L\rangle^{}_{\rm z} = {}_{\rm z}\langle 0_L|
(\hat{\Pi}_{k_0}(0))^2|0_L\rangle^{}_{\rm z} = \hbar/2$. For the twisted field there is no zero-mode, thus no similar separation is needed.
With the factorized initial state (\ref{IS1dec}), the symmetric two-point correlator of the detectors splits into two parts, e.g.,
\begin{eqnarray}
  \langle \hat{Q}_A^2(t)\rangle &\equiv& \langle \hat{Q}_A(t), \hat{Q}_A(t) \rangle \equiv
  {1\over 2} \lim_{t'\to t} \langle \psi(0) | \hat{Q}_A(t)\hat{Q}_A(t') + \hat{Q}_A(t')\hat{Q}_A(t)|\psi(0)\rangle \nonumber\\
  &=&\langle \hat{Q}_A(t), \hat{Q}_A(t) \rangle^{}_{\rm a} +  \langle \hat{Q}_A(t), \hat{Q}_A(t) \rangle^{}_{\rm v}
\end{eqnarray}
where, from (\ref{Qmod}),
\begin{eqnarray}
   \langle \hat{Q}_A(t), \hat{Q}_A(t) \rangle^{}_{\rm a} &=& {\hbar\over 2\Omega_0} \left| q^{A}_A(t) \right|^2,\\
   \langle \hat{Q}_A(t), \hat{Q}_A(t) \rangle^{}_{\rm v} &=& \sum_{n\in {\bf Z}} {\hbar\over 2\tilde{\omega}_n}
       \left|q^{k_n}_A(t)\right|^2. ,
\end{eqnarray}
for the untwisted field, and with $k_n \to k'_n$ for the twisted field.
The uncertainty ${\cal U}\equiv \sqrt{\langle Q_A^2\rangle\langle P_A^2\rangle -\langle Q_A,P_A \rangle^2}$
and the purity ${\cal P}\equiv Tr (\rho^{\rm R}_A)^2 = 1/(2{\cal U})$, where $\rho^{\rm R}_A$ is the reduced state of detector $A$,
will be fully determined by the two-point correlators of the detector
since the quantum state of the combined system is always Gaussian in this linear system.

\subsubsection{UV cutoff}

In Appendix \ref{2ptM2}, the symmetric two-point correlators of a UD' detector at rest in ${\bf R}^1_1$ with the initial state similar to
(\ref{IS1dec}) have been worked out. One can see that $\langle Q^2_A\rangle$, $\langle Q^{}_A, P^{}_A\rangle$, and $\langle P_A^2\rangle$
all suffer from the UV logarithmic divergence. To control, one has to introduce a UV cutoff %, which corresponds to the positive constant
$\omega^{}_M$ in the integrals (\ref{C2n1})-(\ref{I3def}).
The values of $\omega^{}_M$ cannot be too small, or the uncertainty relation of the detector
${\cal U} \ge\hbar/2$ will be violated due to the inconsistency with applying the retarded Green's function in the calculation, similar
to the reason we mentioned in Section \ref{themodel}.

How about the correlators for the UD' detector in ${\bf S}^1\times{\bf R}_1$?

The logarithmic UV divergence from the mode-sum similar to the $I_3$ term in Appendix \ref{2ptM2} will still arise
in calculating the two-point correlators of the detector, and will never decay out due to the echoes.
To get rid of such a UV divergence, again we need to introduce a UV cutoff $n_{\rm max}$ to exclude the modes with $|n| > n_{\rm max}$ for
$k_n$ or $k'_n$ in our effective theory. Again,
to be consistent with the retarded Green's function (\ref{retG}) we applied, the UV cutoff here cannot be too small to violate the
uncertainty relation.

\subsubsection{Linear entropy}
%%%%%%%%%%%%%%%%%%%%%%%%%%%%%%%%%%%%%%%%%%%%%%%%%%%%%%%%%%%%%%%%%%%%%%%%%%%%%%%%%%%%%%%%%%%%%%%%%%%%%%%%%%

\begin{figure}

\includegraphics[width=5cm]{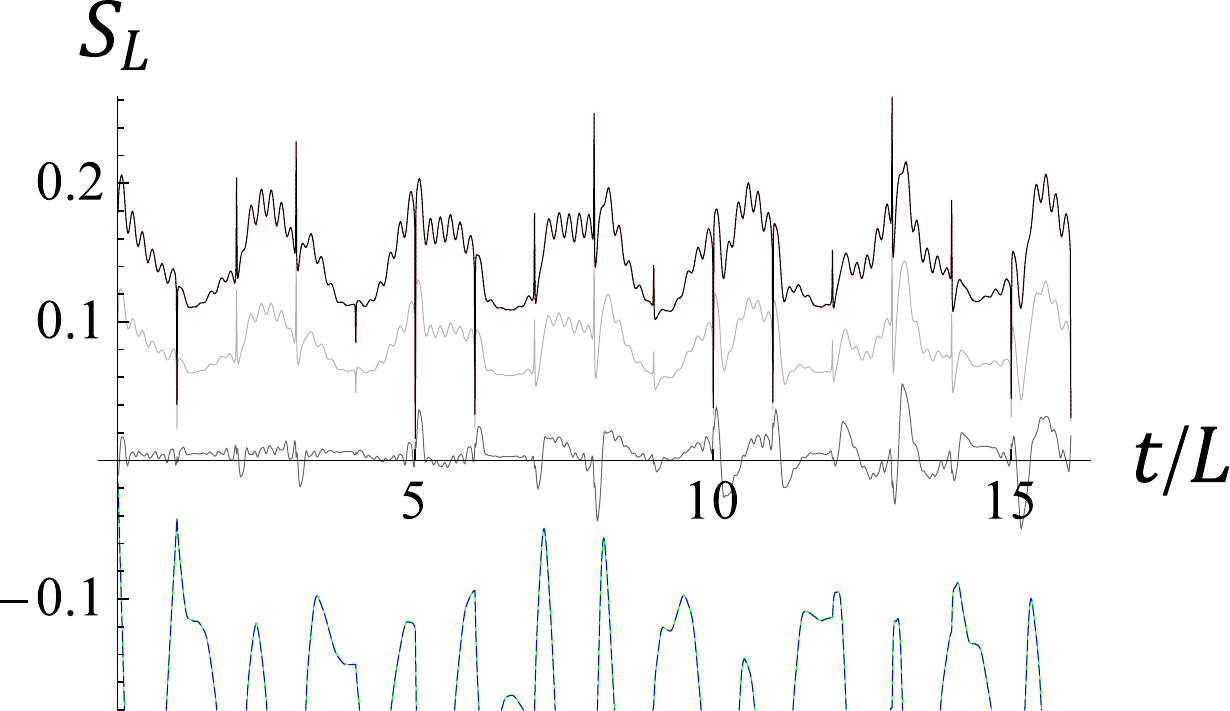}
\includegraphics[width=4.8cm]{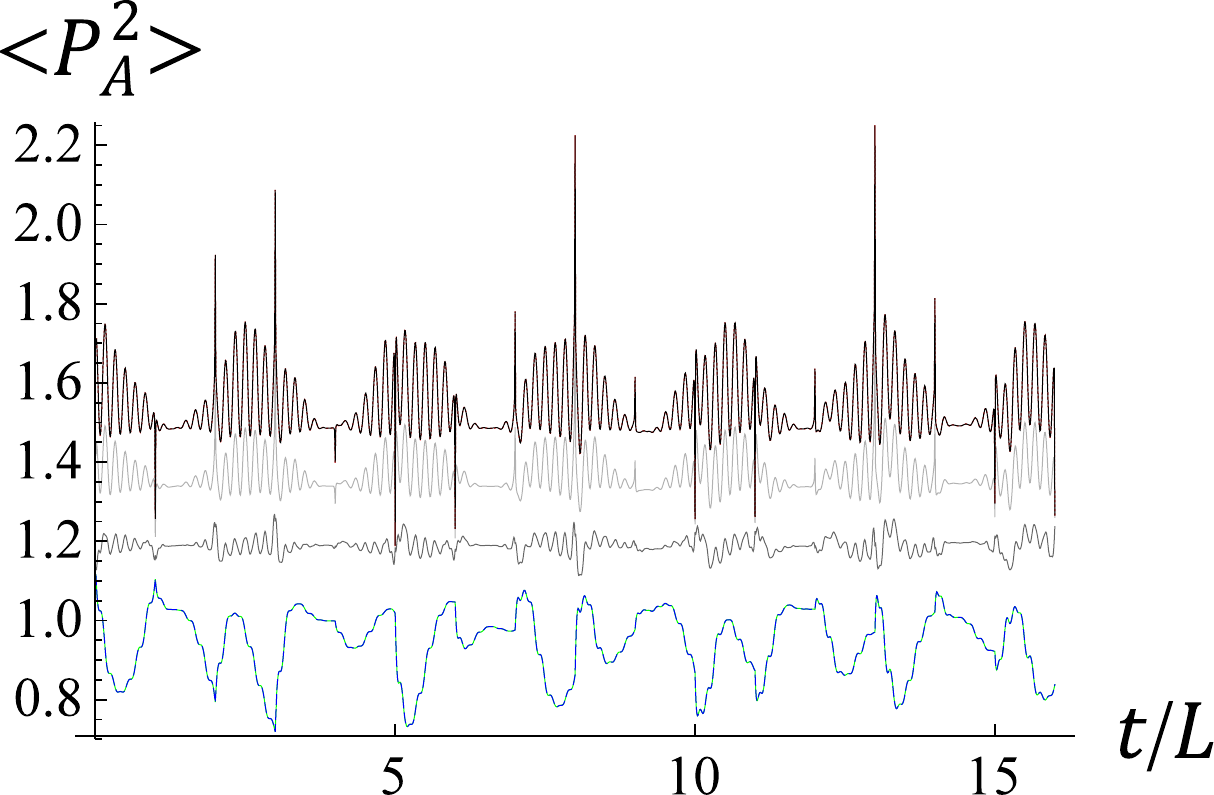}
\includegraphics[width=4.8cm]{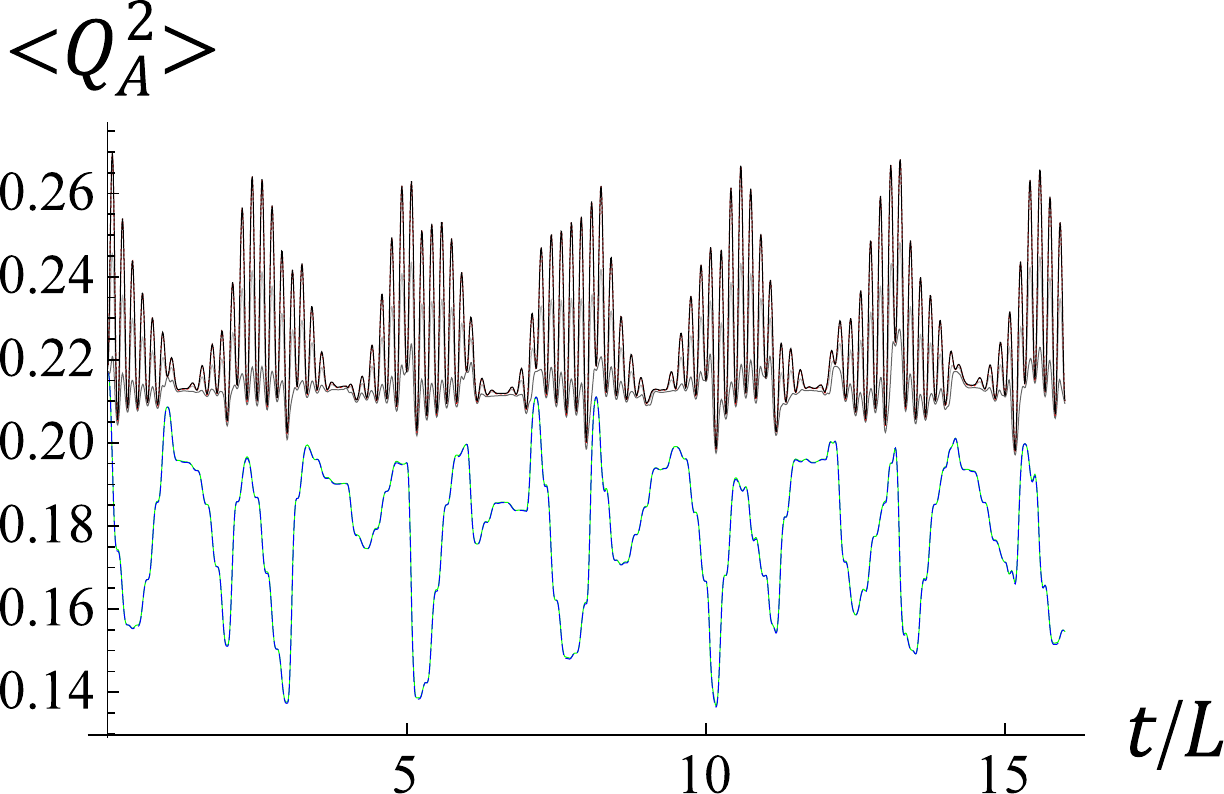}\\
%\hspace{4.8cm}
\includegraphics[width=5cm]{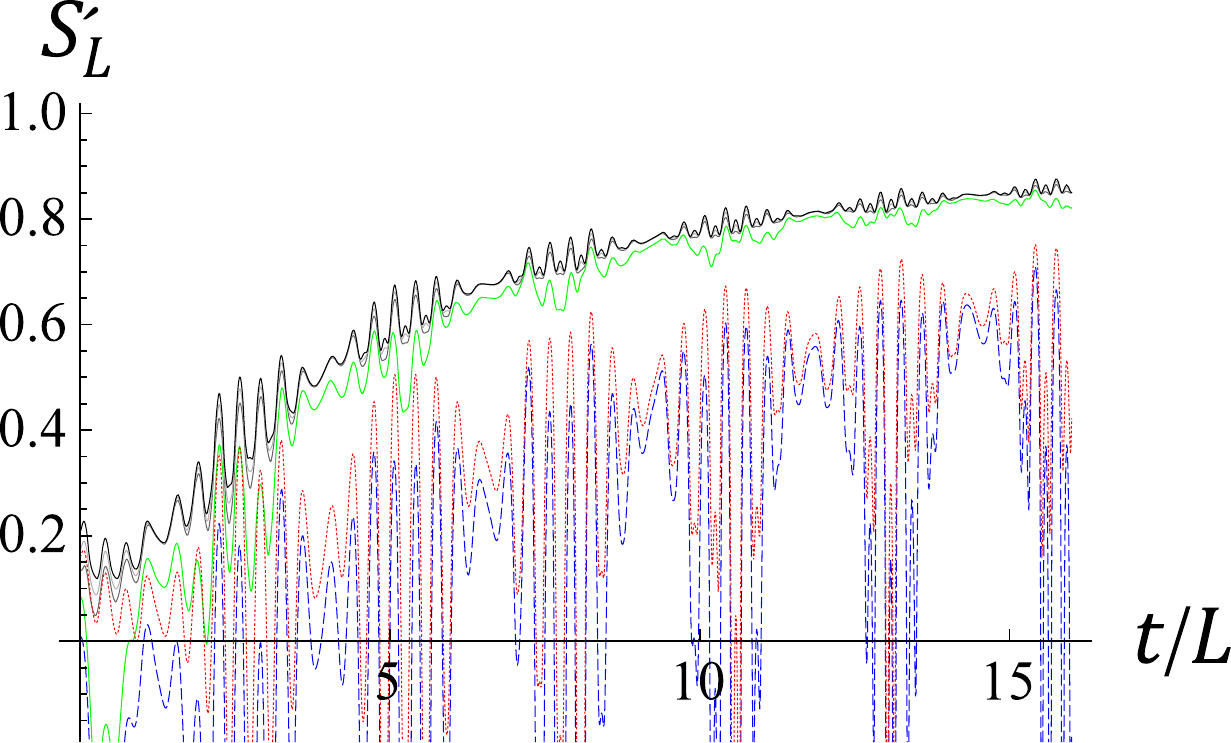}
\includegraphics[width=4.8cm]{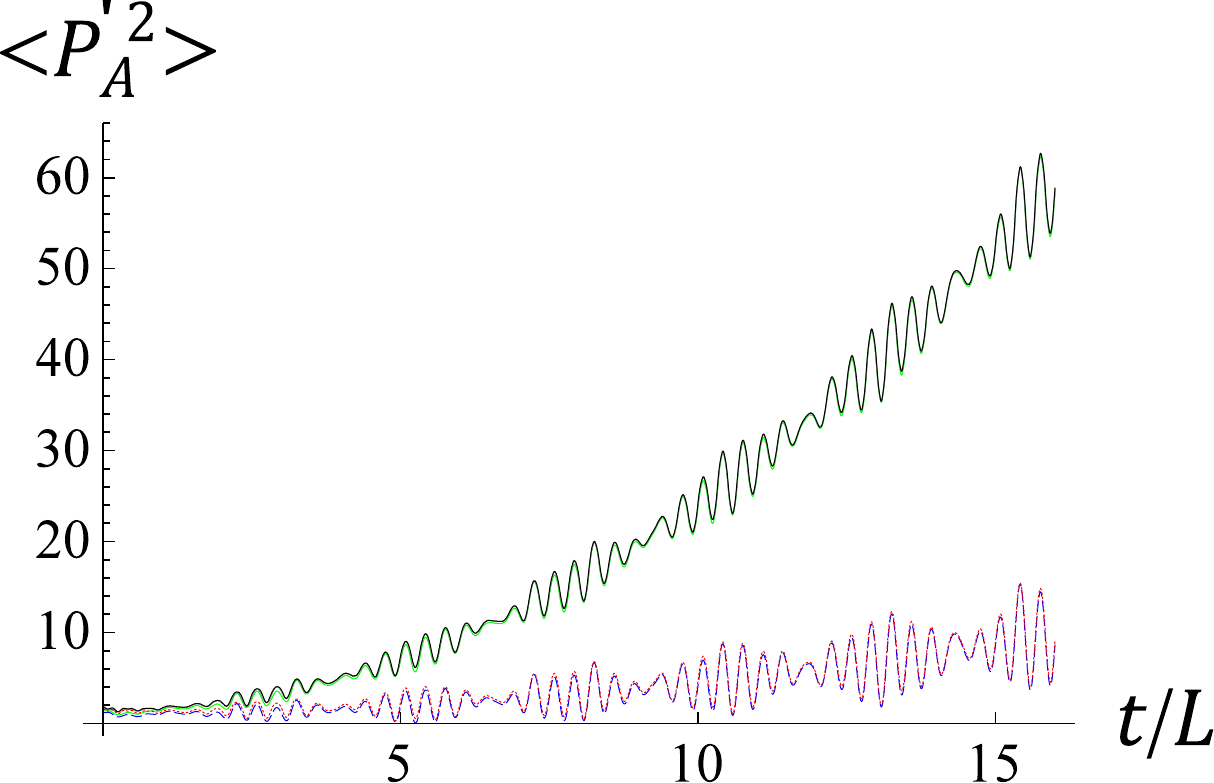}
\caption{Time-evolutions of the linear entropy $S^{}_L$, contributed by the correlators $\langle \hat{P}_A^2(t)\rangle$ and
$\langle \hat{Q}_A^2(t)\rangle$ (upper row) of single detector $A$ in the
untwisted field, with the same parameter values as those in Figures \ref{EmergeEigen} and \ref{mofnr01W23R13}.
Replacing $P^{}_A$ by the gauge-dependent $P'_A$ defined in (\ref{Pdef2}), %, which are not physically measurable,
both $\langle P'^2_A\rangle$ (lower-right) and the corresponding linear entropy $S'_L$ (lower-left) are always increasing at large time
scales. The dark-gray, light-gray, and black curves represent the results contributed by the modes with wave number $k_n = n/R$ from
$|n|=0$ up to $|n|=10$, $100$, and $1000$, respectively. 
One can get a taste of the logarithmic divergence from these three curves, with the magnitudes proportional to
$\log n_{\rm max}$ in the upper-middle plot.
The green curves are contributed only by the modes with $n=0$ and $|n|=3$, the latter are almost on resonance with the detector.
The blue-dashed %($|n|=3$) 
and red-dotted curves %($|n|=1$ to $100$) 
are the green and the black curves with the contributions by the zero mode ($n=0$) removed, respectively.
The difference from other curves with the zero-mode contributions is significant in the lower plots.}
\label{SL1}
\end{figure}

\begin{figure}
\includegraphics[width=5cm]{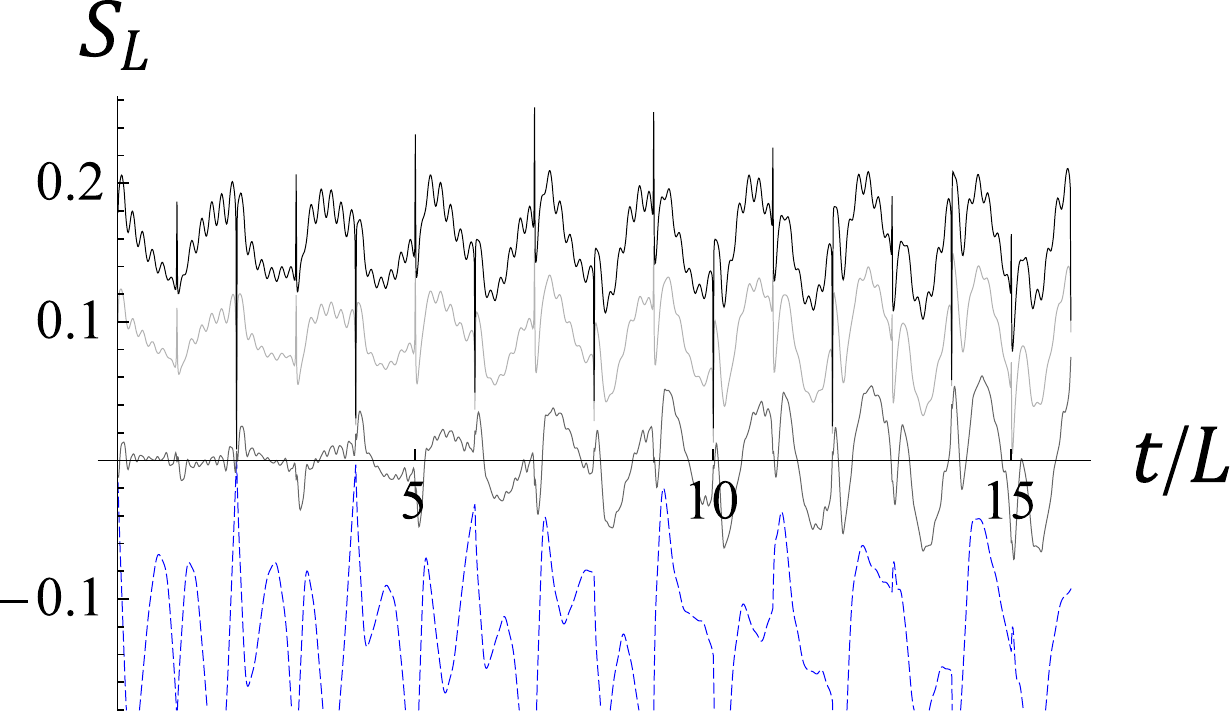}
\includegraphics[width=4.8cm]{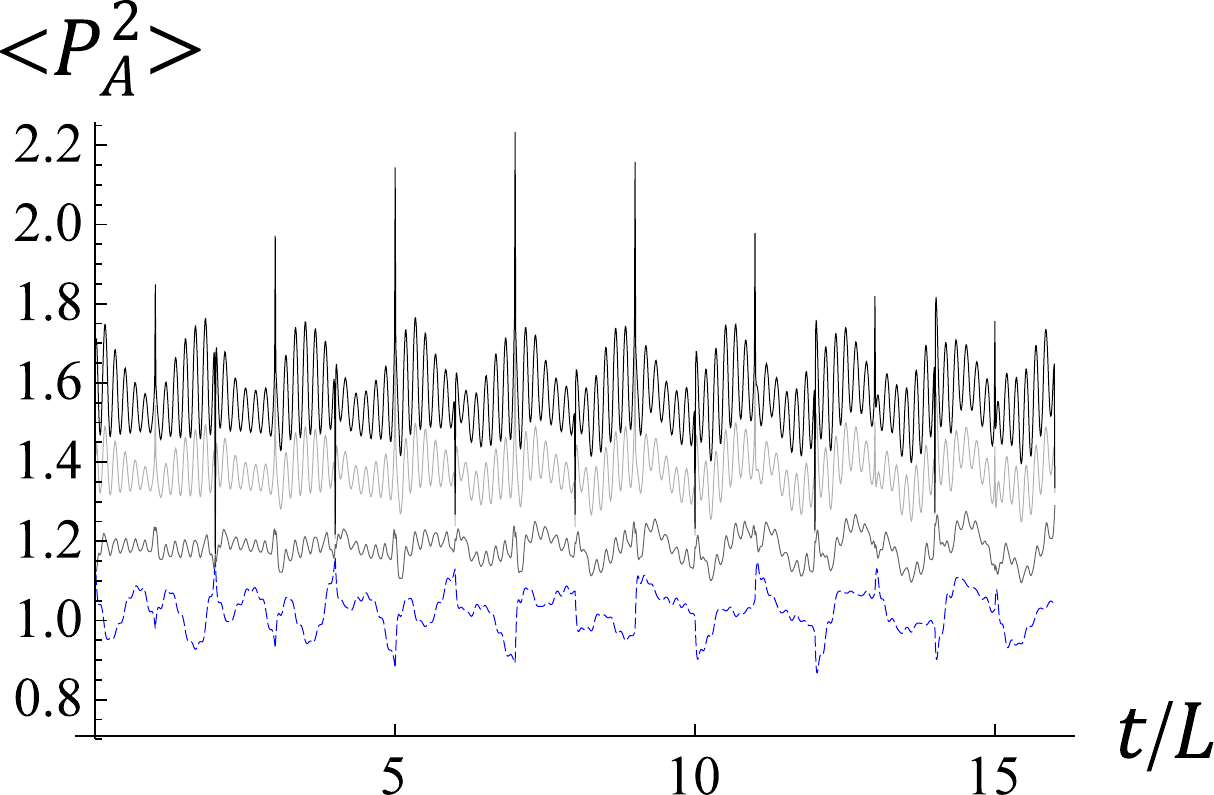}
\includegraphics[width=4.8cm]{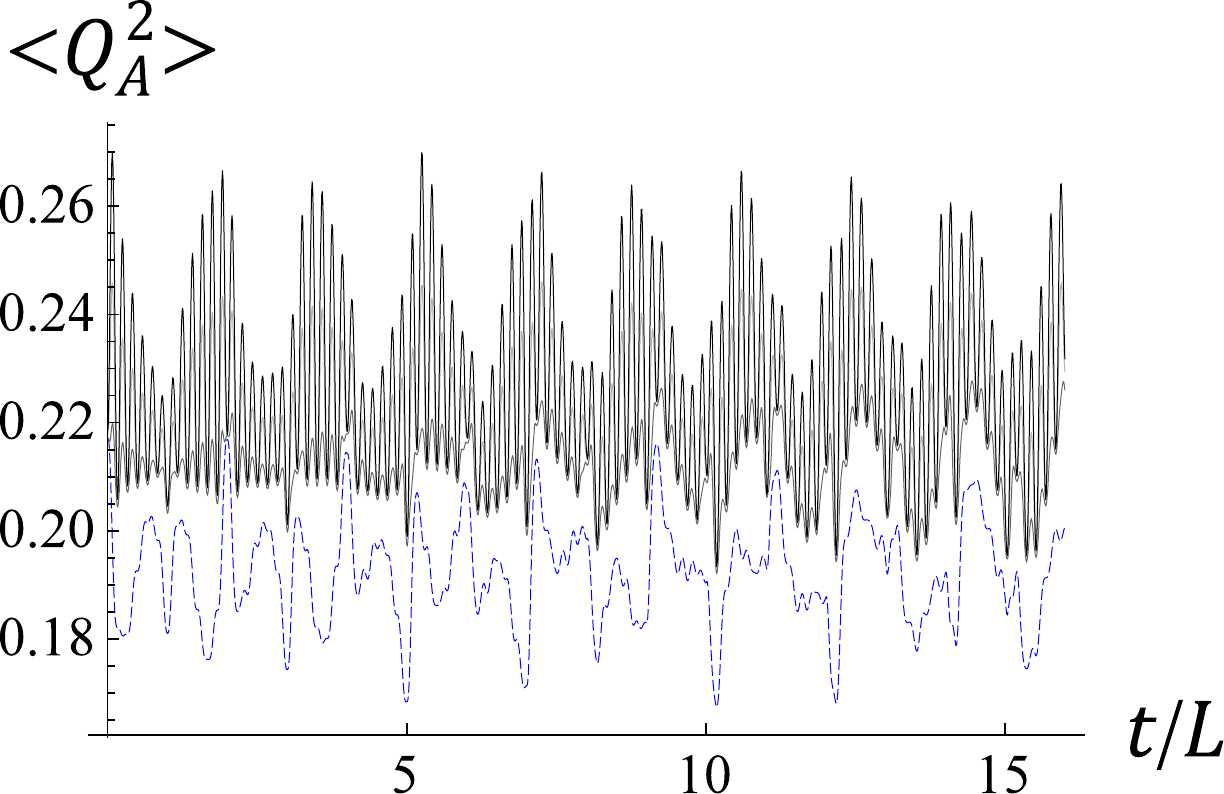}\\
\includegraphics[width=5cm]{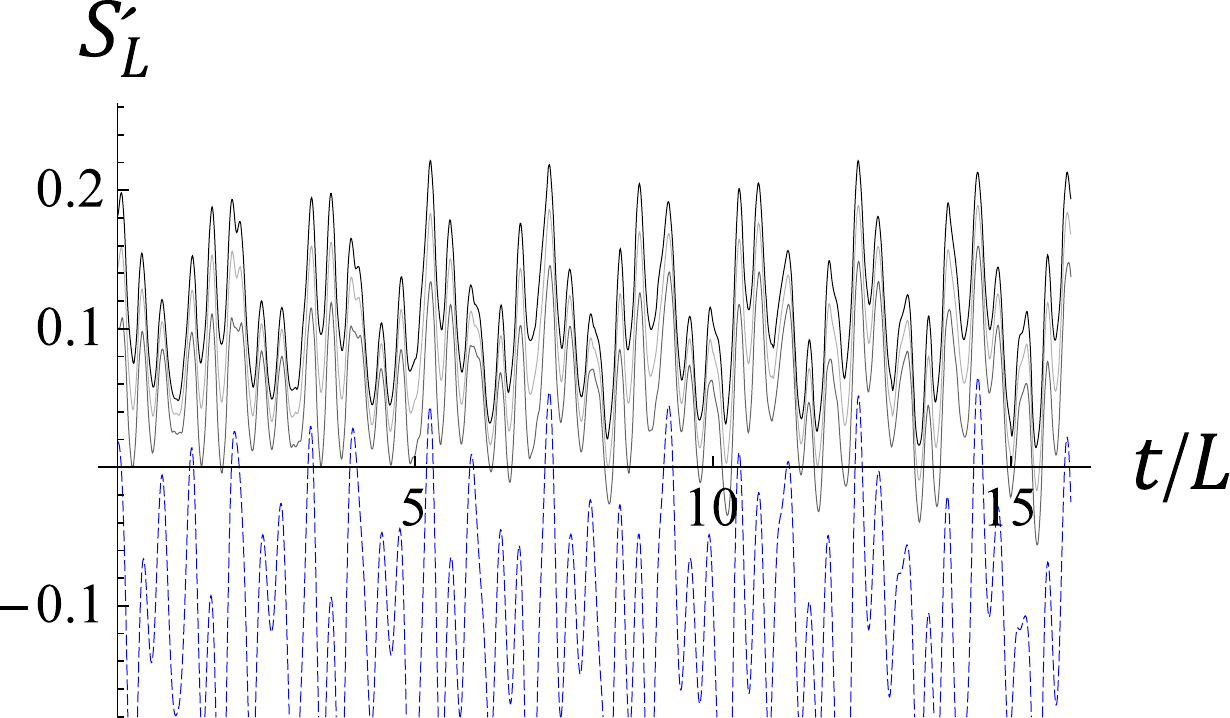}
\includegraphics[width=4.8cm]{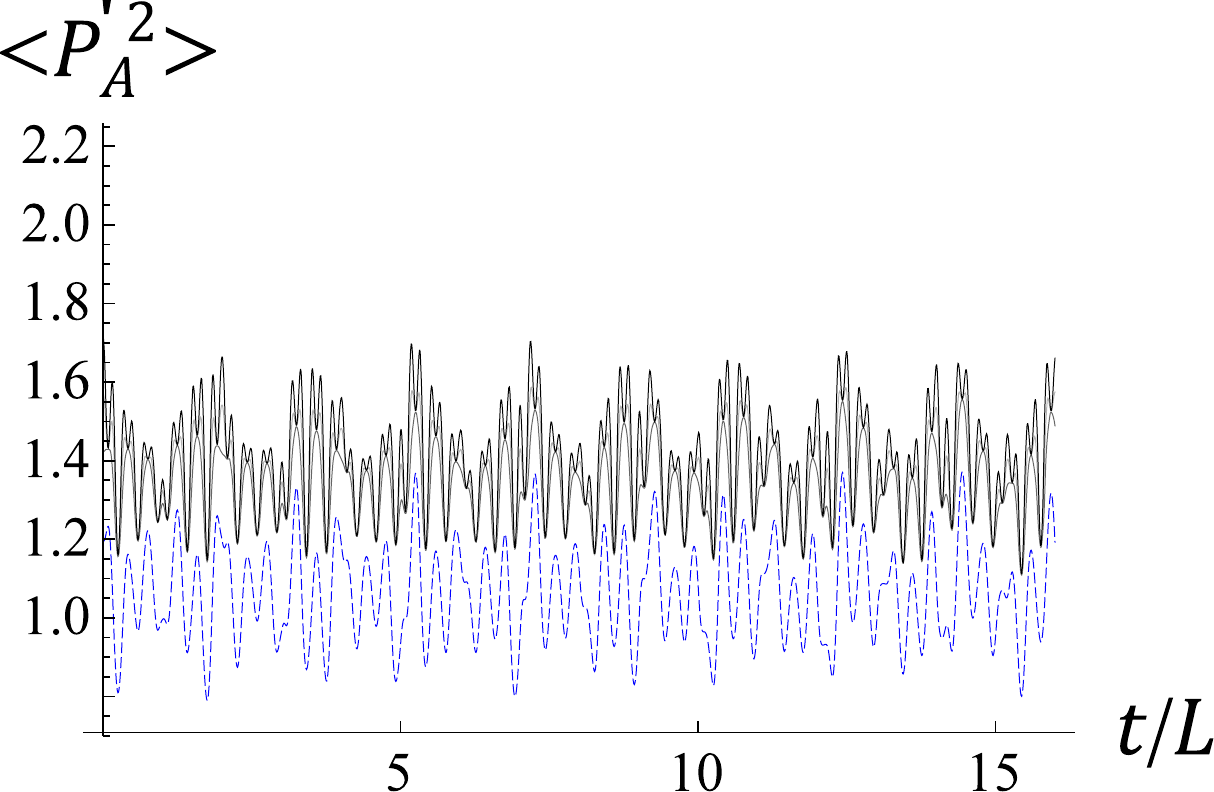}
\includegraphics[width=4.8cm]{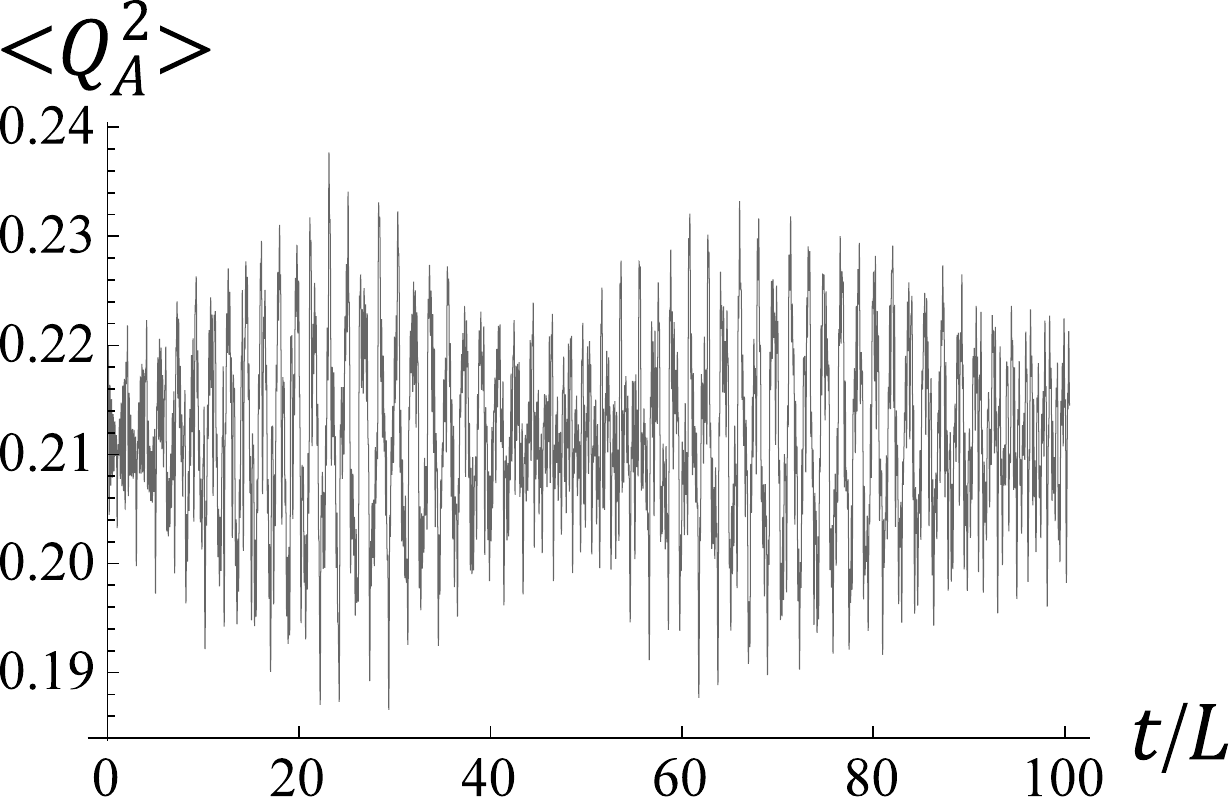}
\caption{
Results similar to those in Figure \ref{SL1} with the same values of the parameters except $\varepsilon=-1$ here for the twisted field.
The dark-gray, light-gray, and black curves represent the results contributed by the modes with $k'_n = [n-(1/2)]/R$ from $|n|=1$ up to
$|n|=10$, $100$, and $1000$, respectively, while the blue-dashed curve is contributed only by the four modes with $|n|=3$ and $4$,
which have the closest frequencies to the detector's. There seems to be no beat with period longer than $L$
in the upper plots. However, the beats with very long period
(about $50 \times L$) emerge in the lower-right plot (contributed by the modes with $|n|=1$ to $10$).
They are produced by the three dominant eigen-modes at eigen-frequencies
$(1.87266, 2.30377, 2.75063)$ [around $(\omega_3, \Omega_0, \omega_4)$]. Indeed, the beat frequency is about
$||2.30377-1.87266|-|2.75063-2.30377|| \approx 0.01575 \approx 2\pi/(48.84L)$ here.
The lower-left and lower-middle plots show the results in terms of the unphysical $P'_A$ defined in (\ref{Pdef2}).
The differences from the results with the kinetic momentum are not as significant as those in the untwisted field:
only the spikes around $t=nL$ in the upper plots are smeared.}
\label{SL1Pi}
\end{figure}

In Figures \ref{SL1}$-$\ref{SL3Pi} we show selected numerical results of the two-point correlators and the detector-field
entanglement in this model with the initial state (\ref{IS1dec}). The degree of entanglement here is characterized by the
linear entropy $S_L = 1 - {\cal P}$: the higher the value of $S_L$, the stronger the entanglement between the detector and the field.

In contrast to the simpler dissipative behavior in Minkowski space, where the time-slices are not compact and the spectrum of the field
modes is continuous \cite{LH07, LH09}, the behaviors of the two-point correlators $\langle \hat{Q}_A^2\rangle$, $\langle  \hat{P}_A^2
\rangle$, $\langle\hat{Q}_A,\hat{P}_A\rangle$, and the linear entropy $S_L$ in ${\bf S}^1 \times {\bf R}_1$ are more complicated.
At early times before the first echo arrives ($t/L<1$), the correlators and $S_L$ do behave similarly to their Minkowskian counterparts
(also see Appendix \ref{2ptM2}). At a large time scale, however, the a-parts of the correlators never decay out, and the beat of the
mode function $q^A_A$ in Figure \ref{EmergeEigen} is obvious in Figure \ref{SL1}, while there is no significant beat in the upper plots
of Figure \ref{SL3}, as indicated in Figure \ref{Specr001W23R2}. This is also true for the twisted field in Figures \ref{SL1Pi} and
\ref{SL3Pi}, though the beats of their mode functions are not shown in this paper.

The behavior of $S_L$ cannot be approximated by including the contributions only by the one
or two field modes nearest to resonance with the detector in the v-part
of the correlators. %, in addition to the a-part of the correlators.
Indeed, the beats in the green and blue-dashed curves in the upper plots of Figures \ref{SL1} to \ref{SL3Pi} are almost gone, since the
contributions of $|q_A^A|^2$ (the upper-left plot in Figure \ref{EmergeEigen}) and $|q_A^{k_3}|^2$ (the second plot from the left in the
upper row of Figure \ref{mofnr01W23R13}) have the same order of magnitude but are out of phase. Moreover,
as we mentioned above, similar few-mode approximations can violate the uncertainty relation in some periods of time in the system's
history due to the inconsistency with the retarded Green function. During these periods, the value of $S_L$ becomes negative and so
unphysical. To get rid of this one has to include enough field modes in the v-part of the correlators
(at least $n_{\rm max} \sim 100$ in Figures \ref{SL1}$-$\ref{SL3Pi}).

The higher-frequency modes are responsible for the small oscillations on top of the beat at the frequency about $2\Omega$ in the weak
coupling limit. As the UV cutoff $n_{\rm max}$ increases,
such small oscillations of $\langle \hat{Q}_A^2\rangle$ and $\langle \hat{P}_A^2\rangle$ will be amplified, while the whole evolution curves of $\langle \hat{Q}_A^2\rangle$, $\langle \hat{P}_A^2\rangle$, and thus $S_L$, will be elevated as well.
Including the higher-frequency modes further helps to resolve the spikes of the evolution curve of $\langle P_A^2\rangle$
occurring around $t=n L$, $n=1,2,3,\ldots$ as shown in the middle plots in the upper rows of Figures \ref{SL1}$-$\ref{SL3Pi}.
These spikes are due to constructive interferences occurring periodically at $t=nL$ when $e^{i\omega_{n'}t}=1$ for all $\omega^{}_{n'}
= |k^{}_{n'}|$ or $|k'_{n'}|$, $n'=1,2,3,\ldots$ such that the mode sum of $\langle P_A^2\rangle$ get an additional logarithmic
divergence. More explicitly, the terms in $\langle \hat{P}_A^2(t) \rangle_{\rm v}$ corresponding to $C^{}_3(t)$ in (\ref{P2vM2}) goes to
the counterpart of $I^{}_3$ in ${\bf S}^1\times {\bf R}_1$ as $t\to nL$ (in ${\bf R}^1_1$ this only occurs as $t\to 0$). While the value
of $\langle \hat{P}_A^2 \rangle$ varies significantly at these moments, once the UV cutoff is introduced, the amplitudes of the spikes
relative to their neighborhoods will be finite. They can be higher or lower than their neighborhood (see the factor of the $C^{}_3$-term
in (\ref{P2vM2})), but will never overwhelm those which have been corresponding to the $I^{}_3$ terms for $t\not=nL$ to make the
corrections to $\langle P_A^2\rangle$ from higher-frequency modes negative.

In the lower-right plot of Figure \ref{SL1Pi}, while the frequency of the detector is near resonant to none of the field mode, the curves
happen to show a beat with a very long period. Such a beat is produced mainly by three eigen-frequencies rather than two. When the free
detector frequency is almost, but not exactly, located at the middle point between two frequencies of the free field modes, and $L$ has
a proper value, the three eigen-modes with eigen-frequencies around these three natural frequencies will dominate.
While the difference between every two dominant eigen-frequencies is $O(1/R)$, which gives no beat beyond the period $L$ of the echoes,
the small difference between 1) the frequency difference of the middle and the lower modes, and 2) the difference of the higher and the
middle modes, stands out and sets a beat frequency.
Note that if the spacing $1/R$ in the frequency spectrum of the free field is too large,
only the eigen-mode with frequency around the free detector's will dominate and so no significant beats will be observed; If
$1/R$ is too small, there will be so many dominant eigen-modes that the evolution becomes complicated.

\subsubsection{Effects of the zero-mode}

For the untwisted field, the curves with or without the zero-mode contribution (e.g. the black and red-dotted curves, respectively, in
Figures \ref{SL1} and \ref{SL3}) are almost indistinguishable in the results with the kinetic momentum $\hat{P}_A$. The zero-mode
contributions to the v-parts of the two-point correlators with $P_A$ are not important here since we have chosen the initial state of the
zero-mode having the minimal uncertainty. Just like other field modes, a slight change of the initial state of the zero-mode will change
the results of the correlators and $S_L$ slightly.

When $S_I$ and $P^{}_A$ are replaced by the gauge-dependent $S'_I$ and $P'_A$, the zero-mode will make $\langle \hat{P}'^2_A\rangle$
growing as $t^2$ indefinitely and so $S_L$ also growing in time at large time scales (see the lower plots of Figures \ref{SL1} and
\ref{SL3}). This behavior, taken on face value, may be construed as the detector-field entanglement increases forever.
However, since $P'_A$ is a gauge-dependent quantity, this ill-behavior is not physically meaningful.
Only the entanglement in terms of the physical $P^{}_A$ matters.

\begin{figure}
\includegraphics[width=5cm]{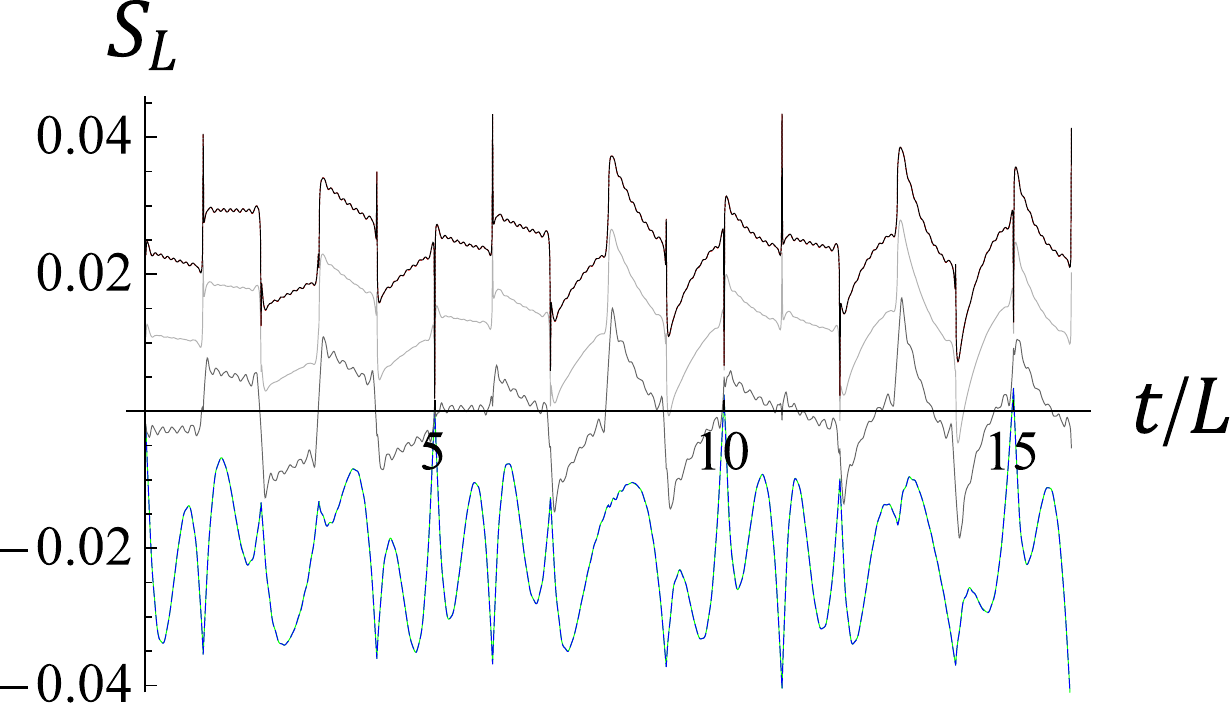}
\includegraphics[width=4.8cm]{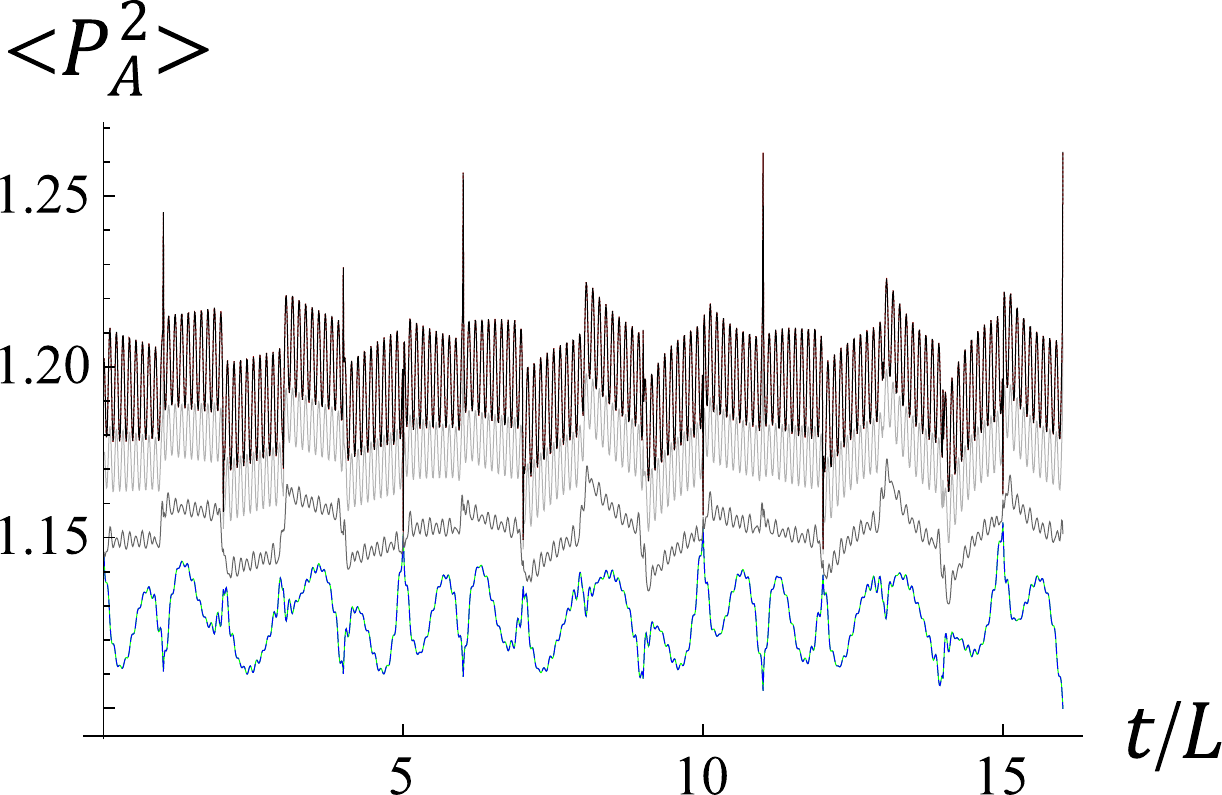}
\includegraphics[width=4.8cm]{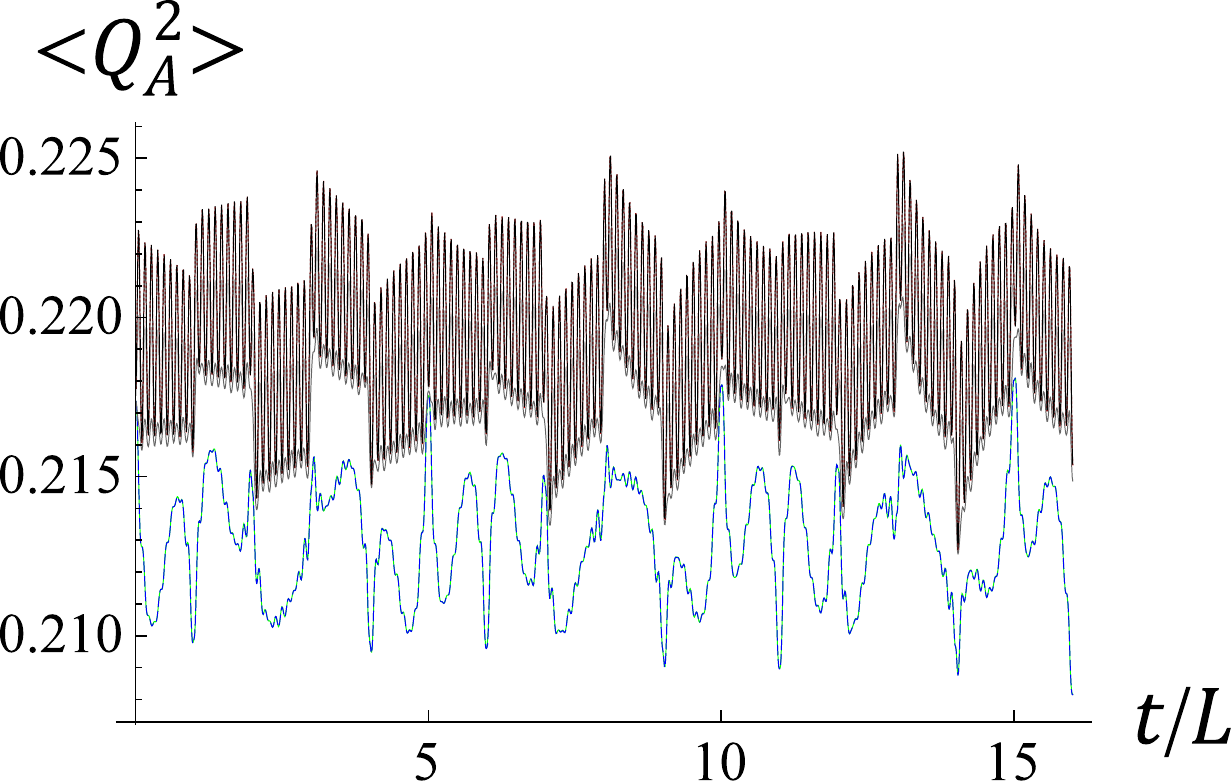}\\
%\hspace{4.8cm}
\includegraphics[width=5cm]{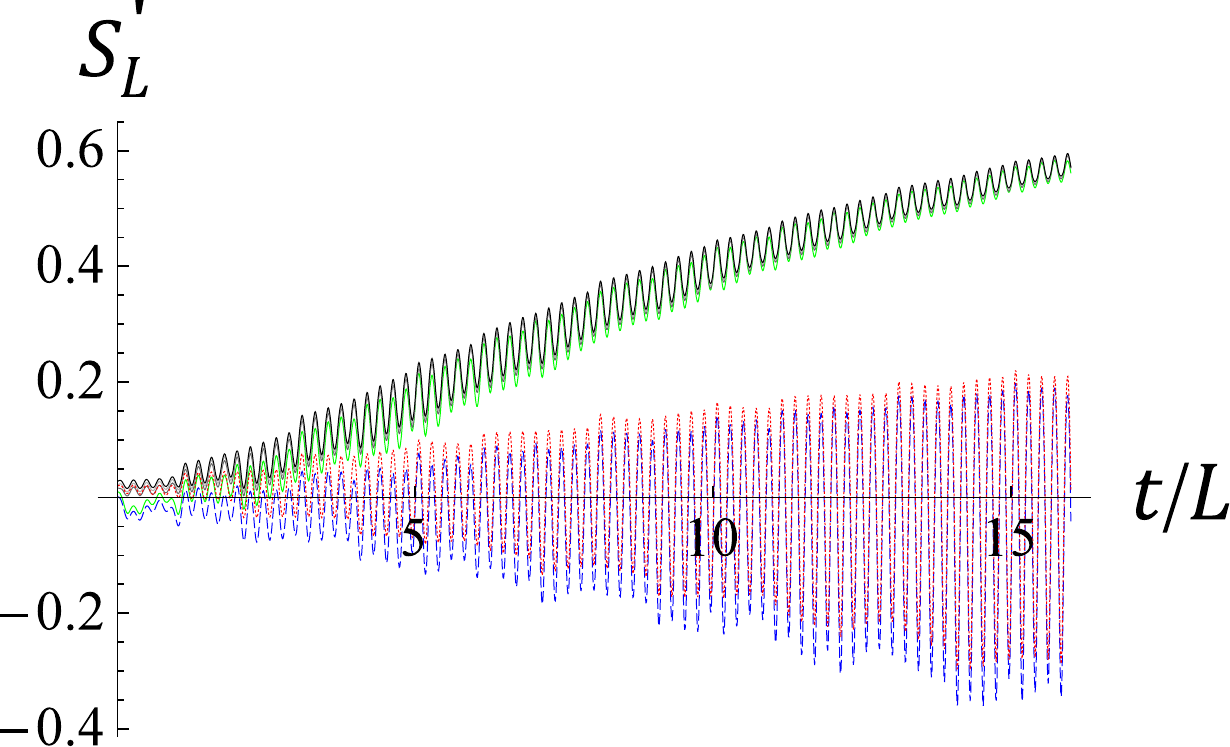}
\includegraphics[width=4.8cm]{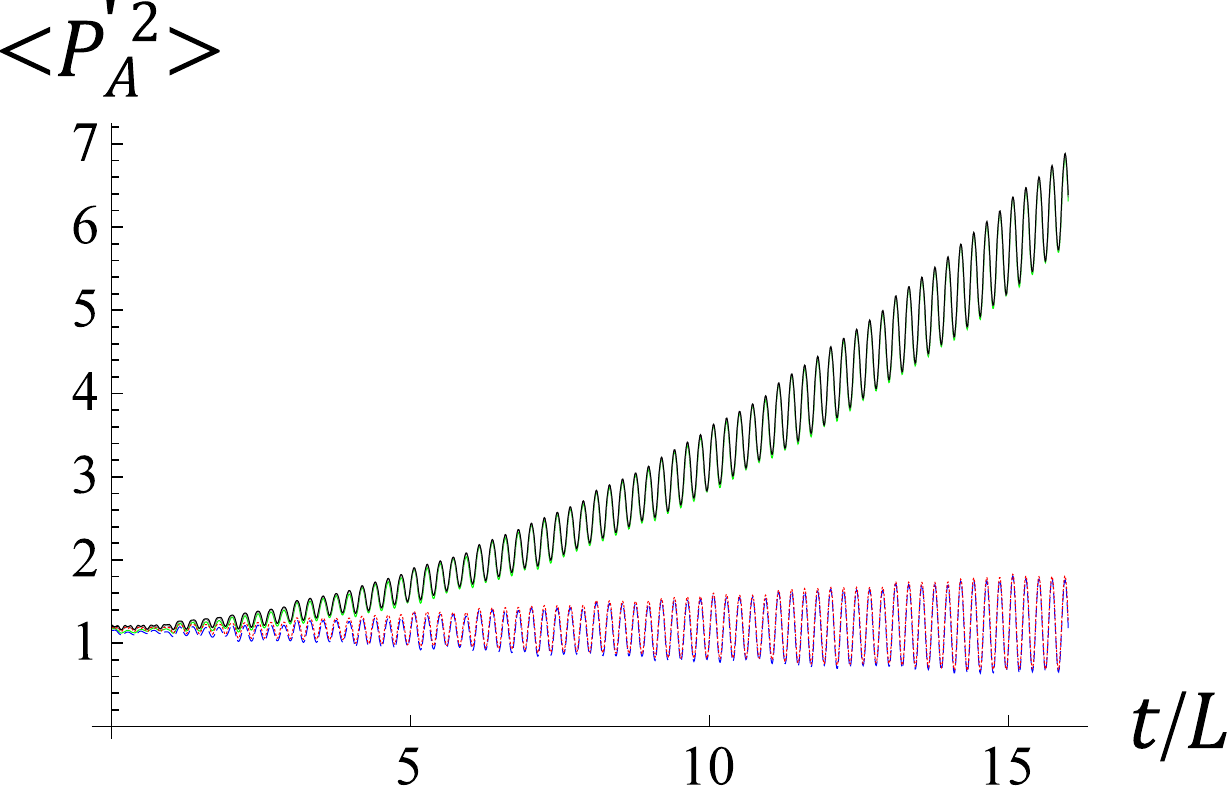}
\caption{Another example in the normal (untwisted) field, for comparison with Figures \ref{Specr001W23R2} and \ref{mofnr001W23R2}.
The upper plots are the results with the physical $P^{}_A$, and the lower plots are those with the unphysical $P'_A$.
The dark-gray, light-gray, and black solid curves represent the results contributed by the modes $\omega_{k_n}$ from $|n|=0$ up to
$|n|=10$, $100$, and $1000$, respectively, while the blue-dashed curve is contributed only by the zero mode and the modes with $|n|=4$
and $5$, which are the modes with the frequencies closest to the detector's. The blue-dashed ($|n|=4, 5$) and red-dotted curves
($|n|=1$ to $1000$) are obtained from the green and the black curves with the contributions by the zero mode removed, respectively.
As in Figure \ref{Specr001W23R2}, no significant beat can be observed here.}
\label{SL3}
\end{figure}

\begin{figure}
\includegraphics[width=5cm]{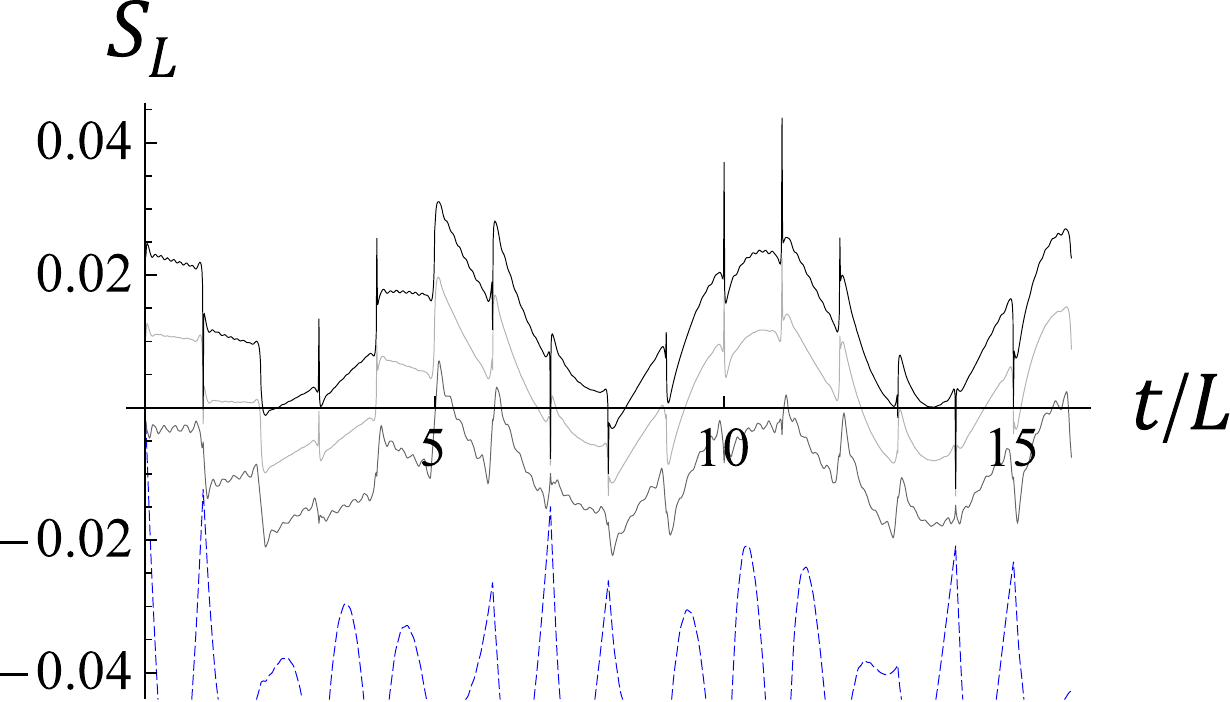}
\includegraphics[width=4.8cm]{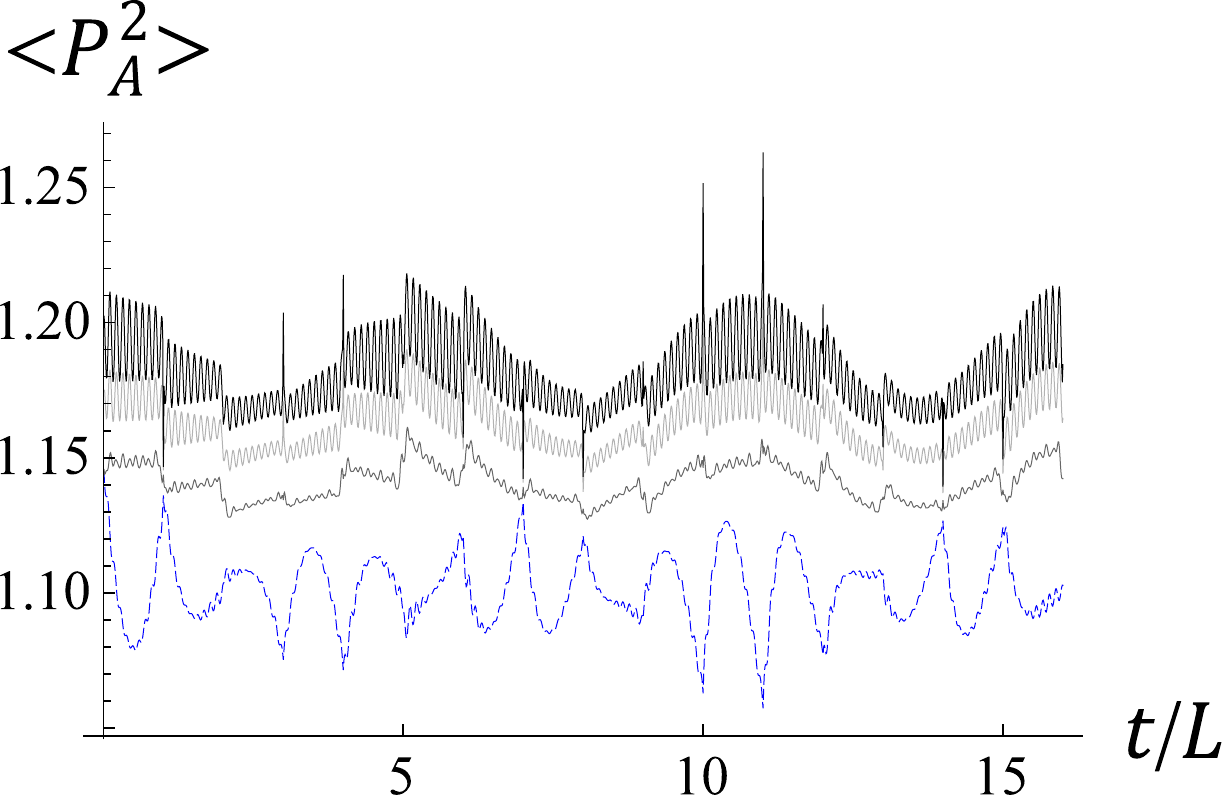}
\includegraphics[width=4.8cm]{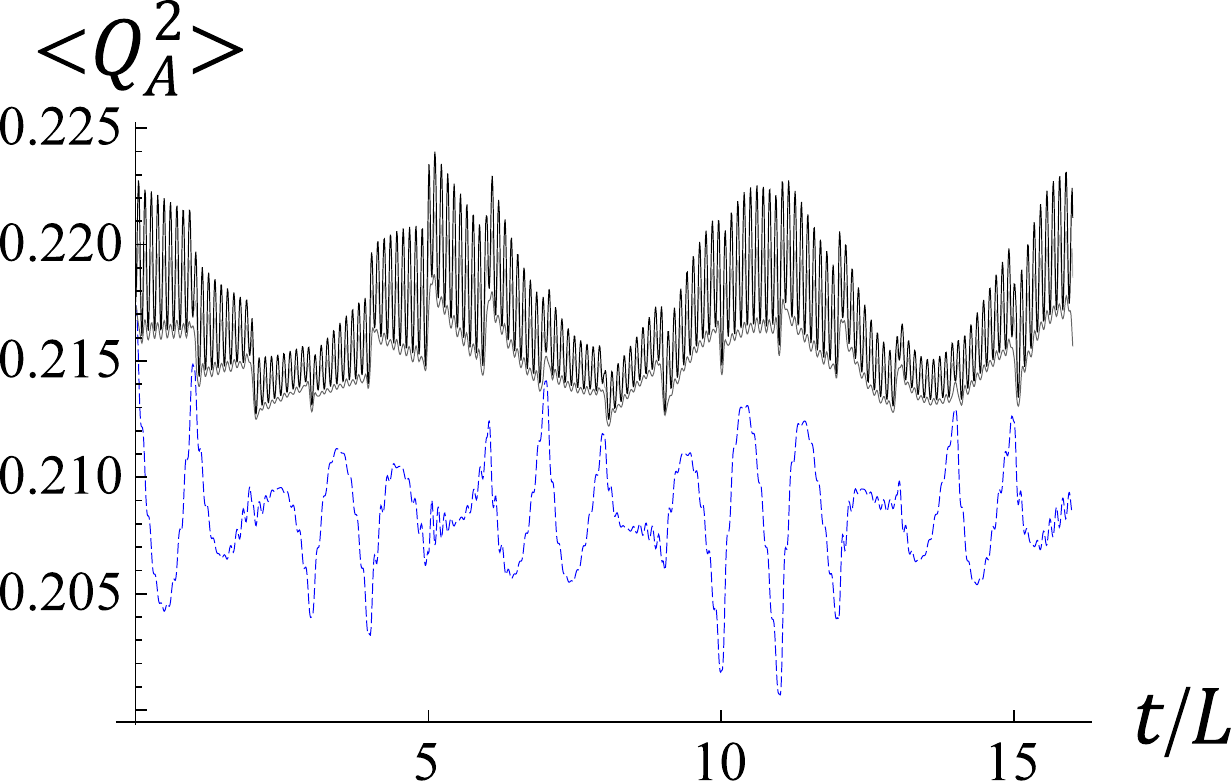}
\includegraphics[width=5cm]{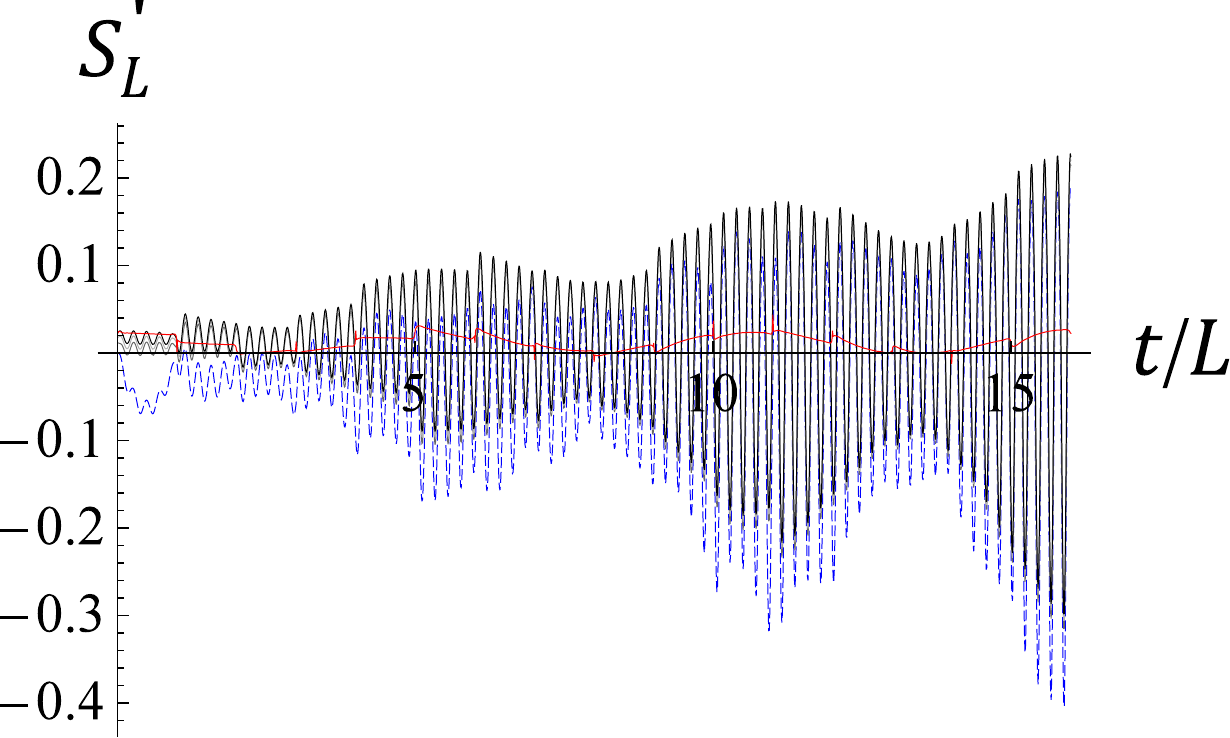}
\includegraphics[width=4.8cm]{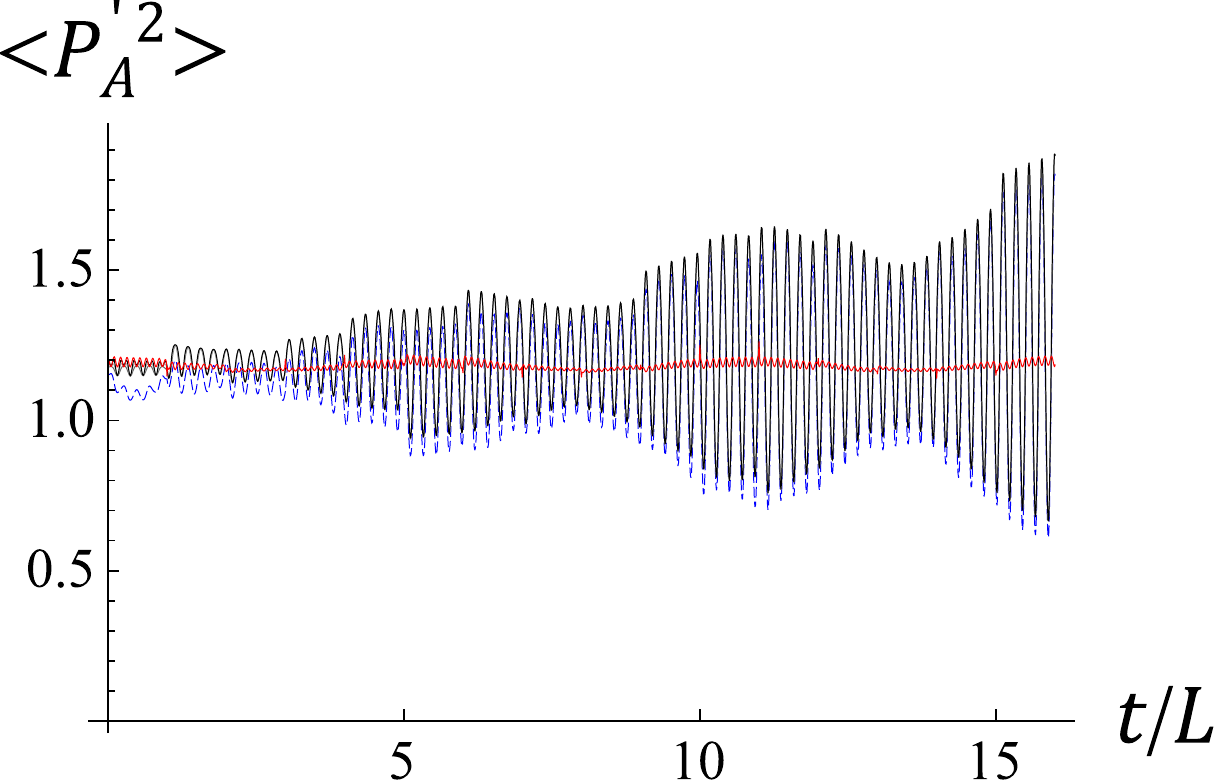}
\caption{The same parameters as those in Figure \ref{SL3} except here the detector is in the twisted field ($\varepsilon=-1$).
(Upper row) The dark-gray, light-gray, and black solid curves represent the results contributed by the modes $\omega_{k'_n} =
|n-(1/2)|/R$ from $n=1$ up to $n=10$, $100$, and $1000$ respectively, while the blue-dashed curve is contributed only by the modes with
$n=5$, whose mode frequency is the closest to the detector's.
In contrast to the case with the untwisted field in Figure \ref{SL3}, here the beating behavior is obvious.
(Lower row) $\langle \hat{P}'^2_A\rangle$ and $S'_L$ contributed by the unphysical
momentum $P'_A$. The red curves represent $\langle \hat{P}_A^2\rangle$ and $S_L$ (the black curves in the upper row) for comparison.}
\label{SL3Pi}
\end{figure}

\subsection{Comparison with perturbative results}

\subsubsection{On the zero mode}

The transition probability of a single UD' detector in the untwisted field from the initial ground state to the first excited state of
the detector, obtained using the time-dependent perturbation theory (TDPT), depend on the initial two-point correlator of the conjugate
momentum of the zero mode, $\langle \hat{\Pi}_{k_0}^2(0)\rangle \equiv {}_{\rm z}\langle 0_L|\hat{\Pi}_{k_0}^2(0)|0_L\rangle^{}_{\rm z}$,
but is independent of the initial correlator of the zero mode itself, $\langle \hat{\Phi}_{k_0}^2(0)\rangle\equiv {}_{\rm z}\langle 0_L|
\hat{\Phi}_{k_0}^2(0) |0_L\rangle^{}_{\rm z}$ \cite{ML14}.
Accordingly the authors of \cite{ML14} claimed that the effect of the zero mode in the UD' detector theory defined in ${\bf S}^1\times
{\bf R}_1$ can be suppressed by choosing a suitable initial state such that $\langle \hat{\Pi}_{k_0}^2(0)\rangle$ is very small. They pointed out that this is not possible for the usual UD detector with $Q\Phi$ coupling in  ${\bf S}^1\times{\bf R}_1$ since in that case
$\langle\hat{\Phi}_{k_0}^2(0)\rangle$ also enters the response function and will become large if $\langle \hat{\Pi}_{k_0}^2(0)\rangle$ is
squeezed (due to the uncertainty relation $\langle \hat{\Phi}_{k_0}^2(0)\rangle$$\langle \hat{\Pi}_{k_0}^2(0)\rangle \ge (\hbar/2)^2$).

From Eqs.(20)-(23) in \cite{LH07}, we have seen that after taking the weak coupling limit $\gamma\ll \Omega$ for the complete expression
of the transition probability, the leading term will be proportional to the $O(\gamma)$ part of $\langle \hat{P}_A^2\rangle + \Omega_0^2
\langle \hat{Q}_A^2\rangle$ ($m_0=1$ here).
If we choose the static initial state so that ${}_{\rm z}\langle 0_L|\{\hat{\Phi}_{k_0}(0), \hat{\Pi}_{k_0}(0)\}|0_L\rangle^{}_{\rm z}=0$,
the zero mode contributions to $\langle \hat{Q}_A^2\rangle$ and $\langle \hat{P}_A^2\rangle$ will be
$({\rm Re}\, q_A^{k_0})^2\langle\hat{\Phi}_{k_0}^2(0)\rangle +({\rm Im}\, q_A^{k_0})^2 \langle\hat{\Pi}_{k_0}^2(0)\rangle$ and
$({\rm Re}\, p_A^{k_0})^2\langle\hat{\Phi}_{k_0}^2(0)\rangle +({\rm Im}\, p_A^{k_0})^2 \langle\hat{\Pi}_{k_0}^2(0)\rangle$, respectively.
When $0 < t < L$, from (\ref{qA00}) one sees that both $q_A^{k_0}$ and $p_A^{k_0} = \partial^{}_t q_A^{k_0}$
are purely imaginary, which implies that ${\rm Re}\, q_A^{k_0}= {\rm Re}\,p_A^{k_0}=0$ for all time according to Eq.(\ref{EOMqp2}). Thus
indeed, the initial variance $\langle \hat{\Phi}_{k_0}^2(0)\rangle$ of the zero mode are totally irrelevant to the two-point correlators of detector $A$ in the untwisted field. The zero-mode contribution to the dynamics of the combined system in our model depends on
$\langle \hat{\Pi}_{k_0}^2(0)\rangle$ only.

\subsubsection{On the validity of perturbative transition probability}

\begin{figure}
%\hspace{-.8cm}
\includegraphics[width=7.2cm]{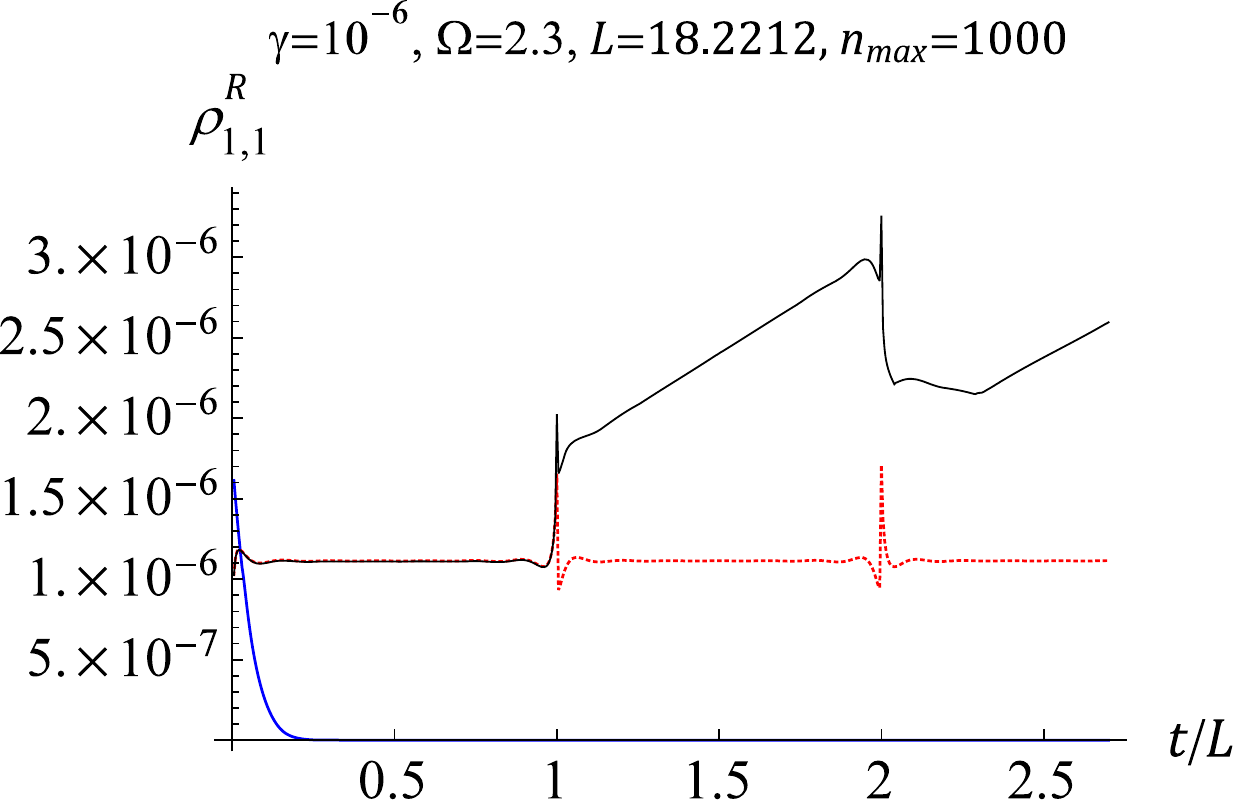}
\caption{Our nonperturbative result for the probability of finding the detector in the first excited state (black), is compared
with (\ref{PrTDPT1}) obtained by first-order TDPT with finite duration of interaction (red dotted) in the perturbative regime
($\gamma \ll \Omega$). One can see that the deviation becomes significant as early as $t\approx L$. For comparison, we also show the
first-order TDPT result (\ref{PrTDPT2}) with infinite duration of interaction weighted by a Gaussian switching function with width $t$
(blue). Note that $L\gamma \ll 1$ even for $t>L$ in this plot.}
\label{CompareTDPT}
\end{figure}

In Figure \ref{CompareTDPT}, we show the transition probability from the ground state $|g\rangle$ to the first excited state
$|e_1\rangle$ of a UD' detector initially in the vacuum state of an untwisted field obtained by our nonperturbative method (black),
and the first-order TDPT result (red dotted) \cite{LH07} given by
\begin{eqnarray}
  \rho_{1,1}^R &\approx&
	  \lambda^2 \left| \langle e_1| \hat{Q}_A(0) |g\rangle \right|^2 \int_0^t dt_1 \int_0^t dt_2 e^{-i\Omega_0 (t_1-t_2)}
	  \partial_{t_1} \partial_{t_2} \langle 0 | \hat{\Phi}^{{}^{[0]}}_{x=0}(t_1)\hat{\Phi}^{{}^{[0]}}_{x=0}(t_2) | 0\rangle \nonumber\\ &=&
		{2\gamma\over \Omega_0 L^2}\left\{ {1-\cos\Omega_0 t\over \Omega_0^2} + 4\pi \sum_{n=1}^{n_{\rm max}} n \left[
		{1-\cos\left( {2\pi n\over L}+\Omega_0\right) t \over \left({2\pi n\over L}+\Omega_0\right)^2} \right] \right\}, \label{PrTDPT1}
\end{eqnarray}
where $\hbar=1$\footnote{Note that it should be $\hbar^{-3}$ instead of $\hbar^{-1}$ in the overall factor of the RHS of Eq.(26) in 
\cite{LH07}.}, $\hat{\Phi}^{{}^{[0]}}_{x}(t) = \sum_{k}  (2 \tilde{\omega}_k)^{-1/2}
\left[\phi^{{}^{[0]}k}_{\; x}(t)\hat{b}^{}_{k} +\phi^{{}^{[0]}k*}_{\; x}(t)\hat{b}_{k}^\dagger \right]$ is the free field operator, and
$|\langle e_1| \hat{Q}_A(0) |g\rangle |^2 = 1/(2\Omega_0)$.
One can see that the TDPT result (\ref{PrTDPT1}) deviates significantly (still $O(\gamma)$) from the nonperturbative result as early as
$t\approx L$ when the first echo returns, though the echoes appear to be higher-order effects in the equation of motion (\ref{EOMqp1}).
For further comparison, we also present the perturbative result (the blue curve in Figure \ref{CompareTDPT})
\begin{eqnarray}
  \rho_{1,1}^R &\approx& {\lambda^2\over 2\Omega_0} %\left| \langle e_1| \hat{Q}_A(0) |g\rangle \right|^2
	  \int_{-\infty}^\infty dt_1 \int_{-\infty}^\infty dt_2 \chi^{}_{t}(t_1) \chi^{}_{t}(t_2) %\times\nonumber\\ & &
		e^{-i\Omega_0 (t_1-t_2)} \partial_{t_1} \partial_{t_2} \langle 0 | \hat{\Phi}^{[0]}_{x=0}(t_1)\hat{\Phi}^{[0]}_{x=0}(t_2) | 0\rangle
		\nonumber\\ &=& {\gamma t^2 \over \Omega_0 L^2} %\left\{ e^{-\Omega_0^2 t^2/2} +
      %4\pi \sum_{n=1}^{n_{\rm max}} n\exp \left[-{t^2\over 2}\left( {2\pi n\over L}+\Omega_0\right)^2\right] \right\},
		\left\{ e^{-t^2\Omega_0^2/(8\pi)} + 4\pi \sum_{n=1}^{n_{\rm max}} n\exp \left[-{t^2\over 8\pi}
		 \left( {2\pi n\over L}+\Omega_0\right)^2\right] \right\}, \label{PrTDPT2}
\end{eqnarray}
with the Gaussian switching function $\chi^{}_{t}(\tau) \equiv 2 e^{-\pi(2\tau/t)^2}$%\pi^{-1/2} e^{-(\tau/t)^2}$
so that the width $t=\int_{-\infty}^{\infty} \chi^{}_{t}(\tau)d\tau = \int_0^t d\tau$ corresponds to the duration of interaction in Eq.
(\ref{PrTDPT1}). We find that the Gaussian switching function greatly suppresses the perturbative transition probability (\ref{PrTDPT2}),
which decays very quickly as the effective interaction time $t$ increases.
Comparing (\ref{PrTDPT2}) with (\ref{PrTDPT1}), one can see that at a fixed $t$, the summand of the summation
$\sum_{n=1}^{n_{\rm max}}$ goes like $n e^{-n^2}$ for large $n$ in (\ref{PrTDPT2}), and $n^{-1}$ in (\ref{PrTDPT1}).
So (\ref{PrTDPT2}) receives much less contributions from the short-wavelength modes than (\ref{PrTDPT1}) does if $n_{\rm max}$
is large enough.

%%%%%%%%%%%%%%%%%%%%%%%%%%%%%%%%%%%%%%%%%%%%%%%%%%%%%%%%%%%%%%%%%%%%%%%%%%%%%%%%%%%%%%%%%%%%%%%%%%%%%%%%%%%%%%%%%%%%%%%%
\section{Two-detector case}
\label{2detectors}

Consider two identical detectors  $A$ and $B$ at rest with natural frequencies $\omega_A=\omega_B=\Omega_0$,
located at $x=z^1_A = 0$ and $x=z^1_B= R\pi = L/2$, respectively, as shown in Figure \ref{setup} (left) and (middle).
Substituting the expansions (\ref{Qmod}) and (\ref{Fmod}) into the Heisenberg equations of motion for the operators yields
\begin{eqnarray}
  & &\left( \partial_t^2 +2 \gamma \partial^{}_t + \Omega_0^2 \right) q^{\mu}_{A(B)}(t) = \nonumber\\ & &
	  -\lambda \partial^{}_t \phi_{z^{1}_A (z^{1}_B)}^{^{[0]}\mu}(t) -{\lambda^2\over 2} \sum_{n'=1}^\infty
    \left\{ \left(\varepsilon^{n'} + \varepsilon^{n'}\right) \theta(t- n' L) \partial^{}_t q^{\mu}_{A(B)}(t- n' L ) +
		\right. \nonumber\\
		& & \left. \left(\varepsilon^{n'} + \varepsilon^{n'+1}\right) \theta \left(t- \left[n' - (1/2)\right]L \right)
		\partial^{}_t q^{\mu}_{B(A)} \left(t- \left[n' - (1/2)\right]L \right)\right\}, \label{EOMqaA0}
\end{eqnarray}
where $\mu = A, B, k_n$ (untwisted) or $k'_n$ (twisted field), $n\in{\bf Z}$,
$\phi_{\;x}^{^{[0]}A}=\phi_{\;x}^{^{[0]}B}\equiv 0$, $\phi_{\;x}^{^{[0]}k}(t)\equiv e^{-i(|k|t - k x)}$ for $k\not=0$
($k$ can be $k_n$ or $k'_n$), and $\phi_{\;x}^{^{[0]}k_0}(t)\equiv 1 - (i/L)t$ for the zero-mode.

From the setups in the extended coordinates
(Figure \ref{setup} (middle) and (right)), one can see that the {\it classical} dynamics for an individual detector and the untwisted
field in the cases with two or more identical detectors would be equivalent to the dynamics of the detector and the untwisted field in the
single detector case: in the two-detector case $Q_A(t)$ is affected by $Q_B(t-(L/2))$ with identical solutions to $Q_A(t-(L/2))$ in the
one-detector case. However, this is not true for the mode-functions in quantum theory. In (\ref{EOMqaA0}), $q_A^A(t)$ is affected by
$q_B^A(t)$ rather than $q_B^B(t)$, and the initial condition for the former is $q_B^A(t<(L/2)) = 0$, which is different from (\ref{qAA0})
for $q_B^B(t<(L/2))= q_A^A(t<(L/2))$.

Suppose the initial state of the combined system at $t=0$ is a factorized state,
\begin{equation}
  \left|\right. \psi (0) \rangle = \left|\right. q^{}_A, q^{}_B \rangle \otimes
  \left|\right. 0_{L}\rangle,
\end{equation}
which is a direct product of a two-mode squeezed state of the detectors $\left|\right. q^{}_A, q^{}_B \left.\right>$,
or in the Wigner function representation,
\begin{eqnarray}
  \rho^{}_{AB}(0) &=& {1\over \pi^2}\exp -{1\over 2}\left[ \beta^2 (Q_A+Q_B)^2 + \beta^{-2} (P_A+P_B)^2 +\right. \nonumber\\ & &\left.
  \alpha^{-2} (Q_A-Q_B)^2 + \alpha^{2} (P_A-P_B)^2\right] \label{IS2dec}
\end{eqnarray}
with constant parameters $\alpha$ and $\beta$ \cite{LH09}, and the vacuum state of the field $\left|\right. 0_{L} \rangle$.
Again, for the untwisted field, we further factorize the field state into $\left|\right. 0_{L} \rangle = \left|\right. 0_{L}
\rangle^{}_{\rm nz}\otimes \left|\right. 0 \left.\right>_{\rm z}$ where $\left|\right. 0 \left.\right>_{\rm z}$ is the initial state of
the zero mode giving ${}_{\rm z}\langle 0_L| \hat{\phi}_{k_0}^2(0) |0_L\rangle^{}_{\rm z} = {}_{\rm z}\langle 0_L| \hat{\pi}_{k_0}^2(0)
|0_L\rangle^{}_{\rm z}= \hbar/2$.

\subsection{Untwisted field}

In the untwisted field, writing $q^\mu_\pm = (q^\mu_A\pm q^\mu_B)/\sqrt{2}$, the equations of motion
(\ref{EOMqaA0}) for the mode functions can be simplified to
\begin{eqnarray}
 & &\left( \partial_t^2 +2 \gamma \partial^{}_t + \Omega_0^2 \right) q^{\mu}_\pm(t) \nonumber\\ &=&
	-{\lambda\over \sqrt{2}}\partial^{}_t \left[ \phi_{\;0}^{^{[0]}\mu}(t) \pm \phi_{L/2}^{^{[0]}\mu}(t) \right]
    -\lambda^2 \sum_{n'=1}^\infty (\pm 1)^{n'} \theta \left(t- n'{L\over 2} \right) \partial^{}_t q^{\mu}_\pm \left(t-n' {L\over 2}\right)
		\label{EOMqpm0} \\
 &=& -{\lambda\over\sqrt{2}} {\cal F}^\mu_\pm (t) \pm\theta\left(t- {L\over 2}\right)
      \left( \partial_t^2 -2 \gamma \partial^{}_t + \Omega_0^2 \right) q^{\mu}_\pm \left(t-{L\over 2}\right),
  \label{EOMqpm}
\end{eqnarray}
where
\begin{equation}
{\cal F}^\mu_\pm (t) = \partial^{}_t\left[\phi_{\;0}^{^{[0]}\mu}(t)\pm \phi_{L/2}^{^{[0]}\mu}(t)\right] -\theta\left(t- {L\over 2}\right)
\partial^{}_t\left[\pm \phi_{\;0}^{^{[0]}\mu}\left(t-{L\over 2}\right) + \phi_{L/2}^{^{[0]}\mu}\left(t- {L\over 2}\right)\right].
\end{equation}
When $0 <t\le L/2$, ${\cal F}^{k_n}_\pm (t) =i|k_n|e^{-i|k_n|t}\left[ 1\pm (-1)^n\right]$ for $k_n\not=0$,
${\cal F}^{k_0}_+ (t) = -2i/L$, and ${\cal F}^A_\pm = {\cal F}^B_\pm={\cal F}^{k_0}_- =0$.
When $t>L/2$, all the components of ${\cal F}^\mu_\pm$ vanish. Compared with
(\ref{EOMqp2}), one can see that the equations for $q^{}_+$ and $q^{}_-$
in (\ref{EOMqpm}) are almost equivalent to the ones for a single detector in the untwisted and twisted fields, respectively, in $S^1$
with circumference $L/2$, except that the force terms ${\cal F}_\pm^{k_n}(t)$ and ${\cal F}_\pm^{k_0}(t)$ driven by vacuum fluctuations
of the field in (\ref{EOMqpm}) are different from their counterpart in (\ref{EOMqp2}).

The above delayed differential equations will be solved together with the conditions that for $0\le t<L/2$, $q_A^B(t) = q_B^A (t) =0$,
$q_A^A(t)=q_B^B(t)$ are given by Eq.(\ref{qAA0}), and $q_A^k(t)=q_B^k(t)$ are given by (\ref{qAk0}) and (\ref{qA00}),
before the mutual influences set in.

\subsubsection{Eigen-frequencies}

Substituting the ansatz $q^{\bf d}_\pm\approx\int d\omega_\pm\tilde{q}^{\bf d}_\pm(\omega^{}_\pm) e^{i\omega^{}_\pm t}$, ${\bf d}=A,B$
into (\ref{EOMqpm}), we find
%or (\ref{EOMqkpm}),
\begin{eqnarray}
   -\omega_+^2 + \Omega_0^2 &=& -2\gamma\omega_+ \cot (\omega_+ L / 4), \label{EFp}\\
   -\omega_-^2 + \Omega_0^2 &=& 2\gamma\omega_- \tan (\omega_- L /4), \label{EFm}
\end{eqnarray}
which are almost the same as (\ref{EFcond}) and (\ref{EFcondTw}) in the one-detector case except $L$ there has been replaced by $L/2$ here.
The solutions of the above two equations are the eigen-frequencies of the mode functions $q^{\bf d}_\pm$.

In weak coupling limit $\gamma\ll \Omega_0$, if $\Omega_0$ is not very close to $2n\pi/L=n/R$, $n\in{\bf N}$, the two detectors with the
same natural frequency $\Omega_0$ will be mixed together by the effective coupling mediated by the field, and so the dominant eigen-modes
will split into $\omega_+ \approx \Omega_0 + \gamma \cot (\Omega_0 L/4)$ from (\ref{EFp}) and $\omega_-\approx\Omega_0 -\gamma\tan
(\Omega_0 L/4)$ from (\ref{EFm}). These two eigen-frequencies can be quite close to each other, and so the beat frequency of
$q^{\bf d}_{A(B)} = (q^{\bf d}_+ \pm q^{\bf d}_-)/\sqrt{2}$ will be approximately
\begin{equation}
    \Delta \approx\gamma\left|\cot {\Omega_0 L\over 4} +\tan{\Omega_0 L\over 4}\right| ={2\gamma\over |\sin(\Omega_0 L/2)|}.\label{2UDbeat}
\end{equation}
When $\Omega_0 \approx (2n+1)\pi/L$,
the above beat frequency will reach the lowest value $\Delta \approx 2\gamma$ in the weak coupling limit.

When the natural frequency of the detectors is very close to the frequency of a free field mode, namely, $\Omega_0 \approx 2 n \pi/L$,
with some positive integer $n$ large enough, the two detectors and the field mode will split into three eigen-modes at frequencies
$2n\pi/L$ and $(2n\pi/L \pm \Delta)$, where $\Delta \approx \sqrt{\gamma/\pi}$ will be the beat frequency in this case.
This is a half of the beat frequency $2\Delta$ we obtained below (\ref{Delbeat}) in the one-detector case. If the eigen-frequency around
$2n\pi/L$ is not exactly at the averaged value of the two eigen-frequencies around $(2n\pi/L \pm \Delta)$, when $L$ has a proper value,
there will be beats with an even longer time scale similar to the one in Figure \ref{SL1Pi}. In other parameter range of $L$,
the beating behavior may disappear or get more complicated.

\subsubsection{Super-radiant mode and instability?}

In Minkowski space, the modes for two detectors similar to $q_+$ here (Eq.(20) in \cite{LH09}) can be interpreted as the super-radiant
modes since the dissipation rate is about twice of the one for the single detector there.
However here in a similar limit, $L \ll \gamma \ll \Omega_0$, expanding $q^{}_+(t-(L/2)) = q^{}_+(t) - (L/2) \partial^{}_t q^{}_+(t) +
\ldots$, Eq. (\ref{EOMqpm0}) gives
\begin{equation}
  {L\over 2} \left[ \left(1 + {\gamma L\over 2} + {L^2\Omega_0^2\over 24}\right)\partial_t^3 q^{}_+
	- \left( 2\gamma + {1\over 4}\Omega_0^2 L\right) \partial_t^2 q^{}_+ +
  \left( {8\gamma\over L} + \Omega_0^2 \right) \partial^{}_t q^{}_+ \right]\approx 0,
\end{equation}
by assuming $q^{}_+$ varies slowly in time.
This implies $\partial^{}_t q_+ \sim e^{(-\Gamma \pm i \omega) t}$ with $\Gamma^2 \equiv (\gamma + \Omega_0 L^2/8)^2 \ll \omega^2 \equiv
\Omega_0^2 + (8\gamma/L)$. $\omega$ is nothing but the lowest eigen-frequency obtained from (\ref{EFp})
in the same limit (where $\cot (L\omega/4) \approx 4/(L\omega)$). So $q^{}_+$ here oscillates in the lowest eigen-frequency rather
than decays in a quicker rate.
The behaviors of such solutions are quite different from the super-radiant modes
in \cite{LH09} since the space here is extremely compact: the size of the space $L$ is the same order of the separation of the
detector $L/2$, both are extremely small.

For two UD detectors at rest in (3+1) dimensional Minkowski space, when the separation is small enough, the detectors will become
unstable \cite{LH09}. Now in an Einstein cylinder, when the value of $L$ is small,
would we get runaway solutions for the mode functions of the detector(s)?

The instability in the analysis in \cite{LH09}
is due to the fact that the amplitude of the retarded fields in (3+1)D Minkowski space diverges as one approaches to the point-source.
However, in our (1+1)D Einstein cylinder the retarded solution of a massless scalar field is regular around the source.
Thus a small value of $L$ in our model will produce no instability. Indeed, in our calculation we always obtain real eigen-frequencies. Imaginary eigen-frequencies never show up when we take $L$ to $0+$.

\begin{figure}
\includegraphics[width=7.2cm]{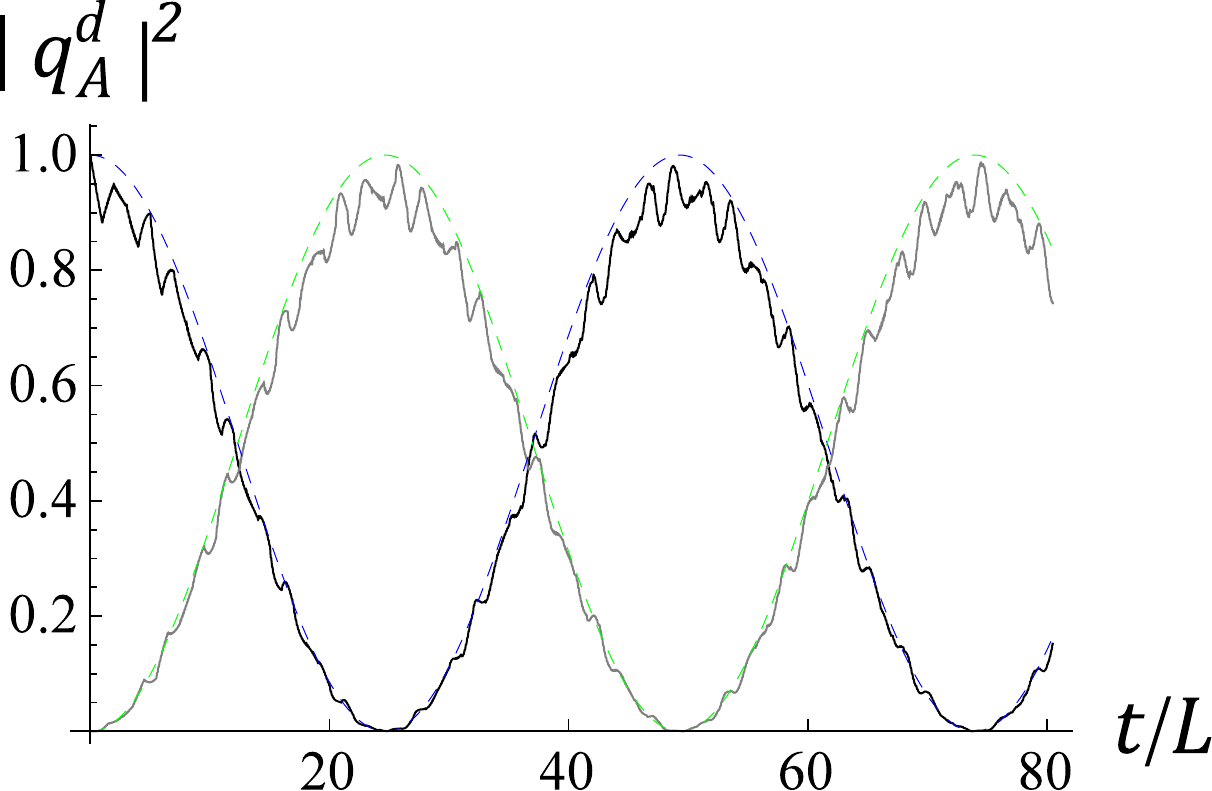}
\includegraphics[width=7.2cm]{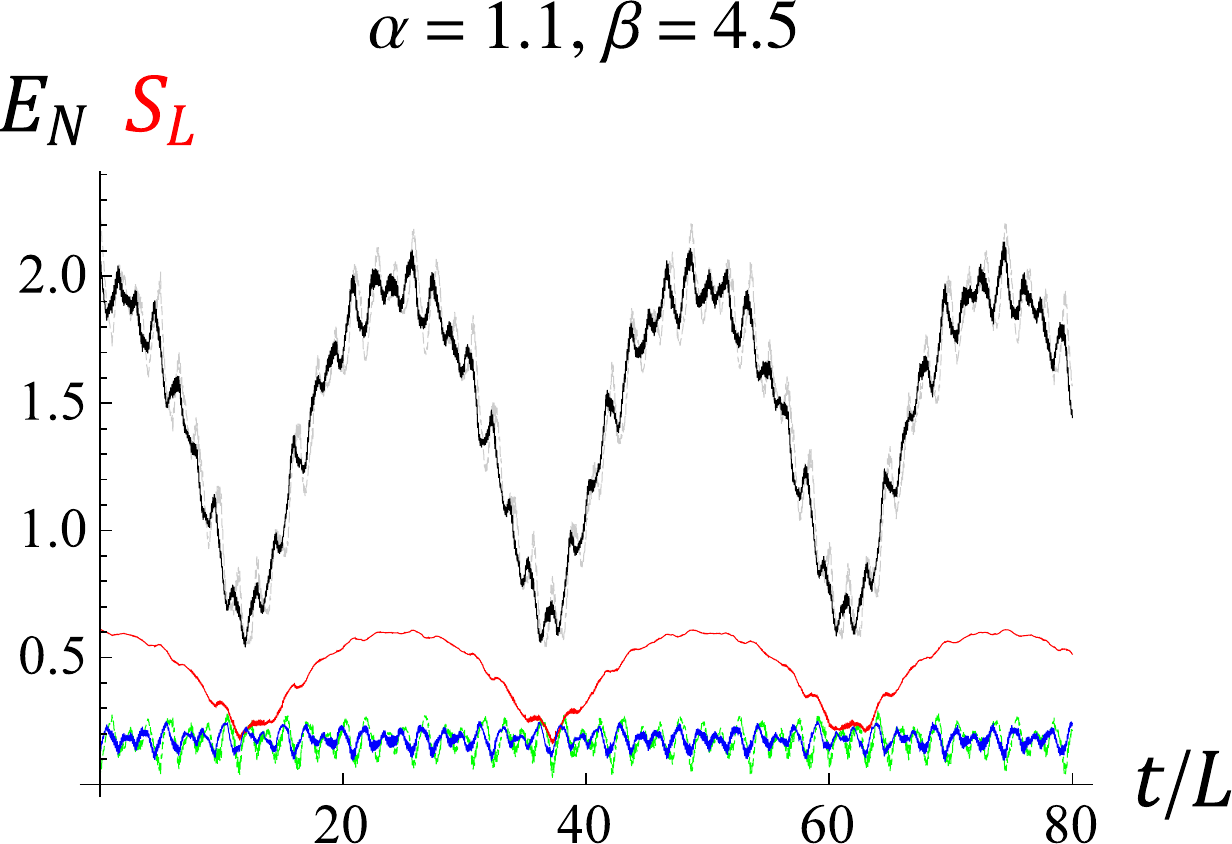}\\
\includegraphics[width=7.2cm]{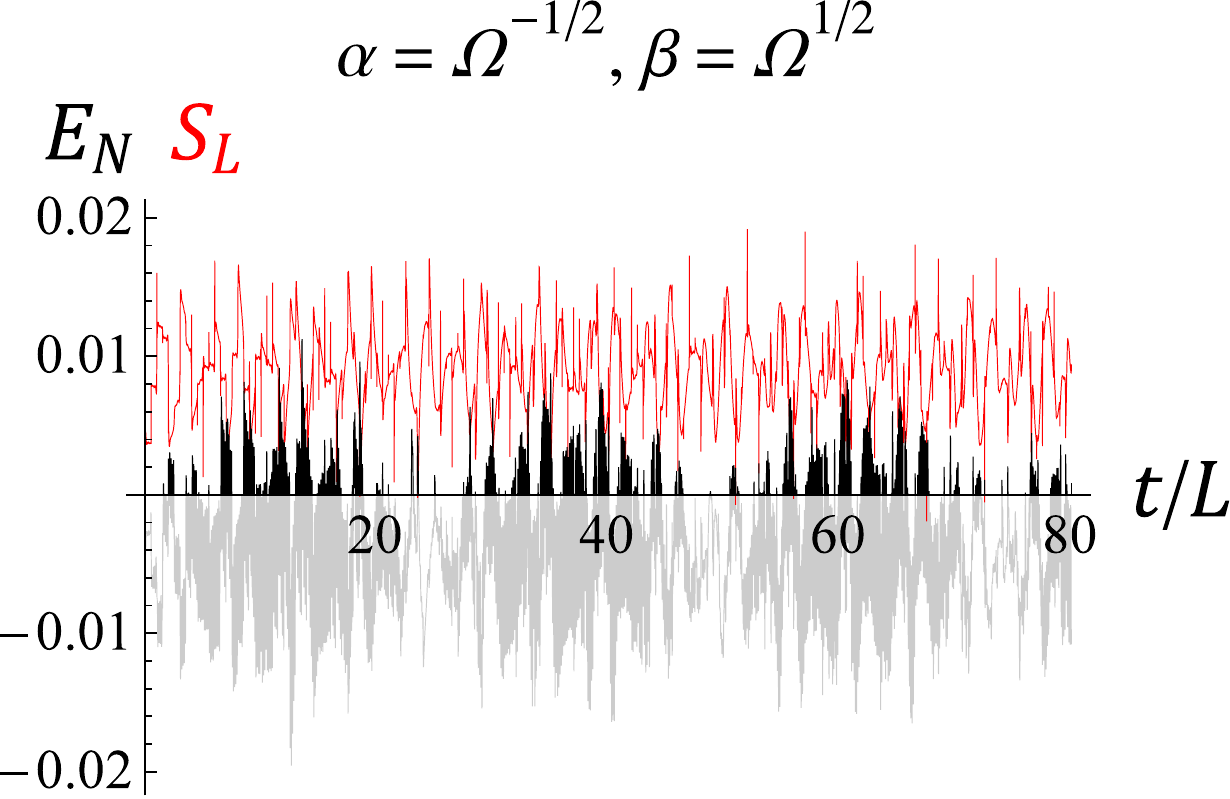}
\includegraphics[width=7.2cm]{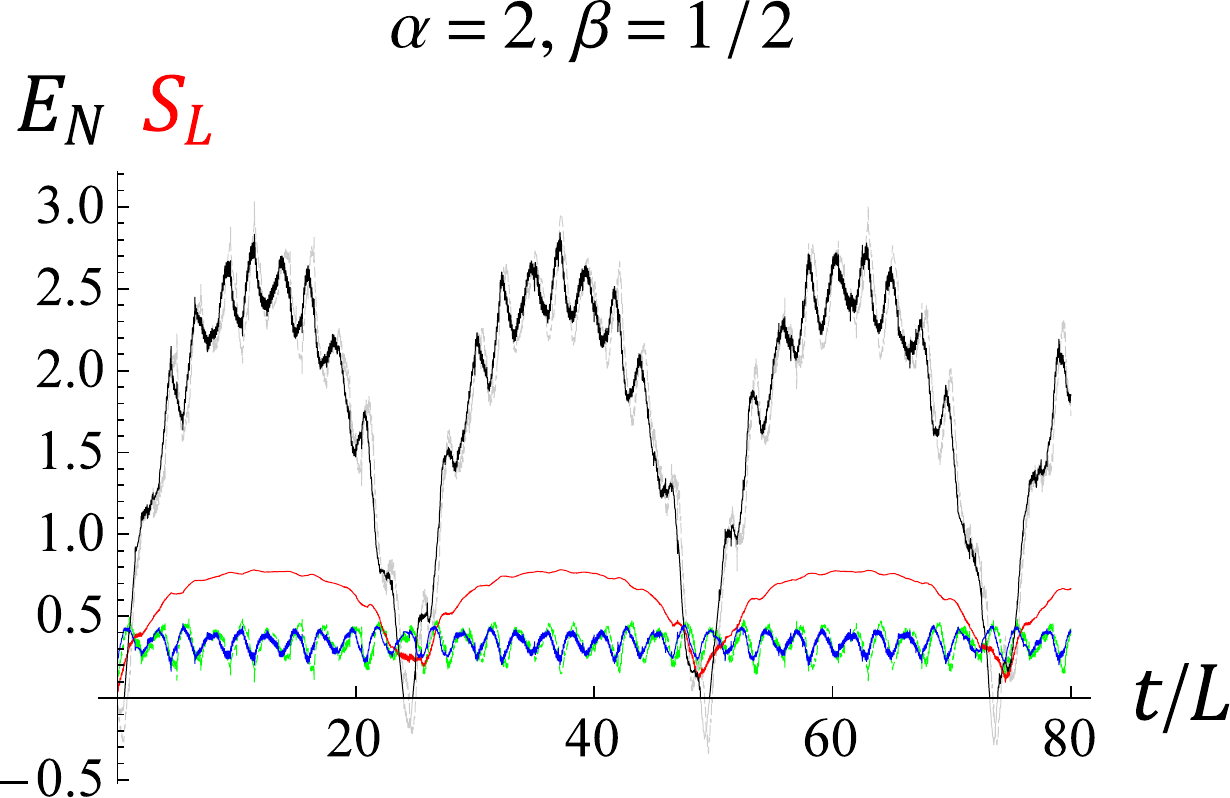}
\caption{Entanglement dynamics of two UD' detectors contributed by the field-modes with $k_n$, $|n|=0, \ldots, 250$, 
in terms of detector $A$'s clock. Here $\gamma = 0.005$, $\Omega = 2.3$, $L = 4\pi$,
and $\alpha$ and $\beta$ are shown in the plot labels. (Upper left) $\left|q_A^A \right|^2$ (black)
has a beating behavior like $(1+\cos t \Delta)/2$ (green-dashed) and $\left|q_A^B \right|^2$ (gray) like $(1-\cos t \Delta)/2$
(blue-dashed) with $\Delta \approx 0.01014 \approx 2\gamma/\sin(\Omega_0 L/2)$ given in (\ref{2UDbeat}).
(Upper-right and lower row) The logarithmic negativity $E_{\cal N}=\max\{ 0,-\log_2 2c_- \}$ (black) of the EnLC
compared with the $E_{\cal N}$ of the EnS (gray dashed), the linear entropy $S_L$ of detector $A$ (red), and
the linear entropies $S^{(2)}_L$ of the detector pair evaluated around the future light cones (blue) and
the hypersurfaces of simultaneity (green dashed) of detector $A$.
The gray part of the curve for the $E_{\cal N}$ of the EnLC represents $-\log_2 2c_-$ at negative values, where $E_{\cal N}=0$.
In the lower left plot, both of the two detectors are initially in their ground states.
One can see that $E_N$ has a beat frequency $2\Delta$, so does
the linear entropy $S_L$ of detector $A$ in the right plots.}
\label{2UDEntDyn}
\end{figure}

\subsubsection{Entanglement dynamics}

Some examples of the entanglement dynamics for identical detectors $A$ and $B$ located at $x=0$ and $L/2$, respectively,
in the untwisted field are illustrated in Figure \ref{2UDEntDyn}.
The degree of entanglement between the two detectors is characterized by the logarithmic negativity \cite{VW02, Pl05} $E_{\cal N} = \max
\{ 0,-\log_2 2c_- \}$ evaluated around the future light cone (EnLC) of detector $A$, which is essential in quantum teleportation from
detector $A$ to detector $B$ \cite{LCH15}: the value of the $E_{\cal N}$ of the EnLC is a monotonic function of the upper bound of the optimal fidelity of coherent-state teleportation \cite{MV09}. Here $c_-$ is the lowest symplectic eigenvalue of the partially
transposed covariance matrix in the reduced state of the detector pair. The higher UV cutoff $n_{\max}$
is introduced, or the larger initial variance $\langle \hat{\Pi}_{k_0}^2 \rangle$ of the zero mode is given, the lower EnLC will be.

In Figure \ref{2UDEntDyn} (upper-left), one can see that $\left|q_A^A \right|^2$  behaves like
$(1+\cos t \Delta)/2$, and $\left|q_A^B \right|^2$  like $(1-\cos t \Delta)/2$ with the beat frequency $\Delta$
given in (\ref{2UDbeat}).
Both $\left|q_A^A \right|^2$ and $\left|q_A^B \right|^2$ have the beat period $2\pi/\Delta$, which is two times longer than the results
in the one-detector case in Figure 2.
In the other plots of Figure \ref{2UDEntDyn} with different choices of the parameters $\alpha$ and $\beta$ for the initial
two-mode squeezed state of the detector pair, the $E_{\cal N}$ of the EnLC of detector $A$ (black curves) in each case shows the same
beating behavior of the mode function squared $|q_{\bf d}^{\bf d'}|^2$, so do the linear entropy $S_L$ of detector $A$.
One can see entanglement creation, sudden-death and revival of the detector pair in the lower plots from the
interplay with the field.

The $E_{\cal N}$ evaluated on the hypersurfaces of simultaneity (constant $t$-slices) (EnS) of detector $A$ (gray dashed curves)
has a similar beating behavior, but their extrema lag behind the EnLC's in the clock of detector $A$, so does the
linear entropy $S_L$ of detector $A$, which is a measure of entanglement between a single detector and the rest of the
combined system (including the field in addition to the other detector).
In the right plots, it appears that $S_L$ of detector $A$
increases as the $E_{\cal N}$ of the ENLC does. However, such a tendency is not clear in the lower-left plot, where detectors $A$ and $B$ are initially in a separable product state of their ground states.

In the right plots we further introduce the linear entropy of the detector pair, $S_L^{(2)} = 1-{\rm Tr} (\rho^{\rm R}_{AB})^2 =
1 - \hbar^2/(4\sqrt{\det{V}})$, where $\rho^{\rm R}_{AB}$ is the reduced state of the detector pair $A$ and $B$, and $V$ is
the covariance matrix of the detector pair. This is a measure of the entanglement between the detector pair and the field, combined
as a bipartite system. Similar to $E_{\cal N}$, the value of $S_L^{(2)}$ can be taken either on the future light cones (blue) or on the
hypersurfaces of simultaneity (green dashed curves) of detector $A$. The extrema of the latter also lag behind the ones of the
former. We find that either way $S_L^{(2)}$ does not follow the largest beats of $E_{\cal N}$. This indicates that the
increased entanglement between detectors $A$ and $B$ are mainly  influenced by the nonlocal correlations in the field, rather than
the correlations between the detectors and the field.

\subsection{Twisted field}
\label{EnABTw}

In the twisted field, $q_{A}^\mu$ and $q_B^\mu$ will not influence each other if detectors $A$ and $B$ are located exactly at
$x=0$ and $x=L/2$ in ${\bf S}^1$. This is because in detector $A$'s point of view, the retarded field from detector $B$ or it's image at
$x=(n'-(1/2))L$, $n'=1,2,3,\ldots$, in the extended coordinates carries a $\varepsilon^{n'}$ factor, but the field from $x=(-n'+(1/2))L$ carries an $\varepsilon^{n'+1}$ factor and cancel the former when $\varepsilon=-1$ (note that $L/2$ is in the domain $(-L/2, L/2]$ for $x$
in the restricted coordinates, but $-L/2$ is not). So the situation of our setup for the twisted field looks rather simple: $q_A^B=q_B^A=0$
for all time, and the equations of motion (\ref{EOMqaA0}) reduce to Eqs. (\ref{EOMqp1}) and (\ref{EOMqp2}) for single detectors with the
subscript ${\bf d}=A$ generalized to ${\bf d}=A, B$ while $\phi_{\;0}^{^{[0]}k_n}(t)$ for ${\bf d}=A$ in Eq.(\ref{EOMqp1}) or
(\ref{EOMqp2}) is replaced by $\phi_{L/2}^{^{[0]}k_n}(t)$ when ${\bf d}=B$.
Their eigen-frequencies are thus identical to those in Section \ref{1DetTwFreq} with the same parameter values.

\subsubsection{Vanishing v-parts of the cross correlators}
\label{VCCorr}

The quantum correlation initially in the twisted field will never be converted to the detector-detector entanglement since
the v-parts of the cross correlators always vanish here. Indeed, writing $\omega'_{n} = (n-1/2)/R$, $n=1,2,3,\ldots$,
so that $k'_n = \omega'_{n} $ for $k'_n >0$, and $k'_{n'}= -\omega'_{n'+1}$ when $k'_{n'}<0$, $n'=0,1,2,\ldots$.
From Eq.(\ref{EOMqp2}), one has $q_{A}^{\omega'_{n}}(t) = q_{A}^{-\omega'_n}(t)$ and $p_{A}^{\omega'_n}(t)= p_{A}^{-\omega'_n}(t)$
since $\partial^{}_t \phi_{\;0}^{^{[0]}\omega'_n}(t) = -i\omega'_n e^{-i \omega'_n t} = \partial^{}_t \phi_{\;0}^{^{[0]}-\omega'_n}(t)$.
Similarly, since $\partial^{}_t \phi_{L/2}^{^{[0]}\omega'_n}(t) =  e^{i \omega'_n (L/2)} \partial^{}_t \phi_{\;0}^{^{[0]}\omega'_n}(t) =
-i (-1)^n\partial^{}_t \phi_{\;0}^{^{[0]}\omega'_n}(t) = -\partial^{}_t \phi_{L/2}^{^{[0]}-\omega'_n}(t)$,  one can see that
$q_{B}^{\omega'_n}(t) = i q_{A}^{\omega'_n}(t) =-q_B^{-\omega'_n}(t)$ and $p_{B}^{\omega'_n}(t) = i p_{A}^{\omega'_n}(t)=
-p_B^{-\omega'_n}(t)$. These imply
\begin{eqnarray}
  \langle \hat{\cal R}^{}_A(t) \hat{\cal R'}^{}_B(t') \rangle^{}_{\rm v} &=&
	  \sum_{n\in {\bf Z}} {\hbar\over 2|k'_n|}{\rm Re} \left\{ \left[\varrho^{k'_n}_A(t)\right]^* \varrho'^{k'_n}_B(t') \right\}
		\nonumber\\ &=&
	\sum_{n >0 } {\hbar\over 2 \omega'_n}{\rm Re} \left\{ \left[\varrho^{\omega'_n}_A(t)\right]^* \varrho'^{\omega'_n}_B(t') +
	    \left[\varrho^{-\omega'_n}_A(t)\right]^* \varrho'^{-\omega'_n}_B(t')\right\} \nonumber\\ &=&
	\sum_{n >0 } {\hbar\over 2 \omega'_n}{\rm Re} \left\{ \left[\varrho^{\omega'_n}_A(t)\right]^* \left(\varrho'^{\omega'_n}_B(t') +
	     \varrho'^{-\omega'_n}_B(t')\right)\right\} =0,
\end{eqnarray}
where $({\cal R}, \varrho), ({\cal R'},\varrho') = (Q,q)$ or $(P,p)$.

\begin{figure}
\includegraphics[width=7.2cm]{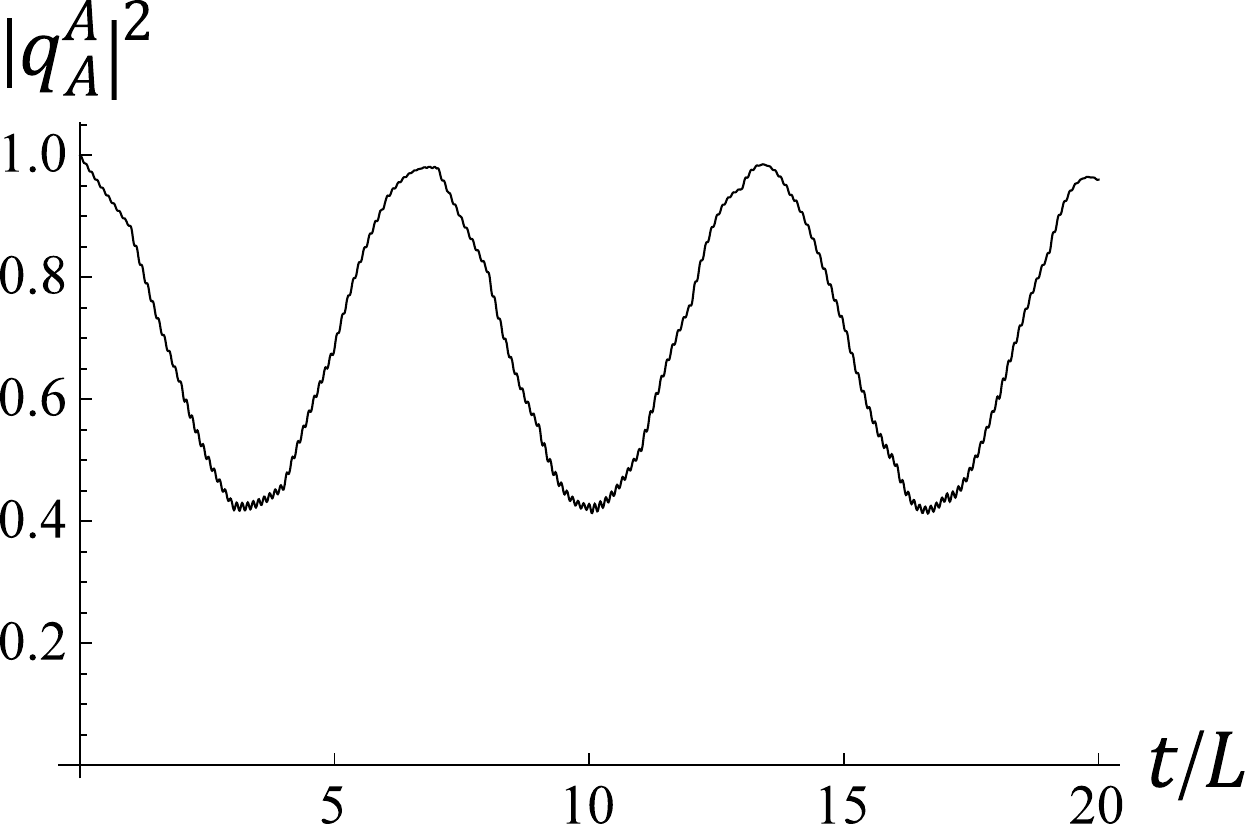}
\includegraphics[width=7.2cm]{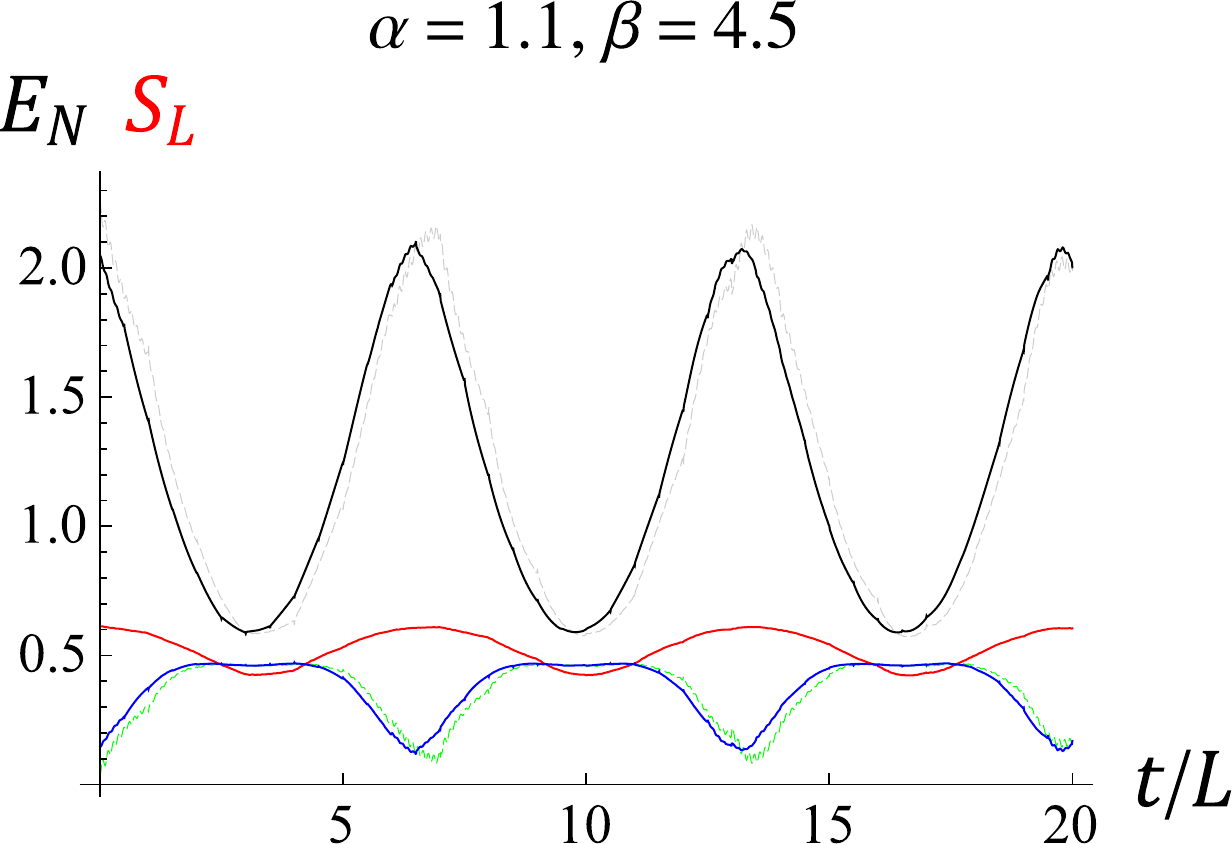}\\
\includegraphics[width=7.2cm]{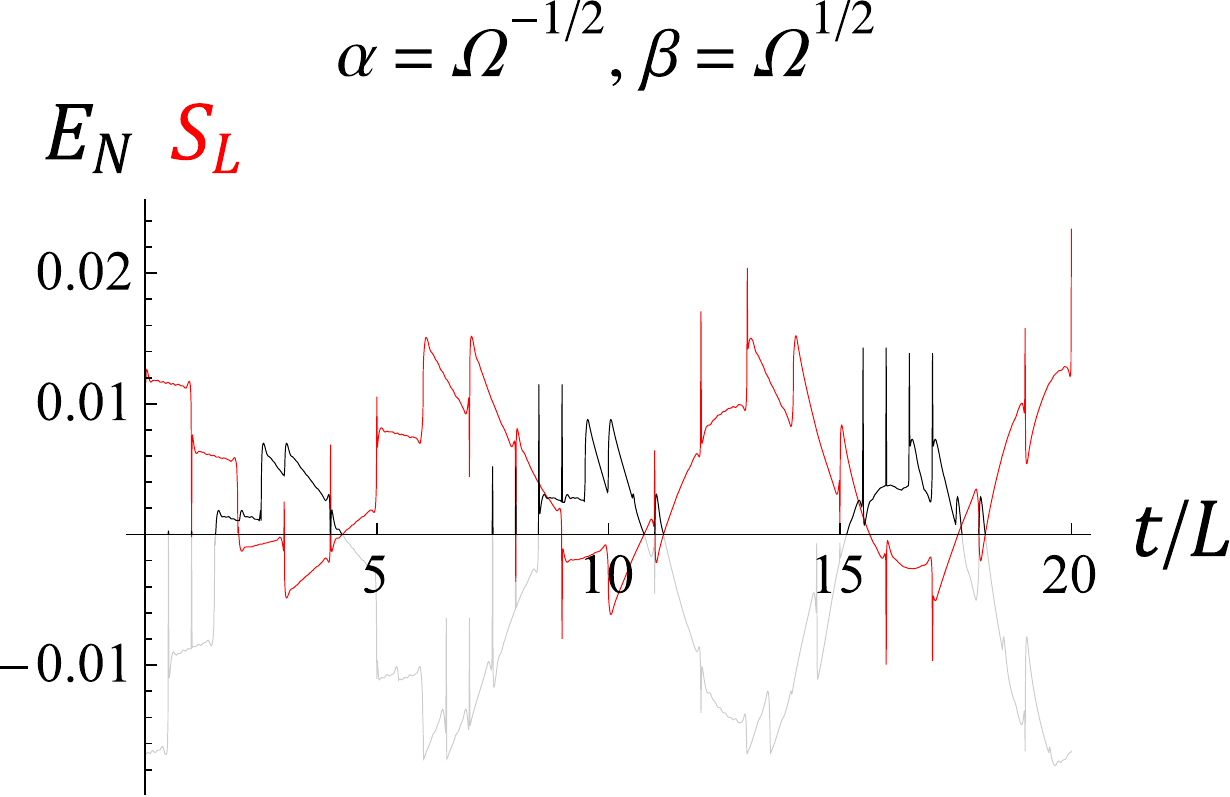}
\includegraphics[width=7.2cm]{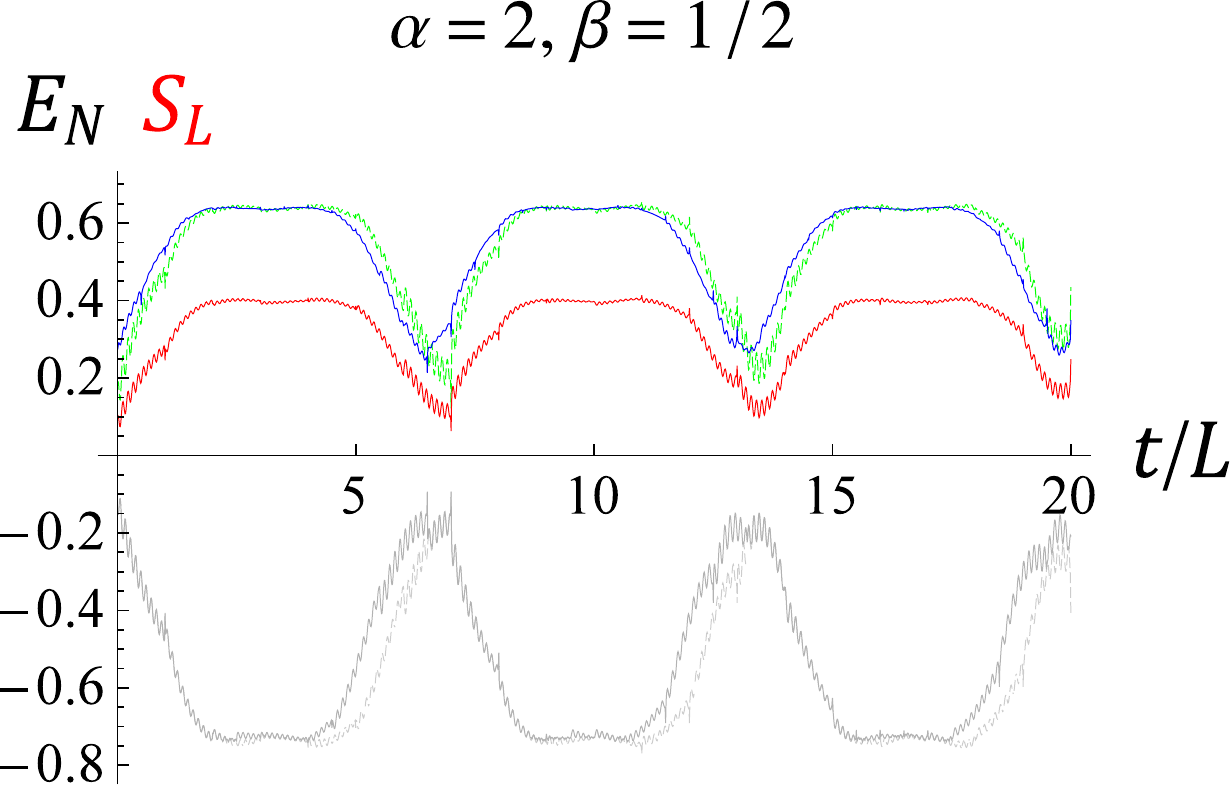}
\caption{
Entanglement dynamics of two UD' detectors in the twisted field, contributed by the field-modes with $k'_n$, $|n| = 1,
\ldots, 1000$. All the parameters have the same values as those in Figure \ref{2UDEntDyn} except $\varepsilon=-1$.
(Upper-Left) The evolution of $\left|q_A^A \right|^2$. ($\left|p_A^A \right|^2 \approx {\Omega^2}\left|q_A^A \right|^2$ with this weak
coupling.) The beat frequency is about $0.0747$, %029$,
which is the frequency difference of the two dominant eigen-modes obtained from (\ref{EFcondTw}).
(Upper-right and lower row) The curves represent the quantities similar to those in Figure \ref{2UDEntDyn}.
One can see that $E_{\cal N}$ of the EnLC (black) has the same beating behavior as the mode functions.
Note that the entanglement creation in the lower-left plot for the case with the initial state as the ground states of the
detectors are not reliable since the linear entropy $S^{}_L$ of detector $A$ (red) is negative whenever the $E_{\cal N}$ of the EnLC
(black) is positive.}
\label{2UDEntTw}
\end{figure}

\subsubsection{Entanglement dynamics}

Some examples for the dynamics of the EnLC between the two detectors in our setup in the twisted field are shown in Figure \ref{2UDEntTw}
with the same values of the parameters as those in Figure \ref{2UDEntDyn} except $\varepsilon=-1$.
One can see that the entanglement dynamics in this setup have the same beating behavior as $|q_A^A|^2$ rather than the
combination of $|q_A^A|^2$ and $|q_A^B|^2$ for the untwisted field.
The beat frequency is the frequency difference between the dominant eigen-modes of $|q_A^A|^2$($=|q_B^B|^2$)
around $\Omega$, which can be obtained from (\ref{EFcondTw}).

The linear entropy $S_L$ of detector $A$ in the upper-right plot of Figure \ref{2UDEntTw} shows the same tendency of
increase or decrease as  $E_{\cal N}$ of the EnLC, very much like the case with the untwisted field (the upper-right of Figure
\ref{2UDEntDyn}). However, in the lower row of Figure \ref{2UDEntTw} the tendency is opposite: $S_L$ increases as the value of
$E_{\cal N}$ or $-\log_2 2c_-$ decreases. If we vary the value of $\alpha$ and $\beta$ continuously, there will be a continuous crossover
between the former and the latter behaviors, where the $S_L$ curvew have valleys both around the peaks and the valleys of $E_{\cal N}$.

Unlike the case with the untwisted field in Figure \ref{2UDEntDyn}, however, here the EnLC and the $S_L^{(2)}$ around the future light cones 
of detector $A$ have opposite behaviors, so do their counterparts evaluated on the hypersurfaces of simultaneity. 
%the EnS and the linear entropy $S_L$ of detector $A$. 
(Again, the extrema of the $E_{\cal N}$ of the EnS and the $S_L^{(2)}$ on the hypersurfaces of simultaneity lag behind those 
%of the EnLC and $S_L^{(2)}$ 
evaluated around the future light cones in detector $A$'s clock.)
This shows a trade-off between the detector-detector entanglement and the field-detector-pair entanglement, due to the lack of
converting the initial nonlocal field-field correlations to the detector-detector correlations in our setup, as we showed earlier
in Section \ref{VCCorr}.

Note that the entanglement creation in the lower-left plot in the case with the initial state as a separable product of the
ground states of the detectors are not reliable here since the linear entropy $S^{}_L$ of detector $A$ is negative whenever the
$E_{\cal N}$ of the EnLC become positive. By extrapolation of $n_{\rm max}$ from our data, we expect that these EnLC will vanish if
we include enough field modes to make $S^{}_L$ of detector $A$ positive at all times (for $n_{\rm max} \sim O(10^4)$, one may have
$S_L>0$ almost at all times except some moments around $t=nL$ when the resonance occurs). This would be another difference from the
cases with the untwisted field as in the lower-left plot of Figure \ref{2UDEntDyn}, where the detector-detector entanglement can be
created by the interplay with the field even if the detector pair was started with the same initial states.

%%%%%%%%%%%%%%%%%%%%%%%%
\section{Summary and Discussion}

\subsection{Twisted field, untwisted field, and zero mode}

The Einstein cylinder, with the topology ${\bf S}^1\times {\bf R}_1$, possesses two inequivalent configuration spaces for a real scalar
field $\Phi_x(t)$ \cite{Is78, DHI79}. The normal (untwisted) field satisfies the periodic boundary condition $\Phi_x(t) =
\Phi_{x+L}(t)$ where $L$ is the circumference of ${\bf S}^1$, and the twisted field satisfies the anti-periodic boundary
condition $\Phi_x(t) = -\Phi_{x+L}(t)$. 

The untwisted massless scalar field contains a zero mode (with $\omega_0 = |k_0| = 0$), which is constant in space \cite{ML14}.
It cannot be simply excluded, otherwise the retarded Green's function of the field will violate causality.
Due to its singular normalization, however, one has to deal with the zero mode separately from other modes. In terms of the
kinetic momenta $P_{\bf d}=\partial_t Q_{\bf d}$ associated with the interaction action $S_I$ in (\ref{SI1}), the effect of the zero mode on the dynamics is similar to those from other modes. If we choose $S'_I$ in (\ref{SI2}) instead, which yields the canonical momenta
$P'_{\bf d}=\partial_t Q_{\bf d} + \lambda \Phi_{z^1_{\bf d}}$, the zero mode turns out to make the two-point correlator of the field
amplitudes and the correlators of the canonical momenta of the UD' detectors grow indefinitely, while the energy of the combined
detector-field system remains constant, and the energy of each subsystem remains bounded. The ill behavior in the theory with $S'_I$ is
simply an illusion when describing the system in terms of the gauge-dependent variables $P'_{\bf d}$.

The dynamics of the detectors in the twisted field is simpler. There is no zero mode, and the correlators of the detectors never grow
indefinitely. Moreover, in the two-detector case, the detector pair in the twisted field will not influence each other if one
detector is located at $x$ and the other is located exactly at ($x + L/2$ mod $L$) in ${\bf S}^1$. In this setup the detector-detector
mode functions $q_{\bf d}^{\bf d'}$ are equivalent to those in the one-detector case, and the quantum correlation initially in the
field will never be converted to the detector-detector entanglement since the v-parts of the cross correlators of the detectors
always vanish here.

\subsection{Eigen-frequencies and beats}

While the evolution curve of a mode function at early times looks quite similar to the curve at late times
at a time scale with only few orders of echoes (see Figure \ref{EmergeEigen}),
as the time scale of observation increases, the eigen-frequencies emerge in the frequency spectrum and the peaks get sharper.

The discrete eigen-frequencies of the detector-detector mode functions $q_{\bf d}^{\bf d'}$ are the solutions of Eqs. (\ref{EFcond}),
(\ref{EFcondTw}), and Eqs.(\ref{EFp}-\ref{EFm}) in the one-detector and two-detector cases, respectively. In the weak coupling limit,
the eigen-frequencies are close to but not exactly the same as the natural frequencies of the free detectors and the free field modes.
The eigen-frequencies of the detector-field mode functions $q_{\bf d}^{k_n}$ are exactly the same as $q_{\bf d}^{\bf d'}$'s,
though the fluctuations of the field working as driving forces in the equations of motion have various frequencies.

In the one-detector case, when the natural frequency of the free detector $\Omega_0$ comes close to the frequency of a free field mode,
they mix and split into two dominant eigen-frequencies, a phenomenon similar to the anti-crossing of energy levels in atomic systems.
These two neighboring eigen-modes produce the beats in the evolution of a mode function. In the weak coupling limit, the beat frequency
is determined by the separation of these two neighboring eigen-frequencies and proportional to the square root of the coupling
strength $\gamma$. The beats enjoy the largest time scale $O(\gamma^{-1/2})$ in the evolution of the mode functions in our compact space
${\bf S}^1$, which is quite different from the largest time scale $O(\gamma^{-1})$ for the detectors in Minkowski space.

When $\Omega_0$ is not close to any frequency of the field modes, there may not be significant beats. However, if $\Omega_0$ is located around, but not exactly at the middle point of two free field mode in the spectrum, these frequencies may mix and split into three dominant eigen-modes, which may produce the beats at a very large time scale. When this happens the beat frequency will be about the
difference of the frequency difference between the higher and middle eigen-modes, and the frequency difference between the middle and the
lower eigen-modes.

In the cases with two identical UD' detectors, there will be at least two dominant eigen-frequencies and so the mode functions always
have significant beats, even when $\Omega_0$ is not close to any frequency of the field modes. The beat frequency ranges from $2\gamma$
to $\sqrt{\gamma/\pi}$ in the weak coupling limit.

The beating behavior of the mode functions in turn affects the behavior of the correlators. In the one-detector case,
the detector-field entanglement characterized by the value of the linear entropy, which is a function of the two-point correlators
for the Gaussian states, gets the same beating features. This may be interpreted as periodic recoherence.
In the two-detector case, entanglement dynamics between the two detectors also exhibit the same beating feature as in the mode functions.
With a suitable choice of the parameter values one can observe a sequence of sudden-death and revival of quantum entanglement
with a (beat) frequency one (twisted field) or two times (untwisted) of the beat frequency of the detector-detector mode functions.
The beats at large time scale and the resonances around $t=nL$ distinguish the dynamics of a detector-field system in
${\bf S}^1\times{\bf R}_1$ from the dissipative behavior of the same system living in Minkowski space ${\bf R}^1_1$.

\subsection{Discreteness of field spectrum}

Around the source point, the retarded Green's function of a field in Minkowski space looks the same as the one in any locally flat
spacetime of the same dimension. Thus in a classical theory, if we couple a source such as a detector with the field in a very short
duration, the classical physics around the interaction region in two different locally-flat spacetimes will be identical. In a quantum
theory, however, different field spectra give different Wightman functions and produce different vacuum fluctuations. In particular,
the field spectrum in a non-compact space is continuous, while the one in a compact space or in a cavity is discrete. These differences originating from nontrivial topology or the existence of boundaries can affect the dynamics even for a single detector
\cite{La05, ZBLH13, Edu15}. A comparison of our early-time results for a single detector in the Einstein cylinder with the ones in
Minkowski space is given in Appendix \ref{2ptM2}, where one can see that, while the evolution of the linear entropy of a detector in the
twisted or untwisted field in the Einstein cylinder looks roughly the same as the one in Minkowski space at very early times,
in most of the cases studied the difference becomes significant as the time scale becomes comparable to the crossing time of the compact
spatial dimension for the retarded field.  In those cases where the natural frequency of the detector is in resonance with the frequency
of the untwisted field, this deviation emerges even earlier, as the evolution curve oscillates about the curve for the Minkowski case
with a finite amplitude.

The discreteness of the field spectrum affects not only the self correlators of a single detector, corresponding to the
detector-field entanglement, but also the cross correlators, which is important in the detector-detector entanglement.
Our nonperturbative results indicate that early-time evolutions of quantum entanglement between two detectors in ${\bf S}^1\times
{\bf R}_1$ and in ${\bf R}^1_1$ is small. This is the range where perturbation theory results is expected to hold, as shown in Ref.
\cite{Edu15} in (3+1) dimensions.
In our setup the discrepancy between the early-time results for the EnLC of a detector pair in 1) the twisted and 2) untwisted fields
is more pronounced than the discrepancy between the linear entropies $S_L$ of each single detector in these two fields.
This may be due to the big contrast that the v-parts of the cross correlators of the detectors vanish in the twisted field but
not in the untwisted field.

We end with a comment on the validity of time-dependent perturbation theory applied to entanglement problems in spacetimes with nontrivial topology and boundaries, such as studied here. In Minkowski space, TDPT with finite duration could be valid for the interaction time up to 
$O(1/\gamma)$. In ${\bf S}^1\times {\bf R}_1$, by counting the exponent of the coupling, we can see that the effect of echoes comes from 
higher-order contributions (e.g., the $\lambda^2$ terms in Eq. (\ref{EOMqp1})). Nevertheless, our results indicate that even the first echo 
can have significant effect during $L < t < 2L$. So we know that as early as $t \approx L$ when the first echo returns, the lowest order perturbation result may become unreliable (Figure \ref{CompareTDPT}). One needs to be careful, however, when comparing with the results of 
\cite{Edu15} which uses perturbation theory in (3+1) dimensions with only one spatial dimension being compact. The absence of echoes in the 
results \cite{Edu15} could be an intrinsic limitation of perturbation theory with a finite width of the Gaussian switching function, or that 
the echo effect is diluted by the other two non-compact dimensions. 
These points are worthy of further investigations by practitioners of perturbative switching methods.

\begin{acknowledgments}
%\vspace{1cm}
%\noindent {\bf Acknowledgments}
We thank Rong Zhou for discussions in the initial stage of this work. SYL thanks Feng-Li Lin and Eduardo Martin-Martinez for helpful
discussions. CHC thanks the support from National Center for Theoretical Sciences (South), Taiwan and Center for Theoretical Sciences,
National Cheng Kung University. BLH thanks the hospitality of the theory group of the Institute of Physics at the Academia Sinica, Taiwan
during his visit in the Spring 2013 where part of this work was carried out, and the Center for Field Theory and Particle Physics at Fudan University, Shanghai, China in the summer of 2015 when it was consummated.
This work is supported by the Ministry of Science and Technology of Taiwan (MOST) under Grants No. 102-2112-M-018-005-MY3, No.
103-2918-I-018-004 and No. 104-2112-M-006-015, and in part by the National Center for Theoretical Sciences, Taiwan.
\end{acknowledgments}

%\begin{appendix}
\appendix

%%%%%%%%%%%%%%%%%%%%%%%%%%%
\section{Detector in (1+1)D Minkowski space}
\label{2ptM2}

For a single UD' detector at rest in (1+1)D Minkowski space, described by the action (\ref{theaction}) and (\ref{SI1}), initially in its
ground state and coupled with the Minkowski vacuum of a massless scalar field at $t=0$, the two-point correlators of the detector read
\begin{equation}
  \langle \hat{Q}^2(t) \rangle = \langle \hat{Q}^2(t) \rangle^{}_{\rm a} + \langle \hat{Q}^2(t) \rangle^{}_{\rm v},
	\hspace{1cm}
  \langle \hat{P}^2(t) \rangle = \langle \hat{P}^2(t) \rangle^{}_{\rm a} + \langle \hat{P}^2(t) \rangle^{}_{\rm v}
\end{equation}
with
\begin{eqnarray}
  \langle \hat{Q}^2(t) \rangle^{}_{\rm a} &=& {\hbar e^{-2\gamma t}\over 2\Omega_0 \Omega^2}\left[ \Omega_0^2 +
      \gamma \left( \Omega \sin 2 \Omega t -\gamma \cos 2 \Omega t \right)\right], \label{Q2M2} \\
  \langle \hat{Q}^2(t) \rangle^{}_{\rm v} &=& {2\hbar\gamma\over \pi \Omega^2}\left\{
	    {\cal I}^{}_1 \left[ \Omega^2 + e^{-2\gamma t}(\Omega\cos\Omega t + \gamma\sin\Omega t)^2\right] +
			{\cal I}^{}_3 e^{-2\gamma t}\sin^2 \Omega t \right. \nonumber\\ & & \hspace{1cm}\left.
			-2 \Omega e^{-\gamma t}\left[ {\cal S}^{}_2(t)\sin\Omega t +
			    {\cal C}^{}_1(t) (\Omega\cos\Omega t+\gamma\sin\Omega t)\right] \right\},\\
  \langle \hat{P}^2(t) \rangle^{}_{\rm a} &=& {\hbar \Omega_0\over 2\Omega^2}e^{-2\gamma t}\left[ \Omega_0^2 -
      \gamma \left( \Omega \sin 2 \Omega t +\gamma \cos 2 \Omega t \right)\right],\\
  \langle \hat{P}^2(t) \rangle^{}_{\rm v} &=& {2\hbar\gamma\over \pi \Omega^2}\left\{
	    {\cal I}^{}_3 \left[\Omega^2 +  e^{-2\gamma t} (\Omega\cos\Omega t -\gamma\sin\Omega t)^2\right] +
			{\cal I}^{}_1 e^{-2\gamma t}\Omega_0^4  \sin^2 \Omega t \right. \nonumber\\ & & \hspace{1cm}\left.
			-2 \Omega e^{-\gamma t}\left[ {\cal S}^{}_2(t)\Omega_0^2\sin\Omega t +
			    {\cal C}^{}_3(t) (\Omega\cos\Omega t-\gamma\sin\Omega t)\right] \right\},
	 \label{P2vM2}		
\end{eqnarray}
and $\langle \hat{Q}(t), \hat{P}(t)\rangle = \partial_t \langle \hat{Q}^2(t)\rangle/2$. Here $\gamma \equiv \lambda^2/4$,
$\Omega \equiv\sqrt{\Omega_0^2 -\gamma^2}$, and
\begin{eqnarray}
  {\cal C}^{}_{2n+1}(t) &\equiv& \int_0^{\omega^{}_M} {\omega^{2n+1}\cos\omega t d\omega\over\left| (\gamma +i\omega)^2+
	    \Omega^2\right|^2}\nonumber\\ &\stackrel{\omega^{}_M\to\infty}{\longrightarrow}&
		{\rm Re} {i(\Omega+i\gamma)^{2n}\over 4\gamma\Omega}\left\{ \cos T \left[ {\rm Ci}(-T) + {\rm Ci}(T) \right]
		   +2 \sin T \, {\rm Si}(T) \right\},\label{C2n1}\\
	{\cal S}^{}_{2n}(t) &\equiv&\int_0^{\omega^{}_M} {\omega^{2n} \sin \omega t d\omega\over \left| (\gamma + i\omega)^2+\Omega^2\right|^2}
	  \nonumber\\ &\stackrel{\omega^{}_M\to\infty}{\longrightarrow}&
		 {\rm Re} {i(\Omega +i\gamma)^{2n-1}\over 4\gamma\Omega}\left\{ \sin T \left[ {\rm Ci}(-T) + {\rm Ci}(T) \right]
		   -2 \cos T \, {\rm Si}(T) \right\},	\\
	{\cal I}^{}_1 &\equiv& \int_0^{\omega^{}_M} {\omega^{} d\omega\over \left| (\gamma + i\omega)^2+\Omega^2\right|^2} \nonumber\\
	    &\stackrel{\omega^{}_M\to\infty}{\longrightarrow}&
	    {1\over 4\gamma\Omega}\left[ \tan^{-1} {\Omega^2-\gamma^2\over 2\gamma\Omega} + {\pi\over 2}\right]
			= {i\over 4\gamma\Omega}\ln{\gamma- i\Omega \over \gamma+i\Omega},\\
	{\cal I}^{}_3 &\equiv& \int_0^{\omega^{}_M} {\omega^3  d\omega\over \left| (\gamma + i\omega)^2+\Omega^2\right|^2} \approx
	    (\Omega^2-\gamma^2){\cal I}^{}_1 + \ln{\omega^{}_M\over \Omega_0}
			\label{I3def}
\end{eqnarray}
for $\omega^{}_M\gg \Omega$,
with $n=0,1,2,\ldots$, $T\equiv (\Omega+i\gamma)t$, the sine (cosine) integral function Si (Ci), the Euler's constant $\gamma_e$, and
the UV cutoff $\omega^{}_M$. %\gg \Omega$.
All the above two-point correlators of the detector depend on ${\cal I}^{}_3$, and so all explicitly depend on the UV cutoff.
At late times ($t \gg 1/\gamma$), one has $\langle \hat{Q}, \hat{P}\rangle \to 0$ and $\langle \hat{Q}^2 \rangle\to 2\hbar\gamma
{\cal I}^{}_1/\pi$, but $\langle \hat{P}^2 \rangle \to 2\hbar\gamma {\cal I}^{}_3/\pi$ is still cutoff dependent significantly.

The above results are actually identical to the two-point correlators of a UD detector in (3+1) dimensional Minkowski space \cite{LH07}.
%(this was checked using Math 10).
To see this, one may insert $\omega^{}_M = 2\pi\Omega e^{\Lambda_1 + \gamma_e}$ to
the ${\cal I}^{}_3 \Omega^2$ term in $\langle\hat{P}^2(t) \rangle^{}_{\rm v}$, while substitute
$\omega^{}_M = 2\pi\Omega e^{\Lambda_0+\gamma_e}$
to other ${\cal I}^{}_3$'s. Thus we can borrow the interpretation from Ref. \cite{LH07} that
the constant $\Lambda_1$ corresponds to the time-resolution of the detector, and the constant $\Lambda_0$ corresponds to the time scale of
switching-on the interaction at the initial moment.

We have also found that, for the interaction action $S'_I$ in (\ref{SI2}), the correlator $\langle \hat{P}'^2\rangle$ with physically
non-measurable momentum $P' = \partial_t Q + \lambda \Phi_{x=0}$ is both IR and UV divergent.

\begin{figure}
%\hspace{-.8cm}
\includegraphics[width=4.8cm]{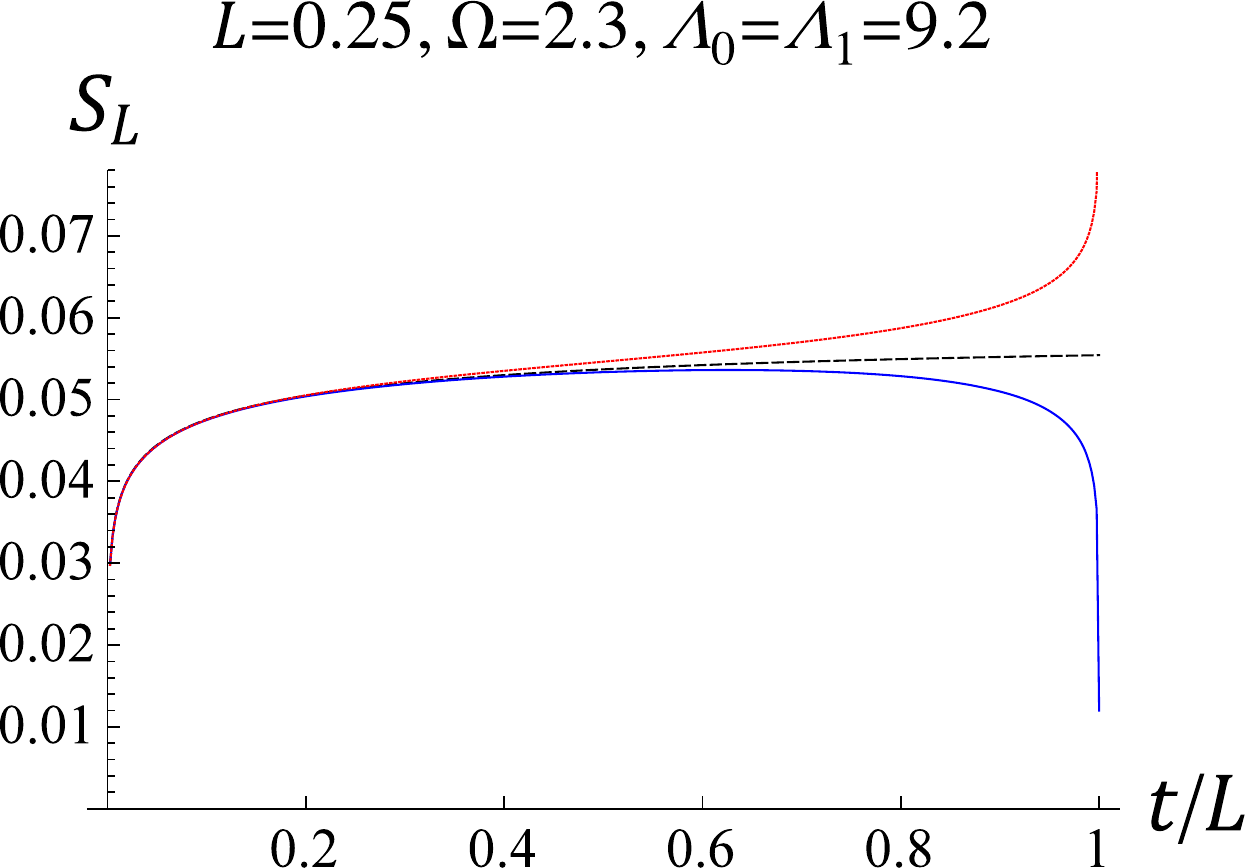}
\includegraphics[width=4.8cm]{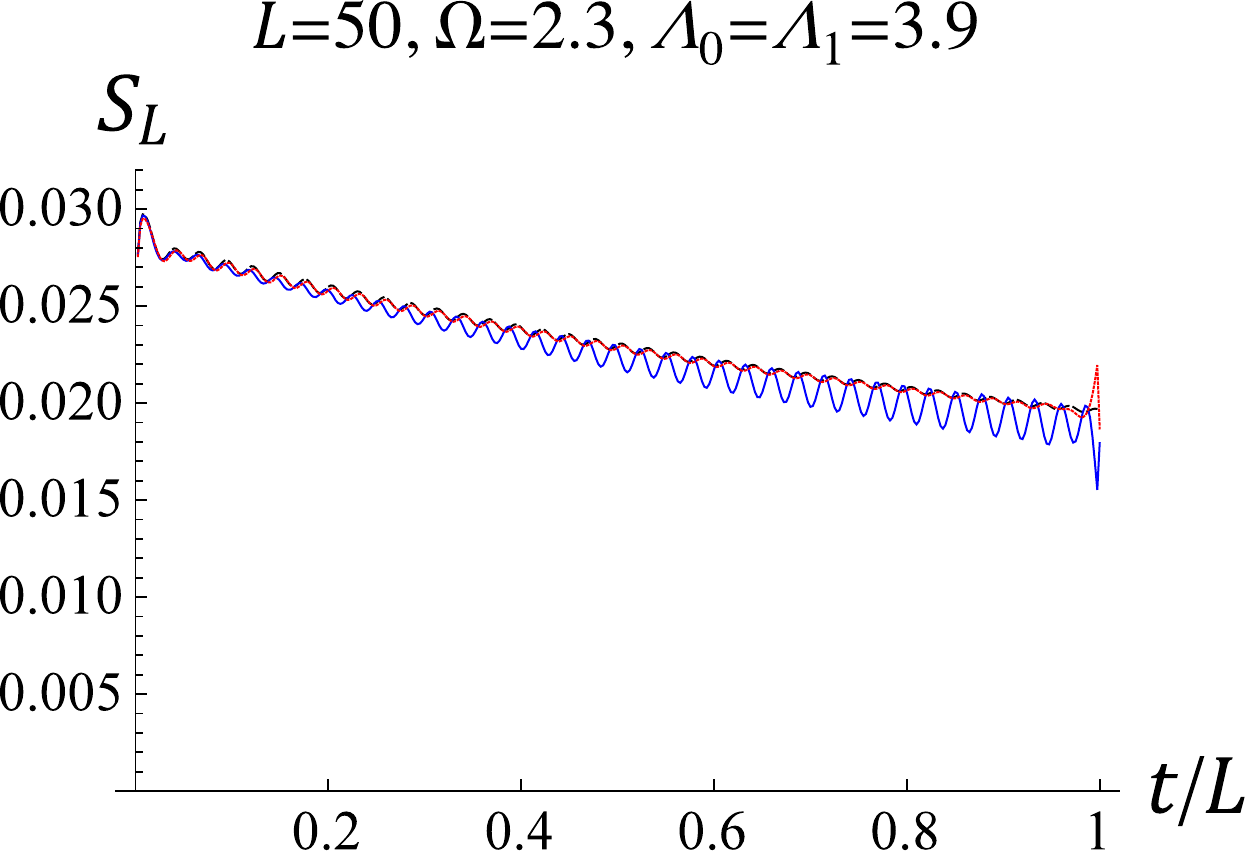}\\
\includegraphics[width=4.8cm]{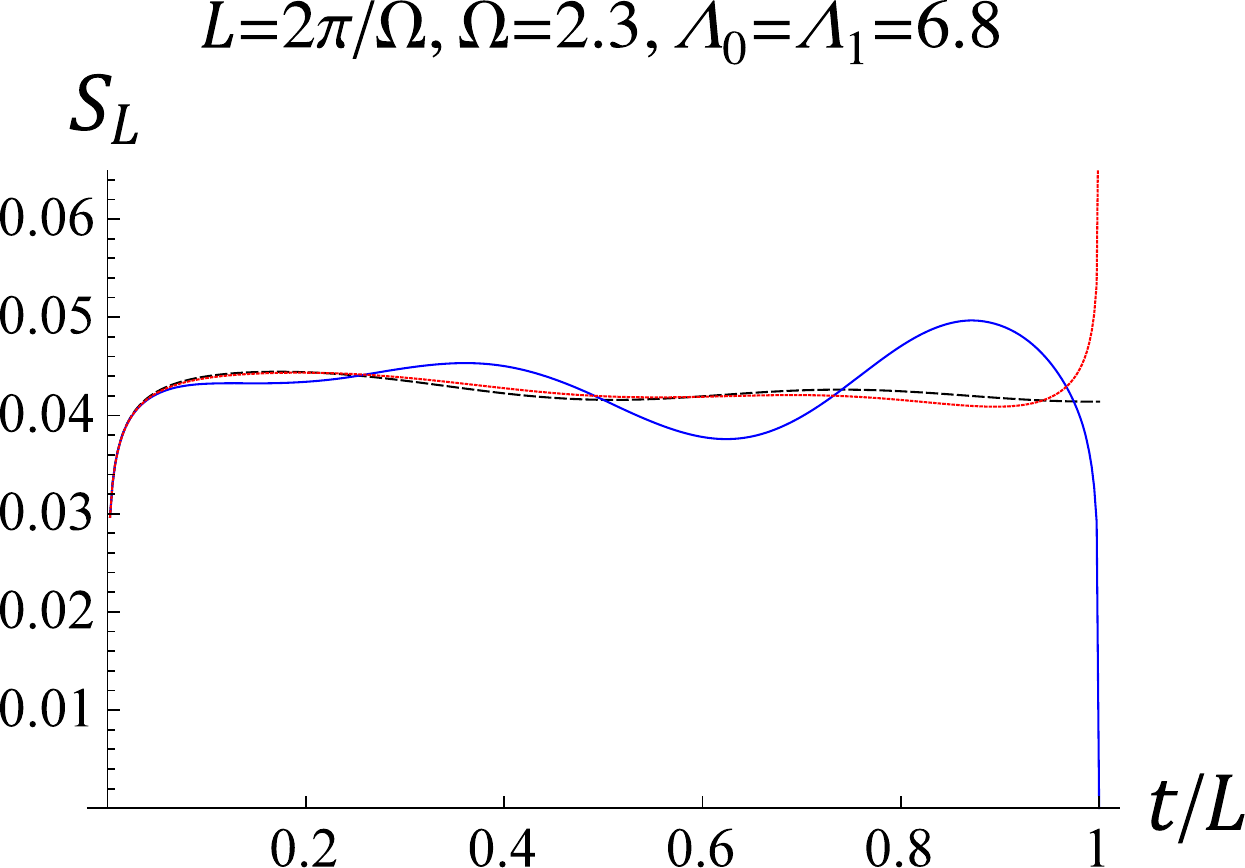}
\includegraphics[width=4.8cm]{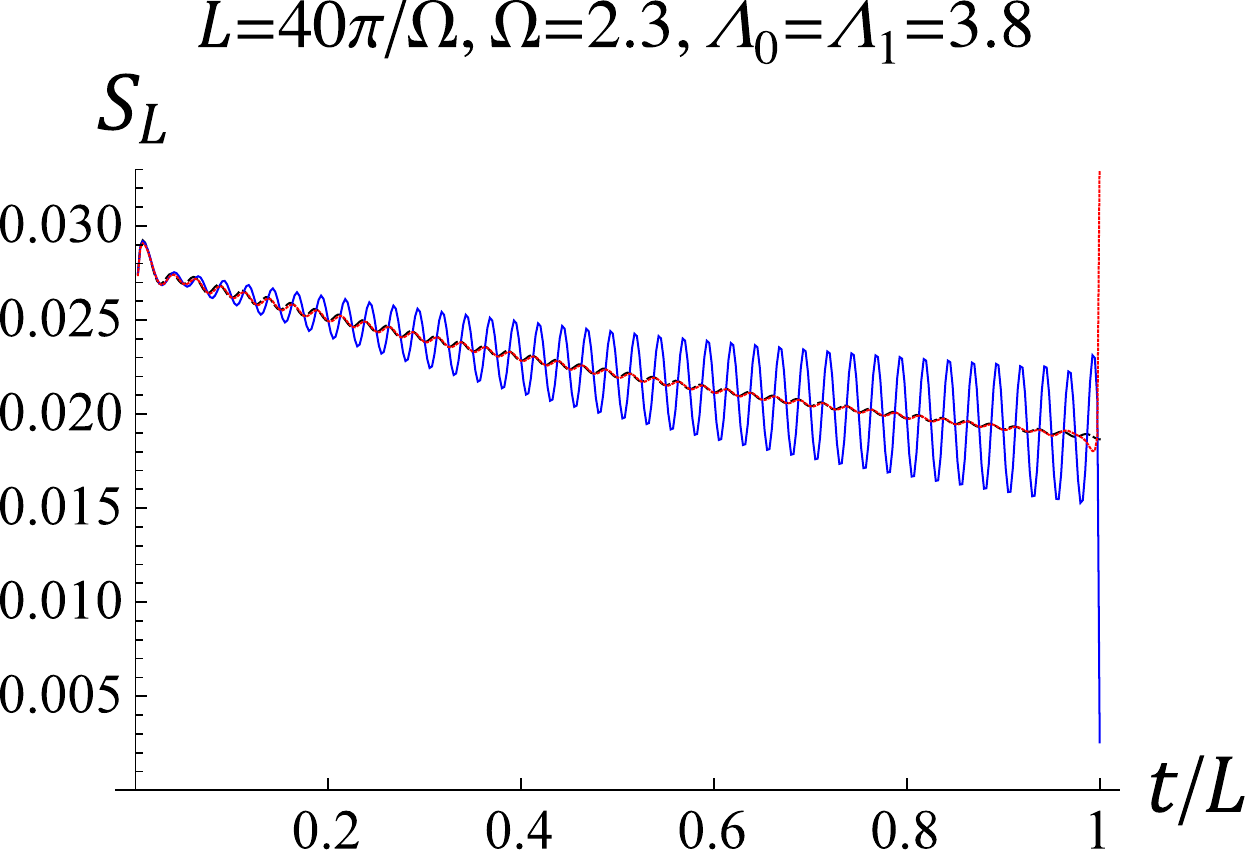}
\includegraphics[width=4.8cm]{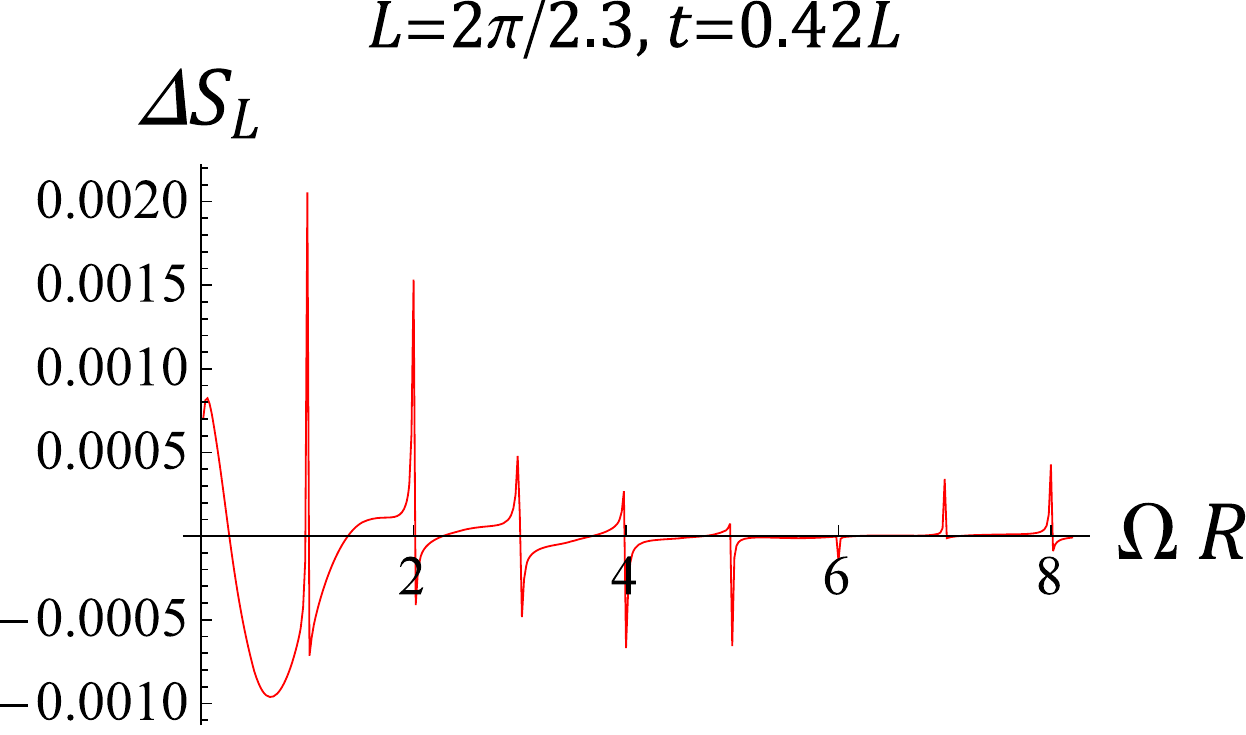}
\caption{Early-time linear entropy $S_L$ of
a UD' detector initially in its ground state and coupled to the untwisted (blue) and twisted (red dotted) fields in
${\bf S}^1\times {\bf R}_1$ at $t=0$, compared with the one in ${\bf R}^1_1$ (black dashed). Here $\gamma=0.01$ and $n_{\rm max}=10000$.
The natural period of the harmonic oscillator is $2\pi/\Omega \approx 2.732$ in these plots.
(Lower right) Frequency dependence of $\Delta S^{}_L \equiv S^{}_L|_{\rm untwisted} - S^{}_L|_{\rm twisted}$, which is the difference of $S^{}_L$ between the linear entropies in the untwisted and the twisted fields at a fixed time $t=0.42L$.
There are spikes around $\Omega = n/R = 2n\pi/L$, $n=1,2,3,\ldots$.}
\label{EarlyTSL}
\end{figure}

In $S^1\times R_1$, the correlators of a single detector at early times before the first echo returns can also be expressed as
Eqs.(\ref{Q2M2})-(\ref{P2vM2}) from (\ref{qAA0}) and (\ref{qAk0}), with the integrals in Eqs.(\ref{C2n1}) to (\ref{I3def}) reduced
to the the Riemann sums of the integrands: $\int_0^{\omega_{\max}} \to \sum_{n=1}^{n_{\max}}$, $\omega \to \omega_n = 2n\pi/L$, $n=1,2,3,
\ldots$, and $d\omega \to \Delta\omega \equiv \omega_{n+1} - \omega_n = 2\pi/L$. In addition, for the cases with the untwisted field, one
needs to include the zero-mode contribution from (\ref{qA00}). While the difference between each integral of (\ref{C2n1})-(\ref{I3def})
in $R^1_1$ and the corresponding discrete sum in $S^1\times R_1$ can be large, most of the differences turn out to cancel in the
correlators. The difference between the correlators with discrete and continuous spectra can be minimized by fine-tuning the parameters
corresponding to the UV cutoff ($\Lambda_0$ and $\Lambda_1$) in the $R^1_1$ case. We find that the correlators and $S_L$ in
$S^1\times R_1$ will not deviate from those in $R^1_1$ significantly until a time scale comparable to $L$.

In Figure \ref{EarlyTSL} we demonstrate some examples of the early-time behavior of the linear entropy $S_L$
for a UD' detector initially in its ground state.
In each plot, the averaged decaying behavior of the three curves at a time scale of $O(L)$ are very similar, while the untwisted field usually gives a greater oscillation at a frequency $\approx 2\Omega$ about a fine-tuned curve for the $R^1_1$ case than the
twisted field does. These oscillations are more significant for large $L$ ($L\gg 2\pi/\Omega$, upper right), and the most significant
on resonance ($\Omega \approx 2\pi n/L$ with integer $n$, lower row). 
Fortunately, the amplitudes of such oscillations are always bounded and will saturate when $t \sim O(1/\gamma)$, provided that
$t$ is still not very close to $L$.
To check this in the on-resonance cases, in the lower-right plot of Figure \ref{EarlyTSL}, we show the frequency dependence of the
$S_L$ curve with the untwisted field subtracted by the one with the twisted field at a fixed time (at each frequency one has some
fine-tuned $S_L$ evolution curves in the ${\bf R}^1_1$ case between the untwisted and twisted results).
Indeed, one can see that, around the resonance peaks, the amplitude of the oscillation at a resonant frequency is not very far from those
at the neighboring frequencies in value.

When approaching $t=L$, in contrast to the curves for the $R^1_1$ cases, the curves for the detectors in the untwisted and
twisted fields in ${\bf S}^1\times {\bf R}_1$ get large but opposite resonant behaviors in time, which have the same origin as those
spikes around $t=nL$ in Figures \ref{SL1}-\ref{2UDEntTw}.

The evolution curves for the probability of finding the same detector in the first excited state, $\rho^R_{1,1}$, look very
similar to the ones in Figure \ref{EarlyTSL}. $\rho^R_{1,1}$ here can be interpreted as 
the transition probability from the initial ground state of the detector to the first excited state in TDPT \cite{LH07}.
We find the values of $\rho^R_{1,1}(t)$ in most of the history in $0<t<L$ are about $O(\gamma)$,
indicating that one should be able to see some hints of the non-perturbative $\rho^R_{1,1}(t)$ curves in the leading order ($\sim
\gamma^1$) of the perturbation theory with a finite duration (cf. Figure \ref{CompareTDPT}).

\end{document}